\def\ifpreprint{\iftrue}
\ifpreprint
\documentclass[12pt,onepage,openany]{memoir}
\else
\documentclass[INR,sample,biber]{nowfnt}
\fi
\usepackage[utf8]{inputenc}
\ifpreprint
\usepackage{fullpage}
\usepackage[mono=false]{libertine}

\usepackage[natbib=true,backend=biber,style=authoryear,mergedate=false]{biblatex}
\usepackage{hyperref}
\usepackage{array}
\fi

\usepackage{epsfig}
\usepackage{multirow}
\usepackage{booktabs}
\usepackage[colorinlistoftodos,disable]{todonotes}
\usepackage[framemethod=TikZ]{mdframed}

\usepackage{comment}
\usepackage{graphicx}
\usepackage{amsmath}
\usepackage{amssymb}
\usepackage{mathabx}
\usepackage{mathrsfs}

\usepackage{imakeidx}
\usepackage{cleveref}
\usepackage[notion,quotation,electronic]{knowledge}

\usepackage{color}

\usepackage{enumitem}
\newlist{inlinelist}{enumerate*}{1}
\setlist*[inlinelist,1]{
  label=(\roman*),
}

\newcommand{\Reals}[0]{\mathbb{R}}

\newcommand{\fdvec}[1]{\mathbf{#1}}

\newcommand{\fdmat}[1]{\mathbf{#1}}

\newcommand{\fdset}[1]{\mathcal{#1}}
\newcommand{\indic}{\mathbb{I}}

\newcommand{\fdprobfn}[0]{\text{P}}
\newcommand{\fdprob}[1]{\fdprobfn(#1)}
\newcommand{\fdcprob}[2]{\fdprob{#1 | #2}}
\newcommand{\mdexpectfn}[0]{\operatorname{E}}

\newcommand{\mdexpectover}[2]{\mdexpectfn_{#1}\left[#2\right]}

\newcommand{\repository}[0]{\fdset{D}}
\newcommand{\doc}[0]{d}

\newcommand{\provider}[1]{p_{#1}}
\newcommand{\groups}[0]{\fdset{G}}

\newcommand{\features}[0]{\phi}
\newcommand{\featuresContent}[0]{\features_{c}}
\newcommand{\featuresMetadata}[0]{\features_{m}}
\newcommand{\featuresUsage}[0]{\features_{u}}

\newcommand{\needs}[0]{\fdset{Q}}
\newcommand{\need}[0]{q}

\newcommand{\request}[0]{\rho}
\newcommand{\requestFeatures}[0]{\request_{\features}}
\newcommand{\requestItems}[0]{\request_{\repository}}
\newcommand{\requestLanguage}[0]{\request_{\ell}}

\newcommand{\implicit}[0]{\request}
\newcommand{\implicitGlobal}[0]{\implicit_{\text{global}}}
\newcommand{\implicitLocal}[0]{\implicit_{\text{local}}}

\newcommand{\relfn}[0]{s}
\newcommand{\relevance}[1]{\relfn(#1)}

\newcommand{\policy}[0]{\boldsymbol\pi}
\newcommand{\targetPolicy}[0]{\policy^{\mathrm{target}}}
\newcommand{\reprvec}[0]{\fdvec{x}}
\newcommand{\repr}[1]{\reprvec_{#1}}
\newcommand{\termRepr}[2]{x_{#1,#2}}

\newcommand{\vocab}[0]{\fdset{T}}
\newcommand{\term}[0]{t}
\newcommand{\termMatrix}[0]{\fdmat{X}}

\newcommand{\evaluationMetric}[0]{\mu}
\newcommand{\systemDecision}[0]{\pi}
\newcommand{\itemUtility}[0]{u}
\newcommand{\discountFactor}[0]{\delta}
\newcommand{\patienceParameter}[0]{\gamma}

\newcommand{\integers}[0]{\mathbb{Z}}
\newcommand{\ranks}[0]{\integers^+}

\newcommand{\prefix}[2]{#1_{\le k}}

\newcommand{\distance}[0]{\Delta}
\newcommand{\dist}[0]{P}
\newcommand{\cdist}[0]{F}
\newcommand{\distOfRank}[1]{\dist_{#1}}
\newcommand{\targetDist}[1][\dist]{#1_{\mathrm{target}}}

\newcommand{\EE}[0]{\operatorname{EE}}
\newcommand{\EEL}[0]{\operatorname{EEL}}
\newcommand{\EED}[0]{\operatorname{EED}}

\newcommand{\CAttn}[0]{\mathcal{A}}
\newcommand{\CRel}[0]{\mathcal{R}}

\newcommand{\Iout}{y}
\newcommand{\Iin}{v}
\newcommand{\IinG}{a}
\newcommand{\IinX}{x}

\newcommand{\scorefn}{\psi}
\newcommand{\score}[1]{\scorefn(#1)}
\newcommand{\decfn}{\delta}
\newcommand{\dec}[1]{\decfn(#1)}
\newcommand{\RVout}{Y}
\newcommand{\RVin}{V}
\newcommand{\Iscore}[1]{\score{\Iin_{#1}}}
\newcommand{\Iscorei}{\Iscore{i}}
\newcommand{\Idec}[1]{\dec{\Iin_{#1}}}
\newcommand{\Ideci}{\Idec{i}}
\newcommand{\distin}{\distance_{\mathrm{in}}}
\newcommand{\distdec}{\distance_{\mathrm{dec}}}

\newcommand{\Gprotected}{\blacktriangledown}
\newcommand{\Gmajority}{\blacktriangleup}
\knowledgestyle*{intro notion}{boldface,index style=textbf}
\knowledgedirective{silent}{autoref,up,md,index style=textbf}
\knowledgedefault{index}

\knowledge{notion}
  |  bias
  |  Bias
\knowledge{notion}
  |  discovery
\knowledge{notion}
  |  fairness
  |  Fairness
\knowledge{notion}
  |  information access system
  |  information access systems
  |  Information access system
  |  Information access systems
  |  information access
  |  iasys
\knowledge{notion}
  |  system
  
\knowledge{notion}
  |  consumer
  |  consumers
  |  user
  |  users
\knowledge{notion}
 | consumer fairness
 | Consumer fairness
\knowledge{notion}
 | provider fairness
 | Provider fairness
\knowledge{notion}
 | user model
\knowledge{notion}
 | turn
\knowledge{notion}
  | session
  | sessions
\knowledge{notion}
  |  provider
  |  providers
\knowledge{notion}
  | subject
  | subjects
  | information subjects
  | subject fairness
\knowledge{notion}
  |  evaluation
  |  Evaluation
\knowledge{notion}
  |  repository
  | corpus
\knowledge{notion}
  |  item
  |  items
  |  document
\knowledge{notion,index=representation!item}
  |  item representation
  |  Item representation
  |  Item understanding
\knowledge{notion,index=representation!user}
  |  user representation
  |  User representation
  |  User understanding
\knowledge{notion,index=embedding!user}
  | user embedding
\knowledge{notion,index=embedding!item}
  | item embedding
\knowledge{notion}
  | information need
  | need
\knowledge{notion}
 | query
\knowledge{notion}
  | results
\knowledge{notion}
  | satisfaction
  | satisfy
  | satisfied
\knowledge{notion}
  | ranking
  | rankings
\knowledge{notion}
  | scoring function
  | score
  | Utility estimates
  | utility estimates
\knowledge{notion}
 | utility
 | relevance
\knowledge{notion}
 | policy
\knowledge{notion}
 | diversity
\knowledge{notion}
 | collaborative filter
 | collaborative filters
\knowledge{notion}
 | maximum marginal relevance
 | Maximum marginal relevance
 | MMR
  
\knowledge{notion,index=harm!distributional}
 | distributional harm
 | Distributional harm
 | distributional harms
 | Distributional harms
\knowledge{notion,index=harm!representational}
 | representational harm
 | Representational harm
 | representational harms
 | Representational harms
\knowledge{notion}
 | individual fairness
 | Individual fairness
 | individually fair
 | Individually fair
\knowledge{notion}
 | group fairness
 | Group fairness
 | group-fair
 | Group-fair
\knowledge{notion}
 | group
 | groups
\knowledge{notion,index=group!dominant}
 | dominant group
 | unprotected group
 | majority group
\knowledge{notion,index=group!protected}
 | protected group
 | protected groups
\knowledge{notion}
 | sensitive attribute
 | sensitive attribute(s)
 | Sensitive attribute
 | protected characteristics
\knowledge{notion}
 | intersectionality
\knowledge{notion, index=bias!societal}
 | societal bias
\knowledge{notion, index=bias!statistical}
 | statistical bias
\knowledge{notion}
 | disparate treatment
 | Disparate treatment
\knowledge{notion}
 | disparate impact
 | Disparate impact
\knowledge{notion,index=parity!statistical}
 | statistical parity
 | Statistical parity
\knowledge{notion,index=parity!error}
 | error parity
 | disparate mistreatment
 | Disparate mistreatment
 | Error parity
\knowledge{notion,index=parity!recall}
 | recall parity
 | Recall parity
 | equality of opportunity
\knowledge{notion,index=parity!predictive value}
 | predictive value parity
\knowledge{notion,index=parity!calibration}
 | calibration parity
\knowledge{notion}
 | we're all equal (WAE)
 | we're all equal
 | WAE
\knowledge{notion}
 | what you see is what you get (WYSIWYG)
 | what you see is what you get
 | WYSIWYG
\knowledge{notion}
 | meritocratic fairness
\knowledge{notion}
 | anti-classification
 | Anti-classification
\knowledge{notion}
 | anti-subordination
 | Anti-subordination
 
\knowledge{notion}
 | multisided platform
 | multisided platforms
\knowledge{notion}
 | disparate effectiveness
\knowledge{notion}
 | exposure
\knowledge{notion}
 | browsing model
 | browsing models
\knowledge{notion}
 | discount function
 | discount model

\knowledge{notion}
 | re-ranking
 | Re-ranking
 | reranking
 | Reranking
 | reranker
 | Reranker

\makeindex[intoc]

\mdfdefinestyle{Highlight}{
    linecolor=blue,
    outerlinewidth=2pt,
    roundcorner=10pt,
    innertopmargin=10pt,
    innerbottommargin=10pt,
    innerrightmargin=10pt,
    innerleftmargin=10pt,
    backgroundcolor=gray!25!white}
	
\newcommand{\highlight}[1]{
\vspace{5pt}
    \begin{mdframed}[style=Highlight]
	    #1
	\end{mdframed}}

\newcommand{\ConceptDef}[1]{\textbf{#1}}

\excludecomment{oldcontent}

\title{Fairness in Information Access Systems}

\ifpreprint
\author{
Michael D. Ekstrand\\
People and Information Research Team\\
Boise State University\\
michaelekstrand@boisestate.edu
\and
Anubrata Das\\
School of Information\\
University of Texas at Austin \\
anubrata@utexas.edu
\and
Robin Burke\\
That Recommender Systems Lab\\
University of Colorado\\
robin.burke@colorado.edu
\and
Fernando Diaz\\
Mila - Quebec AI Institute\\
diazf@acm.org
}
\date{January 2022}

\renewcommand{\maketitlehookb}{
\begin{mdframed}[linecolor=red,roundcorner=10pt,outerlinewidth=2pt]
\textbf{Preprint manuscript} --- this is the author's accepted version of a monograph in \textit{Foundations and Trends in Information Retrieval}.
The final version of record can be found at \url{https://doi.org/10.1561/1500000079}, and further information and supplementary resources
are available at \url{https://md.ekstrandom.net/pubs/fair-info-access}.

Please read and cite the final version as:

Michael D. Ekstrand, Anubrata Das, Robin Burke, and Fernando Diaz. 2022.
Fairness in Information Access Systems.
Foundations and Trends® in Information Retrieval \textbf{16}(1--2) (July 2022), 1--177.
DOI \href{https://doi.org/10.1561/1500000079}{10.1561/1500000079}.
\end{mdframed}
}
\pretitle{\centering\sffamily\bfseries\Large}
\posttitle{\par}
\preauthor{\centering\large \lineskip 0.5em \let\oldand\and \def\and{\vskip 1em}}
\postauthor{\par\let\and\oldand}

\newenvironment{acknowledgements}{
    
    \begin{abstract}
}{\end{abstract}}

\else
\maintitleauthorlist{
Michael D. Ekstrand\\
People and Information Research Team (PIReT)\\
Boise State University\\
michaelekstrand@boisestate.edu
\and
Anubrata Das\\
School of Information\\
University of Texas at Austin \\
anubrata@utexas.edu
\and
Robin Burke\\
Department of Information Science\\
University of Colorado\\
robin.burke@colorado.edu
\and
Fernando Diaz\\
Mila - Quebec AI Institute\\
diazf@acm.org
}

\author[1]{Ekstrand, Michael D.}
\author[2]{Das, Anubrata}
\author[3]{Burke, Robin}
\author[4]{Diaz, Fernando}

\affil[1]{People and Information Research Team (PIReT), Boise State University; michaelekstrand@boisestate.edu}
\affil[2]{ School of Information, University of Texas at Austin; anubrata@utexas.edu}
\affil[3]{That Recommender Systems Lab, University of Colorado; robin.burke@colorado.edu}
\affil[4]{Mila - Quebec AI Institute; diazf@acm.org}
\fi

\begin{document}

\ifpreprint

\maketitle
\newpage

\else
\makeabstracttitle
\fi

\begin{abstract}
    Recommendation, information retrieval, and other information access systems pose unique challenges for investigating and applying the fairness and non-discrimination concepts that have been developed for studying other machine learning systems. While fair information access shares many commonalities with fair classification, there are important differences: the multistakeholder nature of information access applications, the rank-based problem setting, the centrality of personalization in many cases, and the role of user response all complicate the problem of identifying precisely what types and operationalizations of fairness may be relevant.
    
    In this monograph, we present a taxonomy of the various dimensions of fair information access and survey the literature to date on this new and rapidly-growing topic.
    We preface this with brief introductions to information access and algorithmic fairness to facilitate use of this work by scholars with experience in one (or neither) of these fields who wish to study their intersection.
    We conclude with several open problems in fair information access, along with some suggestions for how to approach research in this space.
\end{abstract}

\begin{acknowledgements}
    Michael Ekstrand's contributions are based upon work supported by the National Science Foundation under Grant No. IIS 17-51278.
    Anubrata Das is supported in part by the Micron Foundation, Wipro, and Good Systems\footnote{\url{https://goodsystems.utexas.edu/}}, a University of Texas at Austin Grand Challenge to develop responsible AI technologies. Robin Burke's work was supported by the National Science Foundation under Grant No. IIS 19-11025. We also thank James Atwood, Asia Biega, Yoni Halpern, Matt Lease, Hansa Srinivasan, and the anonymous reviewers for providing additional feedback and suggestions.
\end{acknowledgements}

\ifpreprint
\newpage
\tableofcontents
\fi

\chapter*{List of Key Terms}

\newcommand{\KeyTerm}[1]{

#1 & \kref{#1} & \kpageref{#1} \\
}

\begin{center}
\begin{tabular}{lrr}
\textit{Term} & \textit{Defined in} & \textit{Page} \\
\midrule
\KeyTerm{bias}
\KeyTerm{disparate impact}
\KeyTerm{disparate mistreatment}
\KeyTerm{disparate treatment}
\KeyTerm{fairness}
\KeyTerm{group fairness}
\KeyTerm{individual fairness}
\KeyTerm{information access}
\KeyTerm{information need}
\end{tabular}
\end{center}

The index provides a more comprehensive cross-reference of terms used in this monograph.
\listoftodos

\chapter{Introduction}
\label{sec:intro}

As long as humans have recorded information in durable form, they have needed tools to access it: to locate the information they seek, review it, and consume it.
Digitally, tools to facilitate information access take a variety of forms, including information retrieval and recommendation systems; these tools have been powered by technologies built on various paradigms, from heuristic metrics and expert systems to deep neural networks with sophisticated rank-based objective functions.
Fundamentally, these technologies take a user's \textit{information need} (an explicit and/or implicit need for information for some purpose \citep{Kuhlthau1993-dq}, such as filling in knowledge or selecting a product) and locate documents or items that are \textit{relevant} (that is, will meet the user's need).

Throughout the history of these technologies --- which we treat under the integrated banner of \textbf{information access systems} --- both research and development have been concerned with a range of effects beyond a system's ability to locate individual items that are relevant to a user's information need.
Research has examined the \emph{diversity} and \emph{novelty} of results \citep{Santos2015-pn, Hurley2011-cg} and the \emph{coverage} of the system, among other concerns.  In recent years, this concern has extended to the \emph{fairness} of an information access system: are the benefits and resources it provides fairly allocated between different people or organizations it affects?
Does it introduce or reproduce harms, particularly harms distributed in an unfair or unjust way?
This challenge is connected to the broader set of research on fairness in sociotechnical systems generally and AI systems more particularly \citep{Mitchell2020-mt, Barocas2019-ft}, but information access systems have their own set of particular challenges and possibilities.

Fairness is not an entirely new concern for information access; various fairness problems can be connected to topics with long precedent in the information retrieval and recommender systems literature.
In the context of information retrieval, 
\citet{Friedman1996-mn} and \citet{Introna2000-ov} recognized the potential for search engines to embed social, political, and moral values in their ranking functions.  In order to assess the impact of such values, \citet{Mowshowitz2002-ipm-assessing} developed a metric to measure a search engine's deviation from an ideal exposure of content.   
Although conversations often focus on bias in algorithmic ranking, \citet{Vaughan2007-ox} and \citet{Vaughan2004-fd} note that bias can be introduced because of biased crawling and indexing; in particular, they describe, writing in the 2000s, how Chinese webpages were under-indexed by search engines.  
These observations led to discussion amongst legal scholar about regulation of search engines \citep{Goldman2005-sj,Pasquale2006-mn}.
\citet{Azzopardi2008-et} proposed the notion document \textit{retrievability}  and investigated the skew in this distribution for different retrieval systems.
Work on \emph{popularity bias} \citep{Celma2008-me, Zhao2013-ts, Canamares2018-ki} and rich-get-richer effects \citep{Cho2005-hi}, along with attempts to ensure quality and equity in \emph{long-tail recommendations} \citep{Ferraro2019-gh}, can be viewed as a type of fairness problem: the system should not inordinately favor popular, well-known, and possibly well-funded content creators.
In a group recommendation, one common objective is to ensure that the various members of a group are treated fairly \citep{Kaya2020-lv}.

The work on fair information access that we present here goes beyond these  problems to examine how various forms of unfairness --- particularly those that arise from \emph{social biases} \citep{Olteanu2019-sj} --- can make their way in to the data, algorithms, and outputs of information access systems.
These biases can affect many different stakeholders of an information access system; \citet{Burke2017-ne} distinguishes between \emph{provider}- and \emph{consumer}-side fairness, and other individuals or organizations affected by an information access system may have further fairness concerns.

In this monograph, we provide an introduction to fairness in information access, aiming to give students, researchers, and practitioners a starting point for understanding the problem space, the research to date, and a foundation for their further study.
Fairness in information access draws heavily from the fair machine learning literature, which we summarize in \cref{sec:fairness}; researchers and practitioners looking to study or improve the fairness of information access will do well to pay attention to a broad set of research results.
For reasons of scope, we are primarily concerned here with the fairness of the information access transaction itself: providing results in response to a request encoding an information need.
Fairness concerns can also arise in other aspects of the system, such as the representation and presentation of documents themselves, or in support facilities such as query suggestions \citep{Noble2018-vb}. We provide brief pointers on these topics, but a detailed treatment is left for future synthesis, noting that they have not yet received as much attention in the research literature.
We are also specifically concerned with fairness-related harms, and not the broader set of harms that may arise in information access such as the amplification of disinformation.

Throughout this work, we use the term \intro{system} to describe an algorithmic system that performs some task: retrieving information, recommending items, classifying or scoring people based on their data.
These systems are embedded in social contexts, operating on human-provided inputs and producing results acted upon by humans.
The technical system forms one part of a broader socio-technical system.

\section{Abstracting Information Access}
\label{sec:intro:abstracting}

Our choice to title this article ``Fairness in \emph{Information Access}'' is quite deliberate.
While there is significant technical and social overlap between information retrieval, recommender systems, and related fields, they are distinct communities with differences in terminology, problem definitions, and evaluation practices.
However, there are fundamental commonalities, and they present many of the same problems that complicate notions of fairness, including ranked outputs, personalized relevance, repeated decision-making, and multistakeholder structure.
We therefore refer to them together as \intro{information access systems} --- algorithmic systems that mediate the interaction between a repository of documents or items and a user's information need.

This information access umbrella includes information retrieval, recommender systems, information filtering, and some applications of natural language processing.
In \cref{sec:access}, we present a fuller treatment of this integration and reviews the fundamentals of information access, both to introduce the concepts to readers who come to this paper from a general fairness background and to lay out consistent terminology for our readers from information retrieval or recommender systems backgrounds.

\section{A Brief History of Fairness}
\label{sec:intro:history}

In the pursuit of fairness in algorithmic systems and the society more generally, the authority of Aristotle's citation of Plato ``treat like cases alike'' is a key touchstone: a normative requirement that those who are equal before the law should receive equal treatment \citep{Gosepath2011-iw}.

In more recent scholarship, the study of distributive welfare extends these concepts considerably, recognizing four distinct concepts of fairness: ``exogenous rights, compensation, reward, and fitness.'' \citep{Moulin2004-zf}. 
\textit{Exogenous rights}, as the term suggests, relate to external claims that a system must satisfy: equal shares in property as defined by contract, for example, or equality of political rights in democratic societies.
\textit{Compensation} recognizes that fairness may require extra consideration for parties where costs are unequal --- affirmative action in hiring and college admissions are well-known examples.
\textit{Reward} justifies inequality on the basis of differing contributions: for example, increased bonuses to employees with greater contribution to the bottom line.
Finally, we have \textit{fitness}, the most nebulous category, and the one that many information access systems inhabit. The fitness principle holds that goods be distributed to those most fit to use, appreciate, or derive benefit from them.
It is an efficiency principle, where the fairest use is the one that allocates goods where the distribution achieves the maximum utility.
Fitness has a natural application to information access, as we seek to locate documents and make them visible based on their utility to the user's information need.

U.S. legal theory has developed a rich tradition of anti-discrimination law, aimed at ensuring that people are not denied certain benefits (housing, work, education, financial services, etc.) on the basis of "protected characteristics" (race, color, religion, gender, disability, age, and in many jurisdictions, sexual orientation).
It has given rise to several important concepts, such as the "disparate impact" standard (the idea that an allegedly discriminatory practice can be legally challenged on the grounds that it has disproportionate adverse impact on a protected group, without needing to show intent to discriminate\footnote{Disparate impact is not sufficient basis to \textit{win} a discrimination lawsuit; rather, it is the first step in a multi-stage burden-shifting framework used to decide discrimination cases under the standard. \citet{Barocas2016-wi} provide an overview of the process.}).
\citet{Crenshaw1989-km} points out some of the limitations of this legal framework; in particular, it has often focused on discrimination on the basis of \emph{individual} protected characteristics, and people who have suffered harm as a result of combinations of protected characteristics (e.g. Black women being denied promotions given to both Black men and White women) have difficulty proving their case and obtaining relief.  This theory of particular harms deriving from combinations of characteristics is called \intro{intersectionality}.

Questions of fairness and discrimination have been the subject of significant discussion in many other communities as well.
Educational testing, for example, has several decades of research on the fairness of various testing and assessment instruments; this history is summarized for computer scientists by \citet{Hutchinson2019-th}.
\citet{Friedman1996-mn} provide one of the earlier examples of addressing questions of bias in computer science, pointing out how even seemingly-innocuous technical decisions may result in biased effects when a computing system is used in its social context.
The last ten years have seen significant new activity on fairness in machine learning that forms the primary stream of algorithmic fairness research; in \cref{sec:fairness} we provide an introduction to this literature.

\section{Fairness and Bias}
\label{sec:intro:defs}
\label{sec:intro:fairness-bias}

There are many overlapping terms used to discuss issues of fairness, bias, and discrimination.  While we give a fuller treatment of the vocabulary in \cref{sec:fairness}, we will here introduce how we use these terms in this monograph.  Work we cite may use them differently.

\index{fairness!defined for this work}
\AP When we refer to \intro{fairness}, we are talking about the ways a system treats people, or groups of people, in a way that is considered ``unfair'' by some moral, legal, or ethical standard.
This is typically through effects or impacts that are not experienced in an equitable way, but can sometimes arise through the system's internal operation or representations.  This definition is similar to how \citet{Friedman1996-mn} use the term ``bias''.
There is not one particular definition of what constitutes fairness, as \citet{Selbst2019-hf} and many others have noted; for the purpose of terminology, the important point is that we use the term to refer to normative ideas of what it means to treat people ``fairly'', no matter their source.

\index{bias!defined for this work}
\AP When we talk about \intro{bias}, we are using the term in something closer to its statistical sense: we mean properties of estimators, models, measurements, and data that systematically deviate from their intended ideal target.
As detailed in \cref{sec:fairness:sources}, we share an expansive view of bias with \citet[\S 2.2.1]{Mitchell2020-mt}, noting that these biases can be the kinds of statistical biases familiar to science (systematic discrepancies between data or outputs and the underlying observable world), but they can also be societal biases in the form of systematic discrepancies between the observable world and the arguable ideal world that would arise if society eliminated all forms of illegitimate discrimination.

The key distinction in our work is that we use the term ``bias'' to refer to a fact of the system without making any inherently normative judgment, and ``fairness'' to discuss the normative aspects of the system and its effects.
Some biases are themselves fairness problems; some biases cause fairness problems; some have no effect with regards to the concerns of fairness; and some may be  intentionally introduced to address a fairness problem, often by correcting for another bias.
Most fairness problems arise from biases somewhere in the system, its data, or its evaluation, but we find it useful to distinguish between the technical fact and the moral, ethical, or legal concern.

\section{Fairness and Other Responsibility Concerns}
\label{sec:intro:other-concerns}

Fairness is commonly grouped together with other concerns under the banner of \emph{responsibility} in computing systems.
These concerns include:

\begin{description}
\index{accountability}
    \item[Accountability]
    Research on accountability examines the legal, social, and technical mechanisms by which computing systems and their operators, developers, and providers may be held accountable, usually for the human effects of their systems.  This can connect directly to fairness when considering how to hold organizations accountable for ensuring their systems uphold societal goals to be fair.  Such accountability can be through formal structures, such as applying anti-discrimination law to computing systems, or through informal structures such as applying pressure through publicizing the results of third-party audits.
    
    \index{transparency}
    \item[Transparency]
    Transparency (and its close cousin explainability) seeks to make the operation and results of algorithmic systems scrutable to users, developers, auditors, and other stakeholders so that it can be understood, reviewed, and contested.  This relates to long-standing concern in information access on explanation \citep{Tintarev2007-il}, as well as ideas such as scrutable user models \citep{Kay2002-cy}.
    
    \index{safety}
    \item[Safety]
    Information access systems can be harmful. They can distribute false information, promote fake or dangerous products, and provide support for illegal or malicious activities. These problems have received attention in the research literature, often under the general heading of \textit{adversarial information retrieval}. See related workshops AIRWeb \citep{Fetterly2009-en} and WebQuality \citep{Nielek2016-pt}.
    
    \index{privacy}
    \item[Privacy] Aspects of users' profiles including queries, interaction history, and usage patterns may be highly revealing of sensitive personal information: consider queries about medical symptoms or clicks on web pages for addiction counseling. It follows that information access systems have a duty to protect such information from harmful disclosure. Research on privacy-preserving recommendation seeks technical solutions to this challenge.  \citet{Friedman2015-om} provide a survey of this area.
    
    \index{ethics}
    \item[Ethics]
    Computing ethics is concerned broadly with ensuring that the practice and products of computing adhere to appropriate ethical principles.  The ACM Code of Ethics \citep{ACM_Council2018-lv} specifically calls out non-discrimination, along with attention to potential harms, as an ethical obligation for computing professionals.
\end{description}

The report on the FACTS-IR Workshop on Fairness, Accountability, Confidentiality, Transparency, and Safety in Information Retrieval \citep{FACTS-IR} discusses how many of these concepts play out in information retrieval.  In this work we are concerned with fairness, but bring in other concerns as well when they relate to fairness.

\section{Running Examples}
\label{sec:intro:running}

Throughout this monograph, we will use several examples to motivate and explain the various concepts we discuss.

\paragraph{Job and Candidate Search}
\index{job search}
\index{candidate search}
\index{discovery!jobs}
\index{discovery!job candidates}
Many online platforms attempt to connect job-seekers and employment opportunities in some way.
Some of these are dedicated employment-seeking platforms, while others, such as LinkedIn and Xing, are more general-purpose professional networking platforms for which job-seeking is one important component.

Job-seeking is a multisided problem --- people need good employment and employers need good candidates --- and also has significant fairness requirements that are often subject to regulation in various jurisdictions.
Some of the specific fairness concerns for this application include:
\begin{itemize}
    \item Do users receive a fair set of job opportunities in the recommendations or ads in their feed?
    \item If the system assesses a match or fit score for a candidate and a job, is this score fair, or does it under- or over-estimate scores for particular candidates or groups of candidates?
    \item Do users have a fair opportunity to appear in search lists when recruiters are looking for candidates for a job opening \citep{Geyik2018-lq}?
    \item Do employers in protected groups (minority-owned businesses, for example) have their jobs fairly promoted to qualified candidates?
    \item What fairness concerns come from regulatory requirements?
\end{itemize}

\paragraph{Music Discovery}
\index{music discovery}
\index{discovery!music}
The search and recommendation systems in music platforms, such as Spotify, Pandora, and BandCamp, connect listeners with artists.
These discovery tools have significant impact not only on a user's listening experience and musical enjoyment, but also on artists' financial and career prospects, due both to direct revenue from listening and the commercial and reputational effects of visibility.
Some specific fairness concerns include:
\begin{itemize}
    \item Do artists receive fair exposure in the system's search results, recommendation lists, or streamed programming?
    \item Does the system systematically over- or under-promote particular groups of artists or songwriters through recommendations, search results, and other discovery surfaces \citep{Epps-Darling2020-eh}?
    \item Do users receive fair quality of service, or does the system systematically do a better job of modeling some users' tastes and preferences than others?
    \item Do recommendations reflect well a user's preferences and if not, are there systematic errors due to stereotypes of gender, ethnicity, location, or other attributes?
\end{itemize}

\paragraph{News}
\index{news discovery}
\index{discovery!news}
News search and recommendation influences user exposure to news articles on social media, news aggregation applications, and search engines.
Such influence extends to social and political choices users might make \citep{Kulshrestha2017-cd, Epstein2015-um}.
Additionally, the filter bubble effect \citep{Pariser2011-wu, Van_Alstyne2005-vn} may cause users to be exposed primarily to news items that reinforce their beliefs and increase polarization.
Depending on the journalistic policy of the provider, news platforms may want to facilitate balanced exposure to news from across the social, political, and cultural spectrum, but this may need to be balanced with the need to de-rank malicious and low-credibility sources.

Specific fairness concerns in news discovery include:
\begin{itemize}
    \item Does the system provide fair exposure to news on different topics or affected groups?
    \item Do journalists from different perspectives receive fair visibility or exposure for their content?
    \item Does the system reward original investigators or primarily direct readers to tertiary sources?
    \item Do users receive a balanced set of news content?
    \item Are users in different demographics or locations equally well-served by their news recommendations?
\end{itemize}

\paragraph{Philanthropic Giving}
\index{philanthropic giving}
Online platforms are increasingly a site for philanthropic giving \citep{Goecks2008-nc}, and  therefore recommendation is expected to be an increasing driver of donations. Sites may take an explicitly ``peer-to-peer'' approach to such giving, as in the educational charity site DonorsChoose.org; this results in many possible donation opportunities for donors to select from, requiring recommendation or sophisticated search to help match donors and opportunities. As many philanthropic organizations have a social justice focus, fairness concerns are essential in developing and evaluating their information access solutions, in particular to avoid potential positive feedback loops in which a subset of causes comes to dominate results and rankings.

In philanthropic settings, we would expect fairness issues to include:
\begin{itemize}
    \item Does the system provide fair opportunities for the various recipients / causes to have their needs supported?
    \item Are specific groups of recipients under- or over-represented in the recommendation results?
\end{itemize}

\section{How to Use This Monograph}
\label{sec:intro:reading}

We have written this monograph with two audiences in mind:

\begin{itemize}
    \item Researchers, engineers, and students in information retrieval, recommender systems, and related fields who are looking to understand the literature on fairness, bias, and discrimination, and how it applies to their work.
    \item Researchers in algorithmic fairness who are looking to understand information access systems, how existing fairness concepts do or do not apply to this set of applications, and the things that information access brings to the research space that may differ from the application settings in which fairness is usually studied.
\end{itemize}

Due to our interest in serving both of these audiences, we do not expect our readers to have significant familiarity with either information retrieval or algorithmic fairness, although some background in machine learning will be helpful. We have organized the material as follows:

\begin{itemize}
    \item \textbf{\Cref{sec:access}} rehearses the fundamentals of information access systems.  This will be a review for most information retrieval and recommender systems researchers; such readers should read it for the terminology we use to integrate the fields, but may wish to focus their study energy elsewhere.
    
    \item \textbf{\Cref{sec:fairness}} provides an overview of research on fairness in machine learning generally, particularly in classification.  Algorithmic fairness researchers will likely find this chapter to be a review.
    
    \item \textbf{\Cref{sec:space}} lays out the problem space of fair information access, providing a multi-faceted taxonomy of the problems in evaluating and removing discrimination and related harms in such systems.
    
    \item \textbf{\Cref{sec:consumer,sec:provider}} survey key literature to date (as of 2021) on fairness in information access, with pointers to research working on many of the problems identified in \cref{sec:space}, focused on the two most commonly-studied stakeholders: consumers and providers (with discussion of "subjects" in \cref{sec:provider:subject}).
    
    \item \textbf{\Cref{sec:time}} discusses the need to go beyond point-in-time views of fairness to understand fairness over time how the temporal dynamics of an information access system affect fairness.
    
    \item \textbf{\Cref{sec:future}} looks to future work and provides tips for research and engineering on fair information access.
\end{itemize}

\Cref{sec:space} is the keystone of this work that ties the rest together; subsequent chapters work out details in the form of a literature survey of several of the problems discussed in \cref{sec:space}, and the preceding chapters set up the background needed to understand it.
For readers looking to budget their time, we recommend they ensure they have the necessary background from \cref{sec:access,sec:fairness}, read \cref{sec:space,sec:future}, and read the later chapters that are relevant to their work.

\section{Our Perspective}
\label{sec:intro:perspective}

While we have written this monograph to be useful for researchers approaching the topic of fairness from a variety of perspectives, we think it is helpful to explicitly describe our own perspectives and motivations, as well as the position from which we approach this work and some limitations it may bring.

Information access systems need to meet a variety of objectives from multiple stakeholders.
They need to deliver relevant results to their users, business value for their operators, and visibility to the creators of the documents they present; they often also need to meet a variety of other goals and constraints, such as diversity across subtopics, regulatory compliance, and reducing avoidable harm to users or society.
"Fairness", as we conceive of it and present it in this paper, is not a be-all end goal, but rather another family of objectives to be considered in the design and evaluation of information access systems, and a collection of techniques for enabling those objectives.
It also does not encompass the totality of social or ethical objectives guiding a system's design.
Researchers and developers need to work with experts in ethics, policy, sociology, and other relevant fields to identify relevant harms and appropriate objectives for any particular application; the concepts we discuss will be relevant to some of those harms and objectives.

We also emphasize the importance of starting with a robust \emph{problem framing}: \cref{sec:space} is intended to help readers think about the fairness problem they are trying to solve, and position it in landscape of information access; we have then organized our survey in \crefrange{sec:consumer}{sec:time} around aspects of problem definition, instead of underlying techniques.  Metrics and mitigations are best developed and assessed in the context of a specific, well-defined problem.

Finally, all four authors work in North America, and approach the topic primarily in that legal and moral context.
A Western focus, and particularly concepts of bias and discrimination rooted in United States legal theory, currently dominates thinking and research on algorithmic fairness in general.
This is a limitation of the field that others have noted and critiqued \citep{Sambasivan2020-bo}; our present work acknowledges but does not correct this imbalance.
While we attempt to engage with definitions and fairness objectives beyond the U.S., this article admittedly has a Western and especially U.S. focus in its treatment of the material.
We look forward to seeing other scholars survey this topic from other perspectives.

\section{Some Cautions}
\label{sec:intro:cautions}

We hope that this monograph will help scholars from a variety of backgrounds to understand the emerging literature on fairness in information access and to advance the field in useful directions.
In addition to the general concerns of careful, thoughtful science, work on fairness often engages with data and constructs that touch on fundamental aspects of human identity and experience.
This work must also be done with great care and compassion to ensure that users, creators, and other stakeholders are treated with respect and dignity and to avoid various traps that result in overbroad or ungeneralizable claims.

We argue that there is nothing particularly new about this, but that thinking about the fairness of information access brings to the surface issues that should be considered in all research and development on information systems.

\subsection{Beware Abstraction Traps}
\label{sec:intro:traps}

Our first caution is to beware of the allure of abstraction.
\citet{Selbst2019-hf} describe several specific problems that arise from excessive or inappropriate abstraction in fairness research in general.
Their core argument is that the tendency in computer science to seek general, abstract forms of problems, while useful for developing tools and results that can be applied to a wide range of tasks, can cause important social aspects of technology and its impacts to be obscured.

One reason for this is that social problems that appear to be structurally similar arise from distinct (though possibly intertwined) causes and mechanisms, and may require different solutions.
Sexism and anti-Black racism, for example, are both types of discrimination and fall into the ``group fairness'' category of algorithmic fairness, but they are not the same problem and have not been reinforced by the same sets of legal and social processes.
Discrimination also varies by culture and jurisdiction, and oppression of what appears to be same group may arise from different causes and through different mechanisms in the different places in which it appears.
\citet{Kohler-Hausmann2019-ms} argues that social constructivist\index{social constructionism} frameworks for understanding group identities and experiences imply that even understanding what constitutes a group, let alone the discrimination it experiences, is inextricably linked with understanding how that group is constructed and treated in a particular society --- an understanding that is inherently bound to the society in question, although there may be similarities in group construction in different contexts.

The result is that unfairness needs to be measured and addressed in each specific way in which it may appear.
While general solutions for detecting and mitigating fairness-related harms may arise and be very useful, their effectiveness needs to be re-validated in context for the harms they are meant to address, a point reiterated by \citet{Dwork2018-wf}.

\citet{Hoffmann2019-lg} similarly provides several warnings against overly simple ideas of the harms that can arise from discrimination and bias.
Computational fairness inherits some of these limitations from its reference material, such as limitations of anti-discrimination law; others arise from what \citeauthor{Hoffmann2019-lg}, \citet{Selbst2019-hf}, and others argue are reductionistic operationalizations of rich concepts. \citet{Hoffmann2019-lg} notes in particular---and we agree---that treating categories of personal identity as objective features in a multi-dimensional space (a natural move for computer scientists) obfuscates the role of technical and social systems in enacting and producing such categories. This move also has the effect of reducing "intersectionality" concerns to what can be captured by a subspace projection or similar formal operation, whether or not that corresponds to individual's lived experience.

We believe computing systems in general, and information access systems in particular, have the opportunity to \emph{advance} the the discussion of emancipation and justice, not just bring existing constructs into a new domain.
Information professionals have long been concerned about issues of ethics and justice. Just as two examples, we note that Edmund Berkeley, one of the founders of the Association for Computing Machinery, was an outspoken advocate for the ethical responsibilities of computer scientists as far back as the 1960s \citep{Longo2015-tk}, and the creation of Computer Professionals for Social Responsibility in the mid-1980s \citep{Finn2020-lc}. The call here is to realize that vision fully and for all people affected by information access systems.

\subsection{Beware Limits}
\label{sec:intro:limits}

\index{limitations}
It is crucial to be clear about the limitations of particular fairness studies and methods.
Any work will be limited, if for no other reason than the impossibility of completely solving the problem of discrimination.
Those limitations should not paralyze the research community or keep researchers from doing the most they can to advance equity and justice with the resources available to them; rather, work in this space needs to be forthright and thorough about the limitations of its approach, data, and findings.
Some limitations common to this space include:

\begin{itemize}
    \item Single-dimensional attributes for which fairness is considered, when in reality people experience discrimination and oppression along multiple simultaneous dimensions.
    
    \item Binary definitions of attributes, when in reality many social dimensions have more than two categories or exist on a continuum.
    
    \item Taking attributes as fixed and exogenous, when social categories are complex and socially constructed \citep{Hanna2020-ln}.
    
    \item Incomplete, erroneous, and/or biased data \citep{Olteanu2019-sj,Ekstrand2021-iu}.
\end{itemize}

This is not to say that work on single binary attributes is not useful; research must start somewhere.
But it should not \emph{stop} there, and authors need to be clear about the relationship their work in its broader context and provide a careful accounting of its known limitations.

Some methods are so limited that we advise against their use.
For example, some work on fair information access has used statistical gender recognition based on names or computer vision techniques for gender recognition based on profile pictures.\footnote{We do not provide citations to support the claim that this is in use because our purpose in this paragraph is to critique a general trend, not to focus on any specific paper.  Elsewhere in this monograph, we cite work making use of these techniques where it makes a relevant contribution.}
This source of data is error-prone, subject to systemic biases \citep{Buolamwini2018-im}, reductionistic \citep{Hamidi2018-pt}, and fundamentally denies subjects control over their identities, so we do not consider it good practice.

\subsection{Beware Convenience}
\label{sec:intro:convenience}

Researchers working in this problem space also need to be careful to do the \emph{best} research possible with available resources, and work to expand those resources to increase the quality and social fidelity of their work, and not take the path of least resistance.

One particular application pertains to this article itself and to its proper use and citation.
It is convenient and common practice to cite survey papers to quickly summarize a topic or substantiate its relevance. While we naturally welcome citations of our work, we would prefer to be cited specifically for our contributions to the organization and synthesis of fair information access research.
The purpose of much of this monograph is to point our readers to the work that others have done, and we specifically ask that you \textbf{cite those papers}, instead of --- or in addition to --- this one when that work is relevant to your writing and research.
\chapter{Information Access Fundamentals}
\label{sec:access}

\kl[iasys]{Information access} refers to a class of systems that support users by retrieving items from some large "repository" of data.  The area covers both information retrieval and recommendation systems.  More concretely, the information access problem can be defined as:
\begin{quote}
        Given a "repository" of items and a "user" "information need", present "items" to help the user satisfy that need.  
\end{quote}
The \emph{\kl{repository}} may be the results of a hypertext crawl as in web search, a catalog of products as in commercial recommendation, corporate documents as in enterprise search, or a collection of songs as in music recommendation.  An \emph{\kl{information need}} refers to the latent concept or class the user seeks.  Although unobserved, this need may inferred from explicit expressions from the user (e.g. a keyword query, a question, or a set of example documents) or implicit data about the user (e.g. previously consumed content, time of day).  The \emph{\kl[results]{presentation of items}} refers to the system response and might be a ranked list, a two-dimensional grid, or some other structured output.  Finally, \emph{\kl{satisfaction}} is a measure of the degree to which the system response fulfilled the user's need.  This may be explicit (e.g. ratings, binary feedback) or implicit (e.g. clicks, streams, purchases).

In this section we provide a brief overview of the fundamentals of these systems, both for our readers who are not familiar with information retrieval or recommender systems, and to provide terminology for our integration of the topics.

\section{System Overview}
\label{sec:access:architecture}
\label{sec:access:pipeline}

\begin{figure}[tb]
    \centering
    \includegraphics[width=1\textwidth]{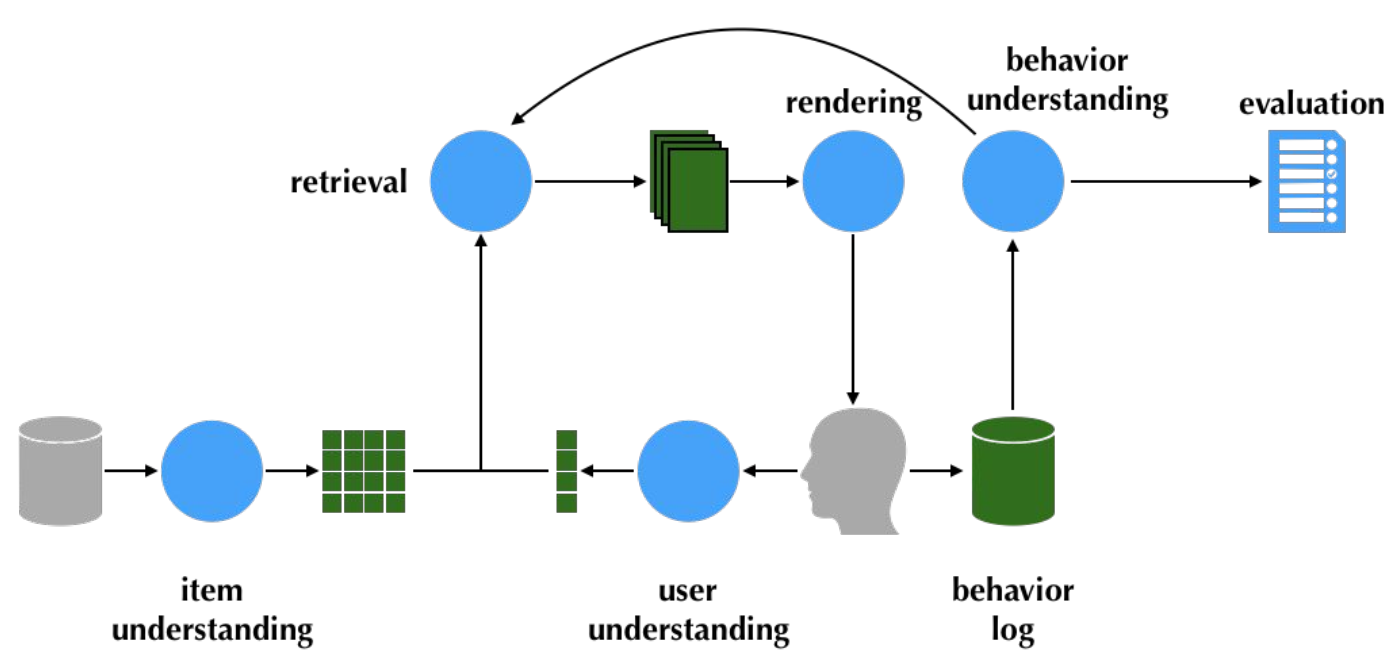}
    \caption[Information access pipeline]{A typical information access pipeline consists of item understanding, user understanding, item retrieval, item rendering, behavior understanding and evaluation.}
    \label{fig:info-access}
\end{figure}

The process of meeting an "information need" involves several steps;  \Cref{fig:info-access} shows one view of the components carrying out these steps and their relationships.  These pieces include:

\begin{itemize}
    \item Understanding (with a computationally-useful representation) the "items" to be retrieved, so they can be connected with information needs.
    
    \item Understanding the "user" and their "information need", so that it can be matched to relevant items.
    
    \item Retrieving items that match the need.
    
    \item Rendering the set of retrieved items for presentation to the user.
    
    \item Understanding users' behavior, particularly in response to presentations of retrieved results, to inform future retrieval and to evaluate the system's ability to meet user needs.
\end{itemize}

\begin{table}[tb]
    \centering\footnotesize
    \begin{tabular}{rl}
         $\doc \in \repository$ & Item in a repository \\
         $\featuresContent(\doc)$ & Content representation of an item $\doc$ \\
         $\featuresMetadata(\doc)$ & Metadata representation of an item $\doc$ \\
         $\featuresUsage(\doc)$ & Usage representation of an item $\doc$ \\
         $\need \in \needs$ & Information needs with the space of possible needs \\
         $\requestFeatures(\need)$ & Feature-based expression of information need \\
         $\requestItems(\need)$ & Item-based expression of information need \\
         $\requestLanguage(\need)$ & Language-based expression of information need \\
         $\implicitGlobal(\need)$ & Inferred global information need  \\
         $\implicitLocal(\need)$ & Inferred local information need \\
         $ r \in \ranks$ & A rank position from the set of possible ranks \\
         $\itemUtility : \repository \rightarrow \Re$ & Item utility function \\
         $\discountFactor : \ranks \rightarrow [0,1]$ & Rank discount function \\
    \end{tabular}
    \caption{Summary of notation for \cref{sec:access}.}
    \label{tab:IA-notation}
\end{table}

\section{Repository of Items}
\label{sec:access:corpus}

\index{document|see {item}}
\index{corpus|see {repository}}
\AP As we have defined it, "information access" is the process of retrieving \intro{items} that are contained in a \intro{repository}.  Terms for this can vary; in text-oriented information retrieval, these are often called ``documents'' in a ``corpus''.
The curation of this repository involves a variety of subtasks and algorithmic research, including content creation, content collection, and content representation.

\index{creation}
Content creation refers to the complex organizational, social, economic, and political dynamics that result in the creation of an item.  This includes items that may be created in relative isolation by a hobbyist (e.g. a self-produced song) or as a result of more complex coordination amongst many contributors (e.g. a major motion picture); together, we call these the \intro{providers} of the item.  Regardless of this variation in apparent scale, such cultural objects are always an artifact of social relationships \citep{Becker1982-zl}.  As such, each item reflects potentially a major stakeholder or stakeholders upon whom a livelihood may rest but also a network of supporting stakeholders, each of whom may have a variety of incentives for participating in the item's creation and consumption.

\index{collection}
Content collection refers to the processes and policies involved in adding or removing items from the repository.  In the simplest case, content collection involves receiving a static archive of items or documents from some external party.  More often, the repository is dynamic, provided as a feed of items to add or remove from the the repository, as with news articles or products.  However, it is often the case that the repository designer has some control over content collection, either because the content must be found (as in hypertext crawling \citep{Pandey2008-fj}), curated (e.g. by removing low quality or adversarial content, or simply caching for performance reasons), or contracted for.  Without loss of generality, we refer to the repository of items at the time when the user engages with the information access system as $\repository$, a set of indices mapping into the items.  

Content representation refers to the information we have about an individual item.  ""Item representation"" in general consists of three parts: content, metadata, and usage data.  In most cases, the content of an item is static and is an artifact created by the author(s) at some fixed point in time.  This might be the text of a academic article, the pixels of an image, or the audio file associated with a song.  
We refer to the content representation of an item $\doc \in \repository$ as $\featuresContent(\doc)$.

\index{metadata}
The metadata of an item expresses information about the content, including when it was authored, the name of the author, the genre of the item, etc.  Metadata may also be inferred by a machine such as with ``learned representations'' or algorithmically-assigned genre labels.  Furthermore, metadata may be dynamic, changing as a function of the world around the item.  For example, this might include frequency of access (outside of a search or recommendation context) or popularity of the author.  We refer to the metadata representation of an item $\doc \in \repository$ as $\featuresMetadata(\doc)$.  

\index{usage}
Finally, usage data about an item refers to the historic interaction between information needs (or their expression) and the item.  This might include, for example, the set of users who consumed the item, the queries for which this item was clicked, etc.  Usage features can be seen as a special type of metadata that have three unique properties: they are often more biased by system decisions; they are often highly suggestive of relevance; and they are updated over time as users interact with the system and its items.  We refer to the usage data representation of an item $\doc \in \repository$ as $\featuresUsage(\doc)$.  

\section{Users and Information Needs}
\label{sec:access:needs}

\AP The system is used by \intro{users} (or \emph{consumers}), who rely on it to meet some \intro{information need}.
When a user approaches the system, they may do so for a variety of reasons with varying degrees of specificity and explicitness.  Their need may isolated (e.g. answering a specific question), a function of mood (e.g. playing a genre of music), or an element of a task the user is involved in (e.g. finding reviews before making a purchase).  Whatever the need, we represent the space of needs as $\needs$, and individual needs as $\need\in\needs$.

\AP An information need can arrive in a variety of ways.   In most information retrieval systems, users can explicitly express their need.  One such family of expressions are ``feature-based'' expressions, where the searcher explicitly describes item representation values likely to be found in relevant items.  For example, a keyword \intro{query} suggests that the items containing the keywords may be more relevant than those not; a faceted query may indicate a time range or class of items more likely to be relevant.  We represent the feature-based expression of an information need $\need\in\needs$ as $\requestFeatures(\need)$.  Alternatively, the user may provide a set of example relevant or non-relevant items
\citep{Smucker2006-of}.  We represent the item-based expression of an information need $\need\in\needs$ as $\requestItems(\need)$.  Finally, the user may express their need by some other means such as a natural language question or description.\footnote{Although, because both natural language questions and text items share a symbolic representation, it is tempting to treat a natural language question as a feature-based expression, the generating processes behind questions and documents are sufficiently different that this would be a mistake.}  We represent this expression of an information need $\need\in\needs$ as $\requestLanguage(\need)$.  

An information need may also be expressed implicitly, and these implicit aspects can be global or local.  An implicit \emph{global} expression reflects relatively stationary properties of the users across access "sessions".  This may include demographic information about the user, their previously accessed items, or information provided or inferred about their preferences.  We represent the global implicit expression of an information need $\need\in\needs$ as $\implicitGlobal(\need)$, which serves as the ""user representation"".  An implicit \emph{local} expression reflects properties of the user or their interactions that  typically vary across access "sessions".  This may include the items viewed or interacted with within the session.  Importantly, this also includes the `surface' of the search or recommendation platform itself, since the type of access or need may be suggested by the user's entry into the system; for example, a ``discovery'' feature versus a ``mood'' feature in a music streaming platform.  Implicit local aspects of the need also includes contextual information such as time and location, which are meaningful parts of the description of many needs. We represent the local implicit expression of an information need $\need\in\needs$ as $\implicitLocal(\need)$.  We note that the distinction between local and global can be fluid, especially in hierarchical information access (e.g. a collection of sessions with a shared goal) \citep{Jones2008-vw}.

\section{Presentation}
\label{sec:access:presentation}

Given an information need and our repository, there are various ways to present ""results"" to users within the constraints and capabilities of the particular user interface.  In this section, we describe several of the types of presentation formats that influence algorithm and system design.

A single \intro{turn} refers to a user approaching the system with a one-shot request and receiving an isolated system response.  A recommendation system home page or isolated search query are examples of this type of presentation context.  The simplest presentation involves the system providing a single item to satisfy the user's information need.  This might occur in limited surfaces like small screens or audio interfaces.  In situations where the interface allows, a ranked list of items can be presented, which the user can serially scan from the top down; much historical work on information retrieval and recommender systems assumes such a layout, and it underpins common evaluation metrics.  A popular way to present image-based results is a two-dimensional grid layout \citep{Guo2020-kl, Xie2019-fd}.  Finally, immersive environments allow for three-dimensional layouts of items \citep{Leuski2000-cd}.

When items have text summaries (e.g. a description or teaser content), they can be presented alongside the item identifier (e.g. a title or URL) to let the user inspect the item without consuming it in its entirety (e.g. reading a whole web page, watching a whole movie).  In some cases, the system can detect a specific part of the content to help the user better make a decision \citep{Luhn1960-ak,Tombros1998-im,Metzler2008-ia}.

\subsection{Interaction and Sessions}
\label{sec:access:interaction}
In reality, almost every information access system involves multiple interactions in order to satisfy a single information need.  Users engage with search systems in \intro{sessions} composed of multiple queries.  Users engage with recommendation systems by navigating through pages, websites, and application interfaces before settling on relevant content.  These settings consist of system decisions (such as those in single turn scenarios)  interleaved with implicit and explicit user feedback, allowing the algorithm to adapt its behavior.  For example, streaming audio systems (e.g. radio-like interfaces) involve a sequence of single item decisions interleaved with user skips or other behavior.  A dialog-based recommendation system similarly exhibits multiple turns before resolution.  At a temporally-extended level, an information need or task may be spread across multiple sessions, such as assembling a bibliography for a class or survey.

\subsection{Rankings}
\label{sec:access:ranking}

No matter the final presentation mode, systems typically operate in terms of \intro{rankings} of items.
The simplest, and historically most common, ranking is to sort items by decreasing relevance according to the probability ranking principle \citep{Robertson1977-ur}.
Items can be ranked in other ways as well (see \ref{sec:access:reranking}); the system may display the items in order in which the underlying algorithm ranked them, or it may rearrange them (e.g. into a grid) or combine the results of multiple rankings (e.g. the rows of rankings display common in video streaming services), but the ranking is the fundamental unit of algorithmic output considered in most information access research, and constitutes the `decisions' that a search or recommendation algorithm typically makes.

\section{Evaluation}
\label{sec:access:eval}
Classic information access \intro{evaluation} of a system decision uses a model of the user interaction together with a item-level utility estimate (e.g. an item label or behavioral signal) to measure the the extent to which the system has \intro{satisfied} the information need \citep{Chandar2020-em}.

There are two forms in which this evaluation can take place: situated and simulated. \textit{Situated evaluation} places the algorithm or system in front of real users, operating the system in the exact environment in which it will be deployed.  

\textit{Simulated evaluation} uses data and algorithms to create a controlled, repeatable simulation of user behavior and metrics.  Offline evaluation, including off-policy evaluation, is an example of this approach.
As with any simulation, the assumption is that the simulator --- or log data --- can be used in a way that predicts performance in a situated setting \citep{Ferro2018-gb}.  Situated evaluation, on the other hand, while considering the state of the world as potentially non-stationary and ephemeral, is more costly in terms of risk to users, time, and engineering effort.

\subsection{Estimating Item Utility}
\label{sec:estimating-item-utility}
\AP Given some expression of an information need and a repository, we would like to estimate the \intro{utility} of each item to the information need \textit{for evaluation purposes}.  We contrast this from utility estimation performed by a system \textit{before} presenting results to users because, in the evaluation context, we can exploit information about the information need and item that may be unavailable at decision time.  This information may be unavailable because of prohibitive labeling cost (e.g. manually rating documents in response to search query) or because the user has yet to observe or experience the item.

Explicit labels refer to utility estimates manually provided by some person. User-based explicit labels can be gathered by providing users with the option to rate items during or after the access or as part of some on-boarding process, as is performed in many recommendation systems.  It is important to understand the user's annotation mindset, especially when gathering labels during an access task.  Are these labels judgments about the quality ``in the context'' or ``in general''?  There is often ambiguity in the precise semantics of a rating of label that can cause confusion in evaluation.  In some situations, primarily in search, the information need and items are easy to understand by a third party and, therefore, ``non-user'' assessors can be employed to gather item quality estimates.  Editorial explicit labels can be gathered using explicit guidelines, reducing some of the labeling ambiguity found in user-based explicit labels.  However, these labels are limited by the interpretability of the information need and relationship of items to it. In many recommendation contexts, we expect utilities estimates to be user-specific and perhaps idiosyncratic expressions of interest and taste to which a third-party evaluator would not have access.

Implicit labels refer to utility estimates derived from observed user behavior believed to be suggestive of utility.  Logged signals like clicks, streams, purchases, and bookmarking all have been found to have this property in information retrieval \citep{Kelly2003-ln}.  These signals depend critically on the domain.  A click may be suggestive of utility in web search but not in news recommendation where the headline alone might satisfy the user's information need.  The precise relationship between post-presentation behavioral signals is often complex deserving its own modeling effort.

Feedback like clicks, streams, or bookmarking all are meant to capture the \textit{instantaneously-measurable} utility of an item and may not provide an accurate reflection of its longer-term utility to a particular task or mood, or the lifetime value of the product to the user.  Clicking or inspecting a document may be suggestive of the utility of an item within the context of an individual ranking or system ordering but may have low utility for a user's higher level goal or task.  As such, when there is measurable, longer-term utility (e.g. task completion, item purchase, user retention), we can attempt to attribute that longer-term utility to earlier system decisions using methods from causal inference, reinforcement learning, and multicriteria optimization \citep{Mann2019-ou, Chandar2020-em}.

\subsection{Evaluating System Decisions}
Although understanding item utility is critical to evaluating an information access system, these items are presented in a structured output, usually a ranked list.  So, while the previous section described how we might estimate an item's utility, we are really interested in measuring the system's ability to provide users with high utility items in the course of their interactions with the system in support of their long term goal.  

\subsubsection{Situated Evaluation}
In an online environment, we can adopt situated evaluation by inspecting interaction data such as short-term and longer-term behavioral signals (e.g. clicks, streams, or purchases).  This data can be used to estimate the utility of consumed items (\cref{sec:estimating-item-utility}) and, by aggregating across users and needs, measure system performance.  In practice, because we are often comparing pairs of systems, these metrics are computed in well-designed A/B tests \citep{Kohavi2020-bu}.

\subsubsection{Simulated Evaluation}
\label{sec:access:eval:simulated}
In offline evaluation, we can simulate online evaluation using a combination of data and user \intro{browsing models}.  Simulation allows for highly efficient testing of algorithms, and avoids any risk to users, who may not be interested in being subject to A/B experiments.  If simulation data and models are freely shared amongst the research community, this allows for standard benchmarking protocols, such as have been adopted in NIST's TREC evaluations.  While the use of labeled data and an evaluation metric are not often considered simulation, one reason we adopt this framing is that it centers the key objective that should guide design decisions in such an evaluation: credibly estimating likely performance with respect to the information-seeking task(s) the system is designed to support.

The data involved in offline evaluation consists of estimated item utility for a set of information needs (\cref{sec:estimating-item-utility}), often called ``qrels'' (for query relevance).  Traditional evaluation such as used in most TREC competitions primarily uses explicit labels from assessors.

Given this data and a system decision (e.g. a ranking), the offline evaluation model simulates the behavior of a user engaging with the system decision.  At the termination of the simulation, a metric value (or values) is returned.  Although many offline evaluation methods can be computed analytically, each encodes a specific information access model, carrying assumptions about user behavior.  

In general, many analytic evaluation metrics involve an inner product between the vector of item utilities at each rank and a rank-discount factor \citep{Carterette2011-rs}.  This can be represented in general form as:
\begin{align}
    \evaluationMetric(\systemDecision) &= \sum_{r=1}^{|\repository|} \discountFactor(r)\itemUtility(\systemDecision_r)
\end{align}
where $\discountFactor : \ranks \rightarrow [0,1]$ is the rank ""discount function"" mapping from possible ranks to the $[0,1]$ interval, $\systemDecision\in S_{|\repository|}$ is the system ranking, and $\itemUtility : \repository \rightarrow \Re$ is a item utility (for this information need).  The implicit user browsing model is encoded in $\discountFactor$, which reflects the probability estimate that a user will inspect an item at a particular rank $r$.  So, for binary relevance, precision-at-$k$ can be defined with:
\begin{align}
    \discountFactor_{\text{P}@k}(r) &= \begin{cases} \frac{1}{k} & r \leq k\\
    0 & r>k
    \end{cases}
\end{align}
This corresponds to the expected utility for a user that randomly selects an item from amongst the top $k$ ranks.  For rank-biased precision \citep{Moffat2008-su},
\begin{align}
    \discountFactor_{\text{RBP}}(r) &= (1-\patienceParameter)\patienceParameter^{r-1}
\end{align}
where $\patienceParameter$ is a hyperparameter. This browsing model models a user who sees an item at rank position $r$ with exponentially decreasing probability.  This is proportional to the expected utility of the last item inspected.

\subsection{Evaluation Encodes Human Values}
Because many modern information access systems optimize performance toward an evaluation metric or utility, system designers should be aware of the variety of personal and societal values imbued in those definitions.  Guidelines for human assessors, heuristics to detect utility in log data, and selection of longer-term utility all are contestable phenomenon and should be interrogated before being incorporated into a production pipeline or presumed to be objective in an academic paper \citep{Stray2020-nk}.

\section{Algorithmic Foundations}
\label{sec:access:algo}

Information access algorithms are responsible for locating the items that will satisfy the user's information need.
These algorithms often work by estimate the utility of the document to an information need through a \intro{scoring function} $\relevance{\need, \doc}$, and using these utility estimates to rank results.
These "rankings" may also be stochastic \citep{Diaz2020-oa}, expressed as a \intro{policy} $\policy(\need)$ defining a distribution over (possibly truncated) rankings of documents.
Deterministic rankings can be treated as a policy placing all probability mass on a single ranking.

Our purpose here is not to provide a comprehensive treatment of information access algorithms but to provide enough information on their key concepts that scholars familiar with machine learning can understand the particularities of information access that are important for understanding how its fairness concerns differ from those of other ML applications.  Readers can find details on algorithmic foundations in a variety of information retrieval textbooks \citep{Van_Rijsbergen1979-sf, Manning2008-fi, Croft2010-yz}.

There are several distinguishing factors between different algorithmic approaches to meeting information needs:

\begin{itemize}
    \item What data about needs and items is used, and how is it represented?
    \item Is utility directly estimated or learned through optimization?
    \item For what objective are utility estimates optimized?
    \item How are utility estimates used to produce the final ranking?
\end{itemize}

For example, many techniques make use of item content $\featuresContent(\doc)$ in some way, but the family of recommendation algorithms called \intro{collaborative filters} ignore document content entirely and use patterns in historical user-item interaction records from $\implicitGlobal(\doc)$ as the sole basis for estimating relevance and ranking items.

\subsection{Vector Space Model}

A long-standing approach to representing items with substantial content information, especially documents and queries is to represent them as vectors in a high-dimensional space.
In the \emph{bag of words} model, this is done by treating each word (or \emph{term} $\term \in \vocab$) as a dimension and locating documents in $|\vocab|$-dimensional space based on the relative frequency with which they use different terms \citep{Salton1975-nb}.
With such an approach, a document $\doc$'s $\featuresContent$ and a need $\need$'s $\requestFeatures$ (and/or $\requestLanguage$) can be represented as vectors $\repr{\doc}$ and $\repr{\need}$, and the system can estimate relevance by comparing the vectors (often using the cosine $\relevance{\doc|\need} = \operatorname{cos}(\repr{\doc}, \repr{\need})$).

The precise definition of the vectors is usually based on the \emph{term frequency}, and --- for document representations --- normalized by the \emph{document frequency} to give greater weight to terms appearing in fewer documents (and thus more useful for locating documents relevant to information needs for which those terms appear in the query).
One common representation is \emph{term frequency --- inverse document frequency}, or TF-IDF:

\begin{equation*}
\termRepr{\doc}{\term} = \operatorname{TF}(\doc, \term) \cdot \operatorname{IDF}(\doc)
\end{equation*}

These vectors form the rows of the $|\repository| \times |\vocab|$ \emph{document-term matrix} $\termMatrix$.  If both document and query vectors are normalized to unit vectors, then the similarity can be estimated with the inner product $\relevance{d,q} = \repr{d} \cdot \repr{q}$.

Vector space models can also be used without query or document content.
Pure collaborative filtering algorithms compute recommendations based solely on users' $\implicitGlobal$ by using a \emph{ratings matrix} --- a partially-observed $|\fdset{U}| \times |\repository|$ matrix recording users' past interactions with items, either through implicit feedback (whether or how frequently the user has consumed the item) or explicit feedback (an ordinal or real-valued preference, such as a rating on a 5-star scale).
Relevance estimation is often done via neighborhoods: finding other users similar to the user needing information and scoring items by a weighted average of these neighbors' ratings \citep{Herlocker2002-tk}, or finding items similar to those rated by the current user \citep{Deshpande2004-ht}.
In both cases, $\implicitGlobal$ consists of the user's past ratings or interactions with items.

One important similarity between the document-term matrix and the ratings matrix is that they are both \textit{sparse} and \textit{incomplete}.  Most documents do not contain most words (sparsity); most documents do not contain all synonyms and paraphrases (incompleteness).  Most users have not consumed most items (sparsity); most users have not provided ratings for all items they have consumed (incompleteness).  

The term-based vector space model can also be integrated with user history for \emph{content-based recommendations}; in this case, a transformation of the items in the user's history, such as the sum or average of their term vectors, is used as a query vector to locate items that match $\implicitGlobal$ on the basis of their associated features or terms instead of user-item interaction patterns.

\subsection{Embedding and Optimizing Utility}
\label{sec:access:embedding}

\AP Two significant developments move beyond the vector space model, and form a key component of modern information access algorithms.
The first is representing (or \emph{embedding}) documents and information needs with a common lower-dimensional space (called a \emph{latent feature space}), resulting in an ""item embedding"".
Introduced for information retrieval as \emph{latent semantic analysis} \citep{Deerwester1990-ma}, one approach is to take the truncated singular value decomposition of the document-term matrix $\termMatrix=\fdmat{D}\Sigma\fdmat{T}$.
The left and right singular matrices of this decomposition provide a low-rank representation of documents ($\fdmat{D} \in \Reals^{|\repository|\times k}$) and a map from term vectors into this vector space ($\fdmat{T} \in \Reals^{k \times |\vocab|}$).
This facilitates compact and efficient document representation and comparison (for example, similar documents can be located by comparing their vectors in $\fdmat{D}$), and allows documents to match information needs that do not contain any of their terms but do contain synonyms.

Variants of this technique have seen widespread use in recommender systems \citep{Koren2009-fo}, where the ratings matrix is decomposed into low-rank representations of users and items.
The sparsely-observed nature of the ratings matrix and the computational complexity of SVD have led to a space of approximation techniques for matrix factorization.
In practice, the user-feature and item-feature matrices are inferred through stochastic gradient descent \citep{Funk2006-tm} or alternating least squares \citep{Takacs2012-mr} to minimize the reconstruction error on the observed values of the ratings matrix.

Learning decompositions through optimization is an instance of the second of these developments: estimating utility through machine learning models.
This is now the basis of most current information access research.
Models can become quite complex and incorporate multiple aspects of items and information needs simultaneously, but their fundamental operation is to learn a function $\relevance{\doc | \need}$, that estimates the item's relvance to the given need $\need$ based on observations, such as search result clicks, purchases, or product ratings.
These estimates can be rating predictions, in the case of recommender systems with explicit ratings, or other estimates such as the probability of the user clicking a result (studied under the label of \emph{CTR prediction}) or the probability that the document is relevant to the user's need (a common framing for search).
Learned utility models can also be implemented directly on a vector-space representation, as in the SLIM technique for learning neighborhood interpolation weights \citep{Ning2011-qu}.

\subsection{User Modeling}
\label{sec:access:user-modeling}

The \intro{user model} is the component of a personalized information access system --- recommendation, personalized search, and other systems that respond to $\implicitGlobal$ containing a user's historical interaction with the system --- that represents the user's preferences for use in the rest of the system.
It is often latent in the system architecture; for example, in the vector space model of collaborative filtering, the user model is just the set or bag of items with which the user has interacted with in the past, and in an embedding-based system it is the user's embedding (and the means by which this embedding is computed).

\AP One early user model for book recommendation by \citet{Rich1979-eh} represented users as probability distributions over a set of \textit{stereotypes} that was incrementally refined through text-based dialogue with the user.
Contemporary systems often learn latent user models from the user's history, typically taking the form of a ""user embedding"" (\cref{sec:access:embedding}).
Since user activity occurs over time and users' preferences are not necessarily stable, some techniques such as that of \citet{Koren2010-oi} decompose user preference into long-term \textit{stable} preference and short-term \textit{local} preference, to separately model the user's persistent taste and ephemeral current interests.

The fundamental idea, common to all these techniques and many more, is that a personalized system will often have a representation of $\implicitGlobal$, computed by a suitable statistical model, that is then used by the final scoring and ranking logic in order to estimate the relevance of an item to the user's information need in accordance with their personal preferences.
Whether this is an entirely separate model, computing and storing user model representations to be used as input data for a ranker, or a sub-component of an end-to-end ranking model depends on the system architecture.

\subsection{Learning to Rank}
\label{sec:access:ltr}

Learned relevance models are not limited to learning pointwise relevance estimates given observed utility signals.
\emph{Learning to rank} \citep{Liu2007-mr} moves past learning to predict pointwise item-need utility and instead optimizes the learning model to rank items consistently with their ability to meet the user's information need.
In the case of binary relevance (the simplest form of utility label), the system learns to rank relevant items above irrelevant items.
Such systems often still learn a scoring function $\relevance{\doc | \need}$, but the function is optimized to produce scores that correctly order documents, regardless of the precise values of those scores, instead of its ability to estimate relevance judgments.

One approach is to optimize \emph{pairwise ranking loss}; a key example of this in recommender systems is Bayesian Personalized Ranking \citep[BPR;][]{Rendle2009-fo}.
Given an information need $\need$, BPR optimizes the probability $\fdprob{s(\doc_+|\need) > s(\doc_-|\need)}$ for a randomly-selected relevant item $\doc_+$ and irrelevant item $\doc_-$ by maximizing $\operatorname{logistic}\left(s\left(\doc_+|\need\right) - s\left(\doc_-|\need\right)\right)$ (this is equivalent to maximizing the area under the ROC curve).
Pairwise loss functions can be applied to a wide range of ranking models (including matrix factorization, neighborhood models, and neural networks) for both recommendation and information retrieval.

\subsection{Re-ranking}
\label{sec:access:reranking}

Many information access techniques do not depend only on a single ranking step.
\intro{Re-ranking} approaches use a base ranking model --- which can be directly estimated, learned from pointwise optimizations, or a learning-to-rank model --- and adjust the outputs to achieve additional goals.

One application of re-ranking is to improve the \intro{diversity} of results.
""Maximum marginal relevance"" (MMR) adjusts the ranking to balance, at each position, maximizing $\relevance{\doc|\need}$ with minimizing the similarity between the new item and previous items \citep{Carbonell1998-qk}.
The idea behind this approach is that relevance is not independent: if one item does not meet the user's need, then another very similar item is also unlikely to meet their need, and therefore the second position should go to an item that is likely to match the query given that the first document did not \citep{Goffman1964-pg}.
\citet{Ziegler2005-zo} provides another approach to diversifying a result list that operates purely over item orderings instead of balancing similarity or relevance scores. As we will see later in this survey, re-ranking is a common approach for improving certain types of fairness in recommendation lists.

Another use for re-ranking is to improve the efficiency of a system facilitating access to a large repository.
Learned relevance or ranking models often have significantly higher computational cost than vector space models, which can be heavily optimized through index structures.
One approach, therefore, is to use a simple first-pass ranker to retrieve a pool of candidate items that is significantly larger than the final result list but much smaller than the repository.
A more complex ranking algorithm, possibly employing modern deep learning models, re-ranks these candidate items to produce the final ranking.

\chapter{Fairness Fundamentals}
\label{sec:fairness}

As noted in \cref{sec:intro:history}, the second decade of the 21st century has brought significant attention to the issue of fairness in computing systems, particularly (but not exclusively) machine learning and statistical tools for use in decision support contexts \citep{Mitchell2020-mt}.
This arises at the intersection of increasing adoption of machine learning technologies in sectors with direct real-world effects such as healthcare, public policy, and law enforcement, and extensive discussions on justice and equity in society at large.
Fairness but one of several dimensions in which the social impact of computing systems are under scrutiny; a community has coalesced around studying ``Fairness, Accountability, and Transparency''\footnote{The FAccT conference (\url{https://facctconference.org}) and related venues.}, and the topic arises in discussion fora on AI ethics and in the various communities working directly on artificial intelligence, machine learning, data mining, and information retrieval, among others.

In this \namecref{sec:fairness}, we provide a brief overview of the landscape of algorithmic fairness: its fundamental concepts and definitions, sources of unfairness or bias, some methods for reducing the unfairness of machine learning systems, and pointers to additional research topics.
We refer readers to papers by \citet{Mitchell2020-mt} and \citet{Barocas2016-wi}, as well as the in-progress book by \citet{Barocas2019-ft}, for more in-depth treatment of these topics.

Algorithmic fairness in general is concerned with going beyond the aggregate accuracy or effectiveness of a system --- often, but not always, a machine learning application --- to studying the distribution of its positive or negative effects on its subjects ("distributional harms") and the ways those subjects are represented by and in the system \citep["representational harms"; ][]{Crawford2017-js}. Much of this work has been focused on fairness in classification or scoring systems; \citet{Mitchell2020-mt} provide a catalog of the key concepts in this space, and \citet{Hutchinson2019-th} situate them in the broader history of fairness in educational testing where many similar ideas were previously developed. There are many ways to break down the various concepts of algorithmic fairness that have been studied in the existing literature, which we summarize in Table \ref{tab:concepts}. 

While ``algorithmic fairness'' is the label used for this research, it does not encapsulate one goal, but rather covers a spectrum of equity concerns.
\citet{Selbst2019-hf} identify a number of ``abstraction traps'' surrounding fairness research, one of which is the \emph{formalism trap}:

\begin{quote}
    Failure to account for the full meaning of social concepts
such as fairness, which can be procedural, contextual,
and contestable, and cannot be resolved through mathematical formalisms.
\end{quote}

This fundamental contextuality and contestability, combined with the incompatibility of disparate notions of fairness \citep{Friedler2021-ns}, imply that universal fairness is not an achievable (or arguably even meaningful) concept.

What can be done, and what much of the research on algorithmic fairness is concerned with, is to identify specific ways in which a system may be \emph{unfair}, and develop tools to measure and mitigate this unfairness.
Determining the ways in which a system may be unfair, and how those failure cases should be assessed and addressed in any particular application, is a domain- and application-specific process that needs to be carried out in consultation with a broad set of stakeholders and subject experts.

There are many different terms that are used, sometimes interchangeably and sometimes with nuanced differences, in scholarship and surrounding discourse on algorithmic fairness.
The terms ``bias'' \citep{Baeza-Yates2018-ob}, ``fairness'' \citep{Dwork2012-ai}, ``discrimination'' \citep{Kamiran2012-lu}, ``equity'' \citep{Katell2020-oz}, and ``justice'' \citep{Lee2019-cp}, among others, are used variously by different authors to discuss the interrelated concerns and phenomena under study.
In much of the literature, terms and concepts are borrowed from legal analysis, particularly the scholarly tradition around U.S. anti-discrimination law \citep{Barocas2016-wi}; \citet{Hoffmann2019-lg} discusses some of the limitations of this trend.
\citet[p. 60]{DIgnazio2020-ad} call for challenging the power structures of data science and its applications, and identify some terms (including ``bias'' and ``fairness'') with perspectives that uphold existing power structures and others (such as ``equity'' and ``justice'') with questioning and dismantling those structures.
Not all authors use terms in the same way, or with the same nuances.

Just as there are many ways in which a system can be unfair, there are also many places in which unfairness can be introduced in any particular system.
In \Cref{sec:fairness:sources}, we provide an overview of how each step of the pipeline may introduce biases that give rise to unfairness.

\begin{table}[tb]
    \centering\footnotesize
    \begin{tabular}{rl}
         $\Iin_i \sim \RVin$ & Input covariates for instance $i$ \\
         $\Iout_i \sim \RVout$ & Observed outcome for instance $i$ \\
         $\Iscorei$ & Computed score for instance $i$ \\
         $\Ideci$ & Decision for instance $i$ \\
         $\IinG_i$ & Sensitive attributes for instance $i$ \\
         $\IinX_i$ & Non-sensitive attributes for instance $i$ \\
         $\distin(\Iin_i, \Iin_j)$ & Distance between observations of instances $i$ and $j$ \\
         $\distdec(\delta(v_i), \delta(v_j))$ & Distance between decisions for instances $i$ and $j$ \\
    \end{tabular}
    \caption{Summary of notation for \cref{sec:fairness}.}
    \label{tab:fair-notation}
\end{table}

In this \namecref{sec:fairness}, we focus primarily on supervised classification problems and  adopt the notation and conventions of \citet{Mitchell2020-mt}.
Given an individual $i$ with observed features $\Iin_i$ and observed outcome $\Iout_i$, modeled as samples from random variables $\RVin$ and $\RVout$, the goal is to learn and evaluate a decision function $\dec{\Iin_i}$.
$\decfn$ is often structured as a (possibly probabilistic) decision process $f$ applied to an underlying score $\scorefn$, such that $\dec{\Iin_i} = f(\score{\Iin_i})$.
The goal is to make ``correct'' decisions, so that $\Iout_i = \Ideci$; we may also wish to accurately estimate probabilities, such that $\Iscorei = \fdcprob{\RVout = 1}{\RVin = \Iin_i}$.

Throughout this \namecref{sec:fairness}, we will use two examples: lending and hiring. Lending is a classic setting for classification: the decision function $\Ideci$ determines whether $i$ will be granted or denied a loan, and the goal is to grant loans to all applicants who will pay them (we assume for the moment no limit on funds to lend). A hiring decision is closely related to the task of job and candidate recommendation introduced in \namecref{sec:intro:running}. A decision might be to choose whether or not to move on a candidate to the next level of review, but the score $\score{\Iin_i}$ could also be used to rank candidates for presentation as recommendations to hiring manager. 

\AP One key concept in fairness, that we discuss in more detail in \cref{sec:fairness:individual,sec:fairness:group}, is the idea of \emph{individual} and \emph{group} fairness.  "Individual fairness" is concerned with treating similar individuals similarly, while "group fairness" is concerned with identifying and addressing differences between groups of data subjects.
These groups are often the kinds of groups treated in anti-discrimination law, such as gender, race, or religion; to represent these groups, we can decompose an individual's covariates $\Iin_i = (\IinG_i, \IinX_i)$, where $\IinG_i$ is the \intro{sensitive attribute(s)} (sometimes called ``protected characteristics'') recording group association and $\IinX_i$ is the other attributes (non-sensitive attributes).

\section{Sources of Unfairness}
\label{sec:fairness:sources}

Before discussing how to measure and precisely define unfairness, we first discuss where it can enter into the system.
The harms that are discussed under the rubric of fairness typically arise from some kind of bias, where observations or outcomes are different --- at least in expectation --- from what they would be if the bias or unfairness were not present.
These biases can arises in many places: in society, in the observations that form our data, and in the construction, evaluation, and application of decision support models \citep{Suresh2019-kt}.

\citet{Friedler2021-ns} use the idea of \emph{construct spaces} and \emph{observation spaces} to define biases in terms of skews in observation and decision-making processes.
The \ConceptDef{construct feature space} (CFS) contains the `true' features that we would use to make decisions in an ideal system, such as the applicant's ability to repay a loan or the job candidate's ability to carry out the duties of a position.

The CFS is unobservable; instead, we have access to the \ConceptDef{observation feature space} (OFS), which is the result of an observation process that results in the input features for the actual decision process ($\Iin_v$); the OFS also contains the observed outcomes $\Iout_i$ for training data. For example, in looking at a loan application, the decision system would see the applicant's current salary, assets, and other financial details and these would form the set of observations on which the decision is made. Similarly, a hiring system might see the candidate's recent employment history, educational credentials, etc. as its OFS.

Decisions are made on the basis of these features, yielding the \ConceptDef{observation decision space} (ODS); the goal is for these decisions to match what they would be if made on accurate and perfect information (without discrimination), the \ConceptDef{construct decision space} (CDS).

\begin{figure}[tb]
    \centering
    \includegraphics[width=\textwidth]{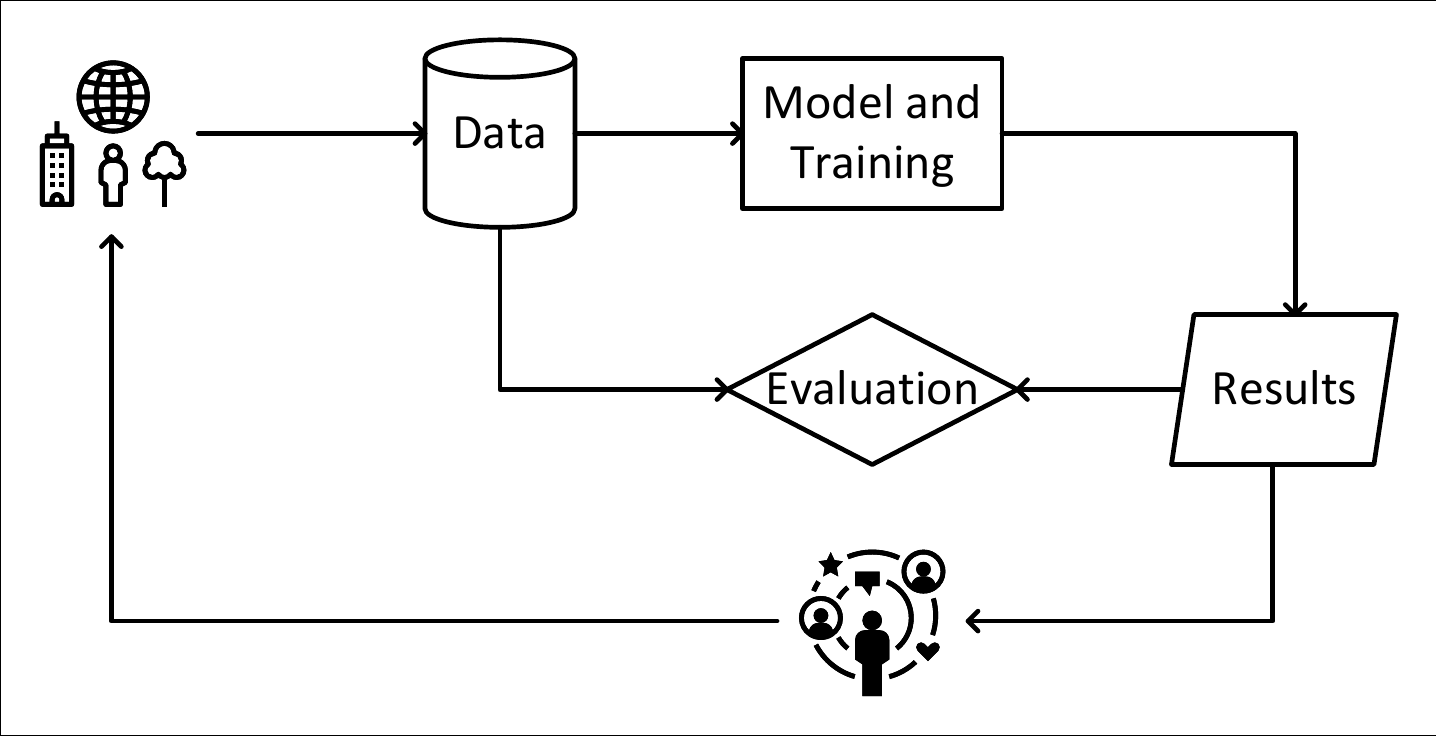}
    \caption[Machine learning pipeline]{A view of the machine learning pipeline.  The environment (natural and social) is observed through data. This data is used to develop and train a model, which produces results of some form, and is used to evaluate those results. The results are also acted on by people, individually or socially, who impact the environment for the next round of modeling.}
    \label{fig:ml-pipeline}
\end{figure}

We focus our own discussion on sources of unfairness on stages of the pipeline or feedback loop in algorithmic decision support.
\Cref{fig:ml-pipeline} shows the various elements of this process, from initial observations to human response the model that affects the natural and social world that will be observed in future iterations; it is the ML equivalent of \cref{fig:info-access}.
Unfairness can arise at any stage in this process, and may be propagated, mitigated, exacerbated, or even re-introduced either at that stage or at further downstream stages.

First, the \ConceptDef{world itself} may be unfair or unjust.
One source of such injustice is historical and ongoing discrimination.
For example, redlining in United States housing policies \citep{Rothstein2017-dc} prevented Black Americans from purchasing homes in wealthier neighborhoods; the neighborhoods where they were allowed to live typically had lower investment in parks, schools, and other amenities to improve quality of life and childhood development.
An entirely accurate survey of family wealth disaggregated by race will reveal significant racial disparities, not because there are innate racial differences in the ability to develop generational wealth (a difference in \citeauthor{Friedler2021-ns}'s \emph{construct space} [\citeyear{Friedler2021-ns}]), but because Black residents were prohibited by private-sector and governmental policy from accessing the same opportunities for wealth-building through home ownership as their White would-be neighbors. Thus, the OFS in a loan decision system, focused on financial indicators, would inevitably look quite different for the average Black applicant as opposed to the average White one. 

Redlining has further follow-on effects that result from things such as disparate quality of education.
The effects of such discrimination also often extend beyond the end of official practice and policy. As with lending, a hiring decision maker looking at educational aspects of candidates' applications may see differences in attainment (OFS) reflective of this history rather than being reflective of differences in ability (CFS) that are relevant for the position.

Group size can also play a role in some types of unfairness \citep{Rolf2021-ny}.
If one group is smaller than another, na\"ive modeling may be more accurate at inferring and predicting things for the majority group, and may be more likely to be incorrect for minority groups.

In addition to capturing bias and injustice in the world, \ConceptDef{data collection} may introduce unfairness into the system.
Sampling strategies determine who is considered for inclusion in the data. Response or submission bias, where some people are more likely to volunteer information or respond to requests than others, can also introduce additional unfairness. 
Selecting and defining variables is crucial as observations or proxies need to be valid and unbiased measurements of the target construct (to avoid \emph{measurement bias}), and their response needs needs to be consistent across different groups in a population \citep[\emph{measurement invariance}, ][]{Steenkamp1998-yj}. For example, it wasn't until September 2021 that US mortgage lender Fannie Mae opted to use on-time rental payments as a measure of creditworthiness, in addition to traditional loan repayment data \citep{Lerner2021-zy}. Collecting only loan repayment data excludes individuals who have historically had less need to borrow or who have religious objections to lending. The selection of particular data as a marker of creditworthiness therefore excludes evidence that might be favorable to a borrower.

The problems above can all introduce bias in the data under its own terms: assuming a set of information goals, the data is biased with respect to its ability to meet those goals.
However, data collection can also be biased by the perspectives that inform what is being measured and how, and how the relevant constructs are defined \citep[``When you measure include the measurer.'', ][]{Hammer2021-ql}.
The decisions that are made in collecting --- from defining its initial goals to defining the variables to record --- reflect the perspectives of the people involved in the process; without broad stakeholder engagement, the data set may be biased \emph{in concept} relative to the needs and goals of a subset of the people it will affect, and without clear documentation of the perspectives and assumptions that went into its design, these biases may be undetectable \citep{Hutchinson2021-vu}. For example, it is well-known that in schools in the US, Black and White students are subject to disciplinary actions at very different rates \citep{Okonofua2015-zv}, and this is at least partly due to the actions of children and youth being perceived differently by teachers and administrators, conditioned by race \citep{Okonofua2016-vn}.

Many variables also require a codebook to determine how the variable is recorded, particularly if it is categorical in nature, and these codebooks reflect specific perspectives in how observations should be recorded.
This becomes particularly salient with sensitive personal attributes such as gender or race; racial categories are often adapted from administrative data collection efforts, but these categorizations vary across time and space and are the results of substantial political processes in determining how to record a fundamentally social construct \citep{Hanna2020-ln}.
Finally, data needs to be collected and interpreted in light of its social and cultural context to avoid unfairly disregarding local knowledge and perspectives.

\index{societal bias|see {bias, societal}}
\index{statistical bias|see {bias, statistical}}
\AP \citet[\S 2.1]{Mitchell2020-mt} provide a statistically-oriented view on the sources we have discussed so far: \intro{societal bias} results in deviations between the ``world as it should and could be'' and the ``world as it is'', and \intro{statistical bias} results in ``systematic mismatch'' between the world as it is and the collected data and observations, including both sampling biases and measurement error.
Redlining is an example of societal bias, because it results in people living in different housing situations than they would if there was no racial discrimination in housing policy (the world as it should be); a systematic mismatch between actual housing situations and their records in the data, for example due to differential reporting, would be a statistical bias.

Machine learning \ConceptDef{models} can also introduce unfairness. Such unfairness may arise from direct use of sensitive attributes (e.g., gender and race) in the model. Models may learn to discriminate indirectly from other proxy variables present in the data.
The objective functions for which a model is optimized further encode specific perspectives about what constitutes a ``good'' model, sharing the respective challenges in determining how to define, collect, and encode data.

Unfairness can arise in \ConceptDef{evaluating} a machine learning algorithm or application in various ways.
All of the problems we have discussed regarding input data apply to evaluation, to the extent that the data is being used to evaluate the model (and it usually is, at least as an initial validation step, even when the final evaluation will involve the results of application deployment).
Further, perspectives captured in the definition of success can skews evaluation outcomes, as success for some stakeholders does not mean success for all \citep{Barocas2021-tv}. 
It is crucial to identify who are the stakeholders of the system, and determine whose utility is reflected in the evaluation metric(s).
This connects to additional questions on the accountability of the system: Who gets to make the decision on who is being served through the design choices? To whom are they accountable?

For example, if a loan decision model is evaluated on the basis of historical data from lenders, it will most likely have thorough information about those individuals to whom loans were given, but little or no information about those whose loans were denied. Thus, the system can learn about false positive decisions but not false negatives, magnifying whatever blind spot caused such errors in the first place \citep{ONeil2017-bf}.

Details in evaluation can also make a difference in the system's fairness \citep{Barocas2021-tv}.
Measuring performance averaged over all subjects will prioritize performance on the majority group, while disaggregating and reweighting performance metrics can favor systems that perform equally well across groups regardless of group size.

Finally, unfairness may be introduced by \ConceptDef{human response} to the system's output: humans and algorithms do not necessarily compose \citep{Srivastava2019-mx}. Model output might influence stakeholders to respond differently, and their response may in some cases be inversely correlated with computational fairness.
For example, \citet{Green2019-sx} found that providing algorithmic risk scores \emph{increased} racial disparity in human assessments of risk in a laboratory setting, even if the scores themselves were racially fair; \citet{Albright2019-cm} found similar results in a study of actual judges' decision-making behavior.
Other social factors may skew human response to a system; for example, community support in assisting loan application and repayment might skew individual's response to loan prediction model outcomes.

As we note, unfairness may enter the system at any of these points, and each requires different interventions and measurements.
Further, bias that is removed at one stage may be re-introduced in another, such as a model re-learning bias from a proxy variable even though the data set removed disparate representation, or a human responding in a biased fashion to unbiased predictions or scores.

\section{Problems and Concepts}
\label{sec:fairness:concepts}

Measuring and mitigating unfairness in a system requires us to identify several things:
\begin{enumerate}
\item Who is experiencing unfairness?
\item How does that unfairness manifest?
\item How is that unfairness determined or measured?  
\end{enumerate}

The answers to each of these questions flow from normative principles that motivate why and how a particular fairness study or intervention is being undertaken.
Being clear about these principles and related assumptions and goals is crucial to doing coherent work, evaluating its ability to meet its goals, and assessing the relevance and appropriateness of those goals to the social problem in question.

\begin{table}[tb]
    \centering\footnotesize
    \begin{tabular}{p{0.30\textwidth}p{0.63\textwidth}}
    \toprule
    Distributional harm
    & Harmful distribution of resources or outcomes.
    \\
    Representational harm
    &  Harm internal or external representation.
    \\
    \midrule
    Individual fairness & Similar individuals have similar experience.
    \\
    Group fairness & Different groups have similar experience.
    \\
    \hspace{1ex}
    Sensitive attribute
    & \hspace{1ex} Attribute identifying group membership.
    \\
    \hspace{1ex}
    \textit{Disparate treatment}
    & \hspace{1ex} Groups intentionally treated differently. \\
    \hspace{1ex}
    \textit{Disparate impact}
    & \hspace{1ex} Groups receive outcomes at different rates.
    \\
    \hspace{1ex}
    \textit{Disparate mistreatment}
    & \hspace{1ex} Groups receive erroneous effects at different rates.
    \\
    \midrule
    Anti-classification
    & Protected attribute should not play a role in decisions.
    \\
    Anti-subordination
    & Decision process should actively undo past harm.
    \\
    \bottomrule
    \end{tabular}
    \caption{Summary of concepts in algorithmic fairness and harms.}
    \label{tab:concepts}
\end{table}

As we noted in the introduction of this \namecref{sec:fairness}, there has been a shift in the field's discourse from pursuing fairness as a potentially-universal goal in itself to the perspective we describe of identifying and addressing specific fairness-related harms.
These harms and approaches for addressing them can be categorized according to a number of concepts, summarized in \cref{tab:concepts}.
As with our notation, we draw heavily from the work of \citet{Mitchell2020-mt} in framing this \namecref{sec:fairness:concepts}, although our organization is somewhat different.

\subsection{Distribution and Representation}
\label{sec:fairness:dist-rep}

The first axis we consider is \emph{harms of distribution} vs. \emph{harms of representation} \citep{Crawford2017-js}. 
\intro{Distributional harms} arise when someone is denied a resource or benefit \citep{Crawford2017-js}; unfairly denying loans, for example, to a group of people.

\intro{Representational harms} arise when the system represents groups or individuals incorrectly, either in its internal representation (e.g. word embeddings that encode stereotyped expectations in the latent embedding space \citep{Bolukbasi2016-uk}) or in how it represents those people to others.
Systems can cause direct representational harm, by misrepresenting people to themselves or others, and representational harms can also cause distributional harms by affecting how the system makes decisions and allocates resources.

Most literature so far on algorithmic fairness considers distributional harms; representational fairness is often introduced as a tool for reducing distributional harms.

\subsection{Individual Fairness}
\label{sec:fairness:individual}

Another axis of unfairness often considered is \textit{individual} versus \textit{group} fairness. \intro{Individual fairness} sets the goal that similar individuals should be treated similarly: given a function that can measure the similarity of two individuals with respect to a task, such as the ability of job applicants to perform the duties of a job, individuals with comparable ability should receive comparable decisions \citep{Dwork2012-ai}.
This fairness concept is grounded in the normative principle that like cases should be treated alike \citep{Binns2020-bm}, a notion of justice that traces back to Aristotle.

Individual fairness is typically operationalized with a task-specific distance metric $\distin$ between individuals and another $\distdec$ between decisions (or decision distributions, in the case of probabilistic decision processes).
Formally, the decision process is deemed fair if similar individuals receive similar decisions:

$$
\distin(\Iin_i,\Iin_j) \le \delta \implies \distdec(\Idec{i}, \Idec{j}) \le \epsilon
$$

There are several important things to note about about this metric:

\begin{itemize}
\item While it constrains decisions on similar instances, it makes no requirements on dissimilar instances: making the same decision for highly dissimilar individuals does not violate individual fairness.
\item It depends on the existence of a fair distance function $\distin$; if the distance function is unfair in some way (e.g. job candidates of the same skill but different races are further apart than candidates with the same race but differing skill levels), individual fairness cannot correct for that.
\item It effectively requires the decision process $\Ideci$ is probabilistic; in fact, individual fairness is impossible to fully satisfy with non-probabilistic discrete decisions \citep{Friedler2016-pb}.
\end{itemize}

The dependence on a fair distance function is particularly important to consider when identifying the assumptions that underly applications of individual fairness.
\citet{Friedler2021-ns} identify axiomatic assumptions for providing different kinds of fairness; one, ``what you see is what you get'' (\intro{WYSIWYG}), is the assumption that available data (the OFS of $\Iin_i$) is an unbiased representation of underlying reality; that is, there is no systemic bias or discrimination that affects the data gathering process.
Under the WYSIWYG assumption, individual fairness is directly applicable: observations $\Iin_i$ can be compared with an applicable distance function, and this suffices to enable the use of the individual fairness constraint to achieve the goal of treating like instances alike.

However, if systemic discrimination is present, either due to a gap between the world's ideal and current states or biases in the data gathering process \citep{Mitchell2020-mt}, then we cannot assume that similarity in the OFS is equivalent to similarity in the CFS, and thus individual fairness is only treating individuals similarly if they are similar in the (biased) observation space; it makes no guarantees about their treatment with respect to their similarity in unobserved construct space.
\citet{Friedler2021-ns} characterized one family of commonly-used assumptions to address such discrepancies as ``we're all equal'' (\intro{WAE}): taking as an axiom the idea that different groups are fundamentally the same with respect to the task, and thus any systematic discrepancy in observation space (e.g. members of different racial groups tending to be dissimilar in the OFS) is the result of discrimination and should be corrected for; \citet[p. 140]{Friedler2021-ns} note that the assumption can be interpreted either as ``members of different groups are the same'' or ``members of different groups should be treated the same for the purposes of our task'', and the resulting math is equivalent.
The WAE view is seldom taken in individual fairness literature, but \citet{Binns2020-bm} argues, and \citet[p. 142]{Friedler2021-ns} concur, that this is not a fundamental limitation of individual fairness, as WAE can be used to construct a distance function that takes group-based discrimination into account, for example by performing group-wise normalization of features.

\subsection{Group Fairness}
\label{sec:fairness:group}

\index{dominant group|see{group, dominant}}
\index{group!unprotected|see {dominant}}
\index{majority group|see {group, dominant}}
\index{group!majority|see {dominant}}

\index{protected group|see{group, protected}}

\intro{Group fairness} is concerned with ensuring that different \intro{groups} have comparable experiences with the system in some way.
The groups in question are often gender, race, ethnicity, religion, and other group associations used in anti-discrimination law, but the goals and definitions of group fairness are not limited to these groups.
As noted previously, group membership is usually formally denoted through a sensitive attribute $\IinG_i$.
Often two groups are considered: a \intro{protected group} $\IinG_i=\Gprotected$ and a \intro[dominant group]{dominant group} (sometimes called the \textit{unprotected group} or \textit{majority group}) $\IinG_i=\Gmajority$, and the goal of the system is to ensure that the protected group is not unfairly (mis)treated.

As with individual fairness, group fairness can flow from either WYSIWYG or WAE assumptions; it is also possible to conceive of WAE as a prior instead of an axiom, but this is seldom done explicitly in the relevant literature.

There are many ways to categorize group fairness constructs; in this \namecref{sec:fairness:group}, we are going to organize them along the lines of legal concepts that inspired them.

\subsubsection{Disparate Treatment}
\label{sec:fairness:disp-treat}

\intro{Disparate treatment} is when members of different groups are intentionally treated differently \citep[\S II.A]{Barocas2016-wi}.
The straightforward way for disparate treatment to manifest is as a direct property of the model, and can be removed by simply not using group membership as an input to the model.

Disparate treatment can also arise through the modelling and feature engineering processes, by selecting features known to correlate with group membership for the purpose of treating groups differently.
When such intention exists, it makes more sense to address it at a process and policy level; its statistical effects will be equivalent to disparate impact.

\subsubsection{Disparate Impact and WAE}
\index{parity|(}
\intro{Disparate impact} is when different groups have different impact from the system's decisions: they experience decisions at different rates.
In algorithmic fairness, this is formalized through \intro{statistical parity} measures.
Given a sensitive attribute $\IinG$, statistical parity is satisfied if decisions are independent of group membership:

$$
\fdcprob{\dec{\Iin} = 1}{a} = \fdprob{\dec{\Iin} = 1}
$$

In the two-group case, this is often defined as parity in outcomes between the two groups:

$$
\fdcprob{\dec{\Iin} = 1}{\IinG = \Gprotected} = \fdcprob{\dec{\Iin} = 1}{\IinG_i = \Gmajority}
$$

Statistical parity measures thus reflect a ``\kl{we're all equal}'' assumption \citep{Friedler2021-ns}; if different groups are fundamentally the same in their loan qualification, then we expect them to repay loans at the same rate, and thus they should receive positive decisions at the same rate.

In U.S. anti-discrimination law, the disparate impact doctrine \citep{Barocas2016-wi} is a test for potential discrimination in regulated decision-making processes such as employment and housing. It is usually operationalized via the ``four-fifths rule''\index{four-fifths rule}; a decision-making process $\decfn$, such as a part of the hiring procedure like a strength test, violates disparate impact under this standard if $\fdcprob{\dec{\Iin} = 1}{\IinG = \Gprotected} < 0.8 \fdcprob{\dec{\Iin} = 1}{\IinG = \Gmajority}$ --- that is, the protected-group pass rate is less than four-fifths of the dominant-group pass rate.

Crucially, however, disparate impact is only one part of a broader scheme for determining whether unlawful discrimination has occurred.  A finding of unlawful discrimination based on disparate impact requires that (1) disparate impact occurs, and (2) it either does not serve a legitimate business purpose, or there would be a less discriminatory way of achieving that business purpose.  This is implemented through a burden-shifting framework:

\begin{enumerate}
    \item The plaintiff shows the challenged practice has disparate impact.
    \item The defendant shows a legitimate business purpose for the practice.
    \item The plaintiff shows a less discriminatory mechanism that would achieve the business purpose.
\end{enumerate}

There is much subtlety in how these rules are implemented and the burden of proof at each stage needed in order for the defendent to be liable for violating anti-discrimination law; \citet{Barocas2016-wi} provide more detail.
The key point for our purposes in this monograph, however, is that statistical parity measures are useful for detecting where discrimination may be occurring, and they are useful objectives in situations where ``WAE'' is the appropriate normative assumption, but they are often not sufficient evidence of discrimination, particularly for establishing liability.

\subsubsection{Error-Based Constructs and WYSIWYG}
\label{sec:fairness:error-wysiwyg}

The next family of group fairness constructs is based on classification or prediction error.
Crucially, these metrics make a "WYSIWYG" (``what you see is what you get'') assumption \citep{Friedler2021-ns}, at least for the outcome variables $\Iout_i$: they assume recorded outcomes are correct and unbiased, and the goal is to use these as a reference point for ensuring that groups are not mistreated.
Methods optimizing these objectives may vary in their assumptions about the bias in $\Iin_i$.

\intro{Error parity}, sometimes called \reintro{disparate mistreatment}, ensures different groups do not experience erroneous decisions at different rates, conditioned on their true outcomes \citep{Zafar2017-qx}. 
In our lending example, if $\operatorname{FNR}_\Gprotected > \operatorname{FNR}_\Gmajority$, then the protected group is more likely to be denied loans that they would pay off.
Fair FNR can be operationalized through an independence constraint:

$$\fdcprob{\dec{\Iin_i}=0}{\Iout_i=1, \IinG_i} = \fdcprob{\dec{\Iin_i}=0}{\Iout_i=1}$$

Similar constraints can be derived for other metrics such as FPR.

\intro{Recall parity}, sometimes called \reintro{equality of opportunity}, ensures that members of different groups are equally likely to receive a favorable positive decision conditioned on positive outcome \citep{Hardt2016-ut}.
Under this objective, the decision process is fair when:

$$\fdcprob{\dec{\Iin_i} = 1}{\Iout_i = 1, \IinG_i} = \fdcprob{\dec{\Iin_i} = 1}{\Iout_i=1}$$

A third category of group fairness objectives that rely on outcomes look at the predictive utility of the decision process. This takes a couple of flavors; we can look at \intro{predictive value parity} in the decision process, and require that decisions for each group have the same positive predictive value \citep{Chouldechova2017-ea}:

$$
\fdcprob{\Iout_i = 1}{\dec{\Iin_i} = 1, \IinG}
= \fdcprob{\Iout_i = 1}{\dec{\Iin_i} = 1}
$$

We can also define similar constructs on any marginal of the confusion matrix \citep{Mitchell2020-mt}; the metrics in this \namecref{sec:fairness:error-wysiwyg} are not an exhaustive list.

We can also look at \intro{calibration parity}, requiring that the underlying scores are equally well-calibrated for each group \citep{Kleinberg2017-ok}:

$$
\fdcprob{\Iout_i = 1}{\score{\Iin_i}, \IinG}
= \fdcprob{\Iout_i = 1}{\score{\Iin_i}}
$$

Finally, in settings where the system is learning stochastic decision policies, such as reinforcement learning and multi-armed bandit scenarios, \citet{Joseph2016-hi,Joseph2018-by} have proposed \intro{meritocratic fairness}, which prohibits the system from preferring a less-qualified candidate over a more-qualified one: if $\Iout_i$ is a continuous measure of qualification and $\Iout_i \ge \Iout_j$, then $\fdprob{\Idec{i}=1} \ge \fdprob{\Idec{j}=1}$.  This construct prevents preference inversions, but does not place any bound on how large the gap in decision probabilities can be; the system can strongly prefer a mildly more qualified candidate without violating meritocratic fairness.

In practice, any of these metrics or objectives knowing the outcome $y$ at the time of decision making. They are useful in training and evaluating supervised classification algorithms under the "WYSIWYG" assumption for the historical training labels, however.

Error-based metrics have intuitive appeal, because they encode a notion of fairness that, on its face, seems quite desirable: that if two people are both qualified for a beneficial decision, their race, gender, or other group membership should not affect the decision process.
They are similar in that respect to individual fairness, using recorded outcomes as the definition of ``similar''.
There are, however, at least three important limitations for the use of these metrics:

\begin{itemize}
    \item "WYSIWYG" is a strong assumption about the accuracy and lack of bias in training labels; in some cases, this assumption amounts to assuming away the problem we are trying to solve.
    
    \item There are fundamental tradeoffs between them.  The Chouldechova-Kleinberg theorem\index{Chouldechova-Kleinberg theorem} \citep{Chouldechova2017-ea,Kleinberg2017-ok} states that it is impossible to simultaneously equalize more than two different error parity metrics unless the underlying base rates are equal or the classifier is perfect.
    Equal FPR, equal FNR, and equal PPV are all desirable properties, but in the presence of unequal base rates and imperfect models, they cannot be simultaneously achieved.
    \citet{Pleiss2017-sg} document incompatibilities between calibration and error-based parity, particularly when base rates are not equal.
    \citet{Friedler2021-ns}, however, note that the "WAE" assumption amounts to assuming the base rates \emph{are} equal, so this set of tradeoffs is no longer in effect; this assumption also often entails biased error in the training and evaluation labels.
    
    \item Observed outcomes, for either training or evaluation, are only available from a subset of the data \citep{Ensign2018-bt}.  In our lending example, repayment data is only available for loans that have been issued --- if the bank does not make a loan, they cannot observe if it is repaid.
    This is very connected to the dynamic in information access that we only obtain feedback on results that are shown to users.
\end{itemize}
\index{parity|)}

\subsection{Awareness, Treatment, and Impact}

An important early result in algorithmic fairness \citep{Dwork2012-ai} is that ``fairness through unawareness'' --- that is, trying to achieve fairness by completely ignoring protected group status --- does not work.
Group identity often correlates with other variables which can serve as proxies for group membership \citep{Feldman2015-ra}; for example, different education or income levels between groups due to societal discrimination.
Ignoring protected group membership can therefore result in models that are unfair under any of a number of definitions.
"Individual fairness" addresses this through the notion of \emph{similarity with respect to task} \citep{Dwork2012-ai}, which may need to compensate for group differences \citep{Binns2020-bm}.
"Statistical parity" addresses this through a "WAE" assumption, resulting in metrics that require groups to experience positive decisions at similar rates.
"Error parity" addresses this through enforcing group parity in decision (in)accuracy, which often requires group labels at least in the evaluation --- if not the training --- stage of model-building.

\citet{Lipton2018-nx} address the general question of whether disparate treatment --- explicitly treating members of different groups differently --- is necessary in order to reduce disparate impact, and argue that disparate treatment is more effective and easier to reason about than more indirect fairness interventions aimed at reducing disparate impact.

\subsection{Motivations and Theories}
\label{sec:fairness:theories}

In addition to different assumptions, and related to different specific goals, fairness objectives can also flow from different fundamental philosophies about the purpose and function of fairness.
In U.S. legal theory, there are --- broadly speaking --- two different motivations for anti-discrimination law \citep{Barocas2016-wi,Xiang2019-kw}.
Under \intro{anti-classification}, the goal of anti-discrimination is to remove the protected characteristic from the decision process: race is not a factor in whether or not someone gets a job or a loan.
The theory of \intro{anti-subordination}, however, says that this does not go far enough, because the effects of past discrimination carry forward in time.
Under this theory, anti-discrimination law and practice need to actively work to reverse the effects of past discrimination and oppression.
Both of these theories can be found motivating fairness literature; sometimes they will converge to some of the same technical constructs, at least as an intermediate step, but they lead to different end goals and sometimes different rhetoric.
The distinction in rhetoric we discussed at the beginning of \cref{sec:fairness} is related to this distinction in policy motivations.

\section{Mitigation Methods}
\label{sec:fairness:methods}

Just as bias can enter the system at different stages in the pipeline (\cref{fig:ml-pipeline}), mitigation techniques can also be applied at different stages.
In this \namecref{sec:fairness:methods} we briefly outline some approaches from the existing machine learning literature; more details can be found in fair machine learning survey papers \citep{Mehrabi2019-no, Caton2020-xu}.

Great care is needed in selecting and evaluating sites of intervention for improving the equity of a system.
Different types and sources of fairness may be best addressed by interventions at different points, but not necessarily at the source.
Further, we cannot assume that fairness composes (e.g. improving the fairness in one stage does not guarantee that downstream stages in the decision-making process do not re-introduce unfairness or introduce new kinds of unfairness); \citet{Dwork2018-wf} argue it is necessary to assess the fairness of the entire system in the context of its actual use and application.

\subsection{Preprocessing}

One intervention site is to remove biases from the input data.
Simply removing  sensitive attributes does not necessarily lead to a non-discriminatory model outcome, as from other features in the dataset might encode information for inferring the sensitive attribute \citep{Dwork2012-ai, Feldman2015-ra}.

\cite{Kamiran2012-lu} propose four different methods for de-biasing data: 

\begin{itemize}
    \item Suppressing sensitive attributes or attributes correlated to the sensitive attributes; this can reduce discrimination in downstream tasks in some cases.
    \item ``Massaging'' the data by altering class labels from negative to positive for sensitive groups and vice versa until discrimination is minimized; this extends another technique by \citet{Kamiran2009-fk}.
    \item Re-weighting the data by carefully assigning weights to certain inputs to reduce discrimination.
    \item Stratified sampling strategies to repeat or skip samples to reduce discrimination selectively.
\end{itemize}

\citet{Feldman2015-ra} propose a repair procedure for data sets by altering observed features ($\IinX$) in the data set to eliminate their utility for predicting sensitive attributes ($\IinG$).  This is a purely correlational or predictive approach, that removes any correlation between sensitive and insensitive attributes regardless of potential causal connections.

\cite{Salimi2019-ij} introduce the idea of \emph{interventional fairness}, using causal directed acyclic graphs to represent functional interactions between variables;  they then divide features into \emph{admissible} and \emph{inadmissible}, where admissible features are those that justifiably influence a decision outcome.  They then define fairness as when all outcomes are causally independent for any combination in the superset of inadmissible variables, and require classifiers to be trained on data sets that satisfy this notion of conditional independence.  They finally repair the database by inserting and deleting certain tuples to ensure it satisfies the conditional independence constraint.

Beyond bias arising from the initial data set, feedback loops can feed model bias back into the data set.
For example, in predictive policing, the only feedback the system receive are influenced by the decision already made by the system, such as which neighborhood to patrol, and crime reported in those neighborhoods only. These feedback loops can result in substantially increased discrimination.  \citet{Ensign2018-bt} document this phenomenon and provide a mitigation strategy that selectively filters the feedback on the model decisions.

\subsection{Representation Learning}

Many machine learning models operate by learning representations that can be used for (possibly multiple) downstream tasks. Imposing fairness constraints on the representations may lead to non-discriminatory output in the downstream tasks \citep{Zemel2013-kq, Madras2018-at, Lahoti2019-qh}. The key idea in fair representation learning is to, as \citeauthor{Zemel2013-kq} put it, ``lose any information that can identify whether the person belongs to the protected subgroup, while retaining as much other information as possible''.

Representational learning maps the data distribution to a latent distribution where the latent distribution satisfies some desired properties. Fair representation learning can be formulated as multi-objective optimization problem, simultaneously minimizing information loss and removing information related to sensitive attributes. Several approaches include probabilistic mapping of the input data to prototypes \citep{Zemel2013-kq}, matrix transformation \citep{Lahoti2019-qh}, and fairness as adversarial objectives \citep{Madras2018-at, Feng2019-vr, Beutel2019-mv}

\subsection{Fairness-Aware Decision Models}

We can also alter the objectives of the decision model itself to include fairness.
This typically takes the form of incorporating one or more of the objectives described in \cref{sec:fairness:concepts} into the model's objectives or training process; they therefore face the tradeoffs and incompatibilities inherent to the various constructs and metrics.
One common approach ins \ConceptDef{regularization}: incorporating one or more fairness objectives penalty terms to the loss function to discourage unfair models.
In many cases, existing loss functions are enhanced with regularization terms in order to strike a balance between non-discrimination and accuracy on the training data \citep{Chakraborty2017-us}.

In \ConceptDef{constrained optimization} approaches, fairness is formulated as a constraint on parts of the confusion matrix at training time \citep{Mitchell2020-mt, Caton2020-xu}. Both regularization approaches and constrained optimization approaches can be unstable, i.e., small changes in dataset might affect the outcome \cite{Friedler2019-aq}.

\ConceptDef{Adversarial learning} can also be applied with an adversarial model attempting to identify unfairness in the primary model's outputs \citep{Celis2019-nc}.
This can also be formulated as a multi-constrained optimization problem \citep{Caton2020-xu}.
\citet{Xu2019-fd} present a more complex adversarial learning approach that attempts to model causal factors with twin generators modeling observed and fair versions of the observed data, with discriminators separating generated from real data and separating the protected and unprotected groups.

For algorithm-in-the loop decision making, \cite{Noriega-Campero2019-sz} propose that decision makers can adaptively collect information to ensure fairness for groups and individuals as necessary. This involves an iterative process between modeling and data gathering or preprocessing, so it is not strictly a model approach; it does, however, implement the observation of \citet{Chen2018-qq} that different causes or types of unfairness may need different interventions, and some can be addressed by collecting more or less data.

\subsection{Postprocessing}

Post-hoc fairness leaves the data and model alone, but post-processes the model outputs in order to provide fairness.
One technique is through \ConceptDef{thresholding}, i.e, using different decision boundaries for different groups to ensure non-discriminatory outcome under some definition \citep{Kamiran2012-lu, Kleinberg2018-gx}.
We treat the postprocessing technique of reranking in more detail later in this monograph, as it is an important tool for fairness-aware information access.

\section{Wrapping Up}
\label{sec:fairness:conclusion}

The algorithmic fairness literature has identified a number of different constructs for measuring and reducing unfairness and discrimination in machine learning systems.
These constructs are not all conceptually or mathematically compatible; different ones flow from different ethical goals and assumptions about the data, its social context, and the harms to be prevented or ameliorated.
For any given application, it is crucial to clearly and precisely describe the problems at play from ethical and/or legal perspectives, and to select or derive constructs that operationalize a suitable set of objectives.
We further cannot assume that any particular fairness objective composes with other parts of the system, or that any particular solution translates cleanly to other problem settings \citep{Selbst2019-hf}. Fairness needs to be defined, assessed, and ensured in a specific problem setting in light of its full sociotechnical context \citep{Dwork2018-wf}.

We have only been able to provide a very brief introduction to algorithmic fairness in this \namecref{sec:fairness}.
We refer readers to \citet{Barocas2016-wi}, \citet{Selbst2019-hf}, \citet{Mitchell2020-mt}, \citet{Dwork2018-wf}, and \citet{Suresh2019-kt} for further study.
\chapter{The Problem Space}
\label{sec:space}

"Information access systems" introduce some fundamental twists to problems of fairness and discrimination, making it difficult to directly apply the fairness concepts for other machine learning settings surveyed in \cref{sec:fairness}.
These difficulties arise from a number of differences, including the addition of multiple classes of stakeholders \citep{Burke2017-ne}, the rivalrous nature of allocating retrieval opportunities \citep{Introna2000-ov, Azzopardi2008-et, Biega2018-zl, Diaz2020-oa}, and the immediacy of the interactive feedback loop \citep{Chaney2018-us, Khenissi2020-wt}; classification-oriented fairness definitions are not necessarily well-suited to assessing these situations.

In this chapter we examine, from several different perspectives, the key considerations involved in applying fairness concepts to information access problems.
This starts with discussing how the classification fairness constructs described in \cref{sec:fairness} break down when applied to information access. We then discuss unique types of harms accruing in such systems, the potential for fairness concerns across multiple stakeholders, the variety of fairness constructs potentially at work, and the connections between these constructs and the broader space of information access research and fairness, accountability, and transparency (FAccT) research.
We note that not every harm we discuss here has been documented in the wild; we provide citations for as many as possible, but ensuring that information access is equitable requires researchers and developers to proactively engage with possible harms, not only the ones already known.
The ACM Code of Ethics \citep[\S 1.2]{ACM_Council2018-lv} states that ``avoiding harm begins with careful consideration of potential impacts on all those affected by decisions'', and it is our goal in this chapter to provide a framework to guide that consideration for fairness-related harms information access.

Fairness is also not a cleanly-defined set of problems with hard boundaries, but rather a lens that encompasses a range of concerns or (potential) harms, as we noted in \cref{sec:fairness}.
Several problems that information access researchers have long considered can be viewed as fairness problems; for example, work on popularity bias in recommender systems \citep{Celma2008-me, Zhao2013-ts, Canamares2018-ki} has the effect of trying to ensure that the system is fair to less-popular items.
There are also many problems, such as incomplete data, that are general problems for information access but take on a fairness dimension when data is missing in a way that disproportionately affects particular people or groups.
We take an expansive view of potential fairness problems in information access, with the aim of promoting a wide range of research and development that identifies and addresses inequity and injustice in information systems and their contexts.

The taxonomy we present in this \namecref{sec:space} is not hierarchical or orthogonal, but is rather a set of overlapping lenses or facets through which fair information access may be viewed.
The linear nature of writing forces us to impose a hierarchical structure on our treatment of the literature, but any classification of such a contestable problem space is necessarily imprecise and imperfect.
We believe, however, that this set of facets is a useful mechanism for understanding the existing literature and for positioning new developments in the broader problem space.
\Cref{tab:harm-summary} summarizes the dimensions we consider, which we will treat in the following sections, after first contrasting information access fairness with the traditional classifier fairness discussed in \cref{sec:fairness}.

\begin{table}[ptb]
\newcommand{\SLine}{\hskip 1em}
\caption[Summary of information access harms]{Summary of dimensions for describing information access harms with pointers to relevant sections where applicable.}
\label{tab:harm-summary}
\centering\small
\begin{tabular}{>{\raggedright\arraybackslash}p{0.7\textwidth}l}
\toprule
\textsf{Category of harm} \\
\SLine Representational harm & \ref{sec:space:misrepresentation}, \ref{sec:provider:representation} \\
\SLine Distributional harm & \ref{sec:consumer:providing-utility}, \ref{sec:provider:exposure} \\
\midrule
\textsf{Stakeholder group} & \ref{sec:space:sides} \\
\SLine Consumers & \ref{sec:space:consumers}, \ref{sec:consumer} \\
\SLine Providers & \ref{sec:space:providers}, \ref{sec:provider} \\
\SLine Subjects & \ref{sec:space:subject}, \ref{sec:provider:subject} \\
\SLine Platform or system \\
\SLine Multiple groups & \ref{sec:space:joint} \\
\midrule
\textsf{Specific types of bias harms} \\
\SLine Direct misrepresentation & \ref{sec:space:misrepresentation} \\
\SLine Unrepresentative list composition & \ref{sec:space:result-sets}, \ref{sec:provider:representation} \\
\SLine Unfair utility & \ref{sec:space:unfair-benefit}, \ref{sec:consumer:utility}, \ref{sec:provider:exposure} \\
\midrule
\textsf{Time scale} \\
\SLine Point-in-time & \textit{most sections} \\
\SLine Evolving over time & \ref{sec:space:time}, \ref{sec:time} \\
\midrule
\textsf{Sources of bias} & \ref{sec:space:pipeline} \\
\SLine Imbalanced user data & \ref{sec:consumer:utility} \\
\SLine Bias in user activity \\
\SLine Bias in user modeling \\
\SLine Bias in content production \\
\SLine Bias in item modeling \\
\SLine Bias in retrieval and ranking models \\
\SLine Bias in evaluation methods \\
\midrule
\textsf{Intervention points} \\
\SLine Data pre-processing \\
\SLine Adjusting models & \ref{sec:consumer:providing-utility}, \ref{sec:provider:rep-ensuring}, \ref{sec:provider:exp-ensuring} \\
\SLine Re-ranking & \ref{sec:consumer:providing-utility}, \ref{sec:provider:rep-ensuring}, \ref{sec:provider:exp-ensuring} \\
\SLine Software process & \ref{sec:consumer:providing-utility} \\
\bottomrule
\end{tabular}
\end{table}

\section{What Breaks in Information Access?}
\label{sec:space:breakdown}

\begin{table}[tbp]
\caption{Comparison between information access fairness and the setting typically assumed in classification-oriented fairness work.}
\label{tab:info-differences}
\centering \small
\begin{tabular}{>{\raggedright}p{0.45\textwidth}@{\hskip 1em}>{\raggedright\arraybackslash}p{0.45\textwidth}}
\textit{Classification}
& \textit{Information access} \\
\toprule
Independent, separate decisions 
& Ranked results, item decisions affect other items \\
\midrule
Subjects receive one decision
& Items subject to repeated decisions over time; users get multiple results \\
\midrule
Decisions independent of user
& Decisions personalized to user \\
\midrule
Target outcome construct independent of user
& Target outcome (relevance) subjective to user \\
\midrule
Data subjects need fair experience
& Multiple stakeholder classes may need fair experience \\
\bottomrule
\end{tabular}
\end{table}

\index{separate, simultaneous, and symmetric}
Unfortunately, the general algorithmic fairness constructs from \cref{sec:fairness}, developed primarily in the context of classification for decision-making, do not apply in a straightforward manner to information access for several reasons.
In many cases, this is because information access violates key assumptions of the existing classification fairness methods, particularly the assumption identified by \citet{Mitchell2020-mt} that the system makes and evaluates its decisions \emph{separately}, \emph{simultaneously}, and \emph{symmetrically}.
In that literature, a classifier is intended to support binary decisions, such as granting or denying a loan; we can try to translate this to information access by saying that the information access system is making decisions about whether and where to display an item in response to an information request.
Information access has several significant differences, summarized in \cref{tab:info-differences}.

\index{rivalrous|see {subtractable}}
\index{subtractable}
First, \textbf{decisions are not independent}, so they cannot be made or evaluated separately.
Information access systems typically rank their results: placing one "item" at the first position in a result list means no other document can occupy that position in that "ranking".
In any given news result list, only one article will appear in first position.
Even if individual items' "utility" estimates are independent of each other, they are ultimately resolved into a single ranking to present in response to a user's "information need".
This ranking may be multi-dimensional, as in a system that arranges rows or carousels of articles, but there is still an ordering and limited opportunity for an article to be in the highest-attention position.

Further, this ranking (or sometimes a set) can be produced in non-obvious ways that depend on more than just relevance.
For example, the principle of \emph{maximal marginal relevance} \citep[MMR; ][]{Carbonell1998-qk} is used to diversify a result list and make it responsive to a range of possible interpretations of a "query".
Under MMR, once an item is placed in the first position, a second document that is very similar to the first --- and just as useful --- may lose the second-place slot in favor of a document that brings more diversity to the crucial early positions of the ranking.
That is, the utility of an item to an information need is \emph{dependent}.
Continuing the news example, once an article has been selected for the first position and its author given priority for potential readership, an MMR approach to topic diversification would likely pick an article on a different story for the second position; so not only does the first-position allocation exclude other authors from that position, it excludes other authors covering the same story from \emph{subsequent} positions in the ranking as well.
We can model result list positions as subtractable or rivalrous goods that are allocated to different documents (and their providers), but this does not immediately resolve the problem; it simply identifies it.

Second, \textbf{decisions are repeated over time}, a violation of the simultaneous evaluation requirement.
Most existing fairness constructs (with the notable exception of literature on fair bandits or fair reinforcement learning) assume that all decisions are made at a single point in time, and do not attempt to account for the system learning and adapting future possible decisions.
Some recent work, such as that of \citet{DAmour2020-zd}, addresses the need for dynamic considerations of fairness but this has not yet made its way into widely-used fairness definitions.
In addition, the same items may be considered for possible ranking multiple times as users interact with the system.
If an item is not given a good position in response to the current information request, that does not preclude it from being given a good position in the next result set.
These changes can come because it is more relevant to the next query, user, and/or context; because the system is engineered to apply some randomness to result rankings even when a query is repeated \citep{Diaz2020-oa, Garcia-Soriano2021-gx}; because the system has learned more about its relevance through user interaction \citep{Glowacka2019-sh}; or some combination of all these.
The system may put one news article at the top of a user's home page, and a different article for a different user or even for the same user's next visit.
Thus, we can speak of fairness \textit{in expectation} based on a system's properties \citep{Diaz2020-oa}, or \emph{amortized} over multiple information requests and results \citep{Biega2018-zl}.
This deviation from the standard classification fairness setting provides significant opportunity to overcome the limitations of non-independent decisions, by providing more opportunity for documents to made visible, but it changes the analysis and design of approaches to ensuring fairness.

Third, \textbf{decisions are personalized to users}.
Returning to the lending example, a classification tool that estimates a potential borrower's risk of default should not return different risk scores or recommended decisions based on which loan officer is currently using the system.
Information access systems, however, are often personalized \citep{Liu2020-is}.
This is especially apparent for recommender systems, where most of the value proposition arises from modeling "users"' particular tastes and identifying products that match them, but many search engines also personalize to their users, as different users have different information needs (e.g. a programmer and a herpetologist likely have different primary interests when searching for ``python'') and different preferences for information sources to address a particular need.
The global implicit aspects of the information need associated with a user mean that two different users with the same explicit dimensions to their need may well be seeking different items.
This applies to many of our examples from \cref{sec:intro:running}; news platforms will tailor news recommendations to the user's interests, professional networking platforms will look for job listings that match a candidate's interests; and music services tailor recommendations and playlists to the user's tastes.
In practice, personalization relates to repeated decision-making, as we can return different documents to different users, but has some of its own distinct implications as well, such as the possibly of unfairness in representations of either user preferences or item characteristics.

Fourth, \textbf{outcomes are subjective}.
While there is a great deal of subjectivity in framing problems and measuring outcomes, typical fairness work assumes an environment in which true outcomes are, in some sense, knowable, at least at some future time.
It is assumed that a loan applicant will or will not repay the loan, and this repayment is independent of the bank employee evaluating their application (although it may, in practice, be affected by lender practices such as payment reminders and late payment policies).
In many information access contexts, however, the utility of an item usually has at least some degree of subjectivity to it: different users may disagree on the relevance of a document to a query, different users have different preferences for songs or appraisal of their quality, and so on.
The goal of the system in the ideal is not to model some external, objective notion of utility, but to model the relevance of the item to a particular user, in a particular context, with a particular query.
The same item may need to receive different scores or ranking decisions in response to different requests either from different users or the same user in different contexts.

Fifth, \textbf{multiple stakeholders have fairness concerns}.
While most problems have multiple classes of stakeholders --- the bank wants to make a net profit, the loan applicant wants a loan, and the community has an interest in residents having access to credit while maintaining low incidence of foreclosure-induced homelessness --- the applications considered in much work on algorithmic fairness have a clear stakeholder class for whom fairness is taken to be important.
We attempt to ensure that lending is fair to loan applicants, but do not spend effort on ideas of fairness to banks; likewise job applicants and the employer.
Information access, however, has multiple stakeholders with salient fairness concerns.
Two particular ones, which we introduced in \cref{sec:intro} and discuss in more detail in \cref{sec:space:sides}, are \emph{"consumers"} (the users of the system who consume information or products) and \emph{"providers"} (people or entities who create or provide the items consumed).
Research (and system deployment) accounting for fairness in information access needs to identify one or more stakeholder groups for whom it will consider fairness, and to clearly document this decision and its implications.
The exact position and fairness requirements of individuals may also change from product to product, even within the same domain or platform; for example, prospective employees are consumers and employers are providers in a recommender for job listings, but the roles are reversed for a job candidate search tool to for use by recruiters.  The same platform may well provide both of these types of recommendations.

These differences --- and likely others --- mean that it is important to study fairness in information access as a distinct problem in its own right.
The extensive work to date on fairness in other algorithmic settings has much to say that can and should inform this work, but we cannot expect methods or metrics from classifier fairness to directly and immediately apply to information access without adaptation to the particularities of information access problems.

\highlight{Machine learning fairness has much to teach us about how to frame, understand, and measure fairness, but the details often do not directly translate to information access problems.
Na\"ive application of classifier fairness constructs to information access often breaks down.}

\section{Kinds of Harms in Information Access}
\label{sec:space:harms}

There are a number of different ways an information access system can harm one or more of its stakeholders, particularly arising from its role as a mediator of users' access to information that may or may not meet their needs.

An unfair distribution of performance --- specifically one that favors over-represented populations --- can systematically hurt the retention of entire subpopulations of users.  An information access system that optimizes for mean performance can improve mean performance metrics (e.g. user satisfaction), at the expense of under-performance on under-represented groups, which are dominated by over-represented groups; metrics may also improve through the attrition of users from an under-represented group.  In simulation experiments, \citet{Hashimoto2018-jx} demonstrate that traditional empirical risk minimization results precisely in these dynamics.  In the context of two-sided recommendation, attrition may occur from content consumers or providers, potentially compounding any effects over time (i.e. attrition of providers can impact the size of the catalog and performance for consumers).

\index{accountability}
Increasingly, especially in protected domains like housing \citep{HUD-disparate-impact} and employment \citep{Raghavan2020-fl, Sanchez-Monedero2020-ai}, legal human rights regulation is being expanded to include algorithmic decision-making, like information access \citep{Wyden2019-nc,Thune2019-ki}.  As such, techniques for auditing and addressing systems for unfairness will become important from a legal perspective.

Journalistic investigation provides an alternative way in which algorithmic unfairness may be surfaced \citep{Diakopoulos2015-mw,Angwin2016-xi,Pelly2018-hc}.  These investigations can provoke regulation and hurt user perception and trust, potentially leading to attrition.

Besides the utilitarian effects on the information access provider, harms include a variety of social externalities.  Metrics such as user satisfaction or retention can ignore the broader impact of mediating information, especially news, which can affect social and political institutions.  While content consumers can be affected by mediation, the livelihood of content producers can particularly be at the whim of algorithmic decision-making, incentivizing classes of content more amenable to distribution than under-represented content.  

These harms suggest that information access system designers should understand the broader implications of the technology they produce.  Indeed, early in the development of information retrieval, \citet{Belkin1976-em} noted the ethical responsibility of information retrieval researchers to avoid political or economic manipulation.  Librarianship provides a professional discipline close to those designing information access systems and a code of ethics that ensures ``equitable services are provided for everyone whatever their age, citizenship, political belief, physical or mental ability, gender identity, heritage, education, income, immigration and asylum-seeking status, marital status, origin, race, religion or sexual orientation'' \citep{IFLA-ethics}.

\highlight{
There are multiple ethical, legal, and business reasons why information access system developers should consider fairness in their system design and evaluation.
}

\section{Fair for Who?}
\label{sec:space:sides}
 
\index{stakeholders|(}
While most of the history of information access research has concentrated on optimizing system performance for user outcomes (see, for example, the many rounds of TREC providing evaluations of IR systems' ability to retrieve relevant results, and the standard practice of assessing predictive or top-$N$  accuracy in recommender systems), there has been a growing acceptance in recent years that, in some contexts, information access systems serve multiple goals and possibly multiple parties, each of which is affected by the system's results and behavior. The integration of the perspectives of multiple parties into recommendation generation and evaluation is the goal underlying the sub-field of \textit{multistakeholder recommendation}~\citep{Abdollahpouri2020-survey, Mehrotra2018-fk}.

Many e-commerce sites operate as \intro{multisided platforms}, a business model analyzed in the economics literature by \citet{Rochet2003-zm} and \citet{Evans2011-zy}. One important finding is that different applications require different distributions of utility. In many multisided platforms, there is a `subsidy side' of the transaction where one set of parties uses the platform at a reduced cost or no cost. For example, users of the OpenTable restaurant reservation service do not directly pay for reservations; instead, restaurants pay for each reservation made~\citep{Evans2016-ah}.

In information access as well, the outcomes of the system may be biased towards one group of parties for similar reasons. In addition, the need for personalization may vary across systems and between stakeholders. For example, in an e-commerce site, product suppliers will usually not care about the characteristics of consumers: as long as products are surfaced to likely buyers --- either through recommendations or search results --- they will be satisfied with the behavior of the system. Online advertising is different: typically an ad campaign is targeted towards a particular audience, so a recommended ad placement is only considered successful if the ad matches the user's interests, to the extent they are known, and the user matches the target audience towards which the campaign is oriented.

Any system may have a multiplicity of stakeholders who are impacted by its decisions. In the case of fairness, the most salient ones will often be those noted above: consumers and providers. It may well be the case that these stakeholders individually have no interest in fairness; they may be primarily interested in the best outcomes for themselves. Fairness is usually a constraint that the system creators impose on outcomes, either to satisfy their own organizational mission or to meet the demands of yet other less-proximal stakeholders, such as government regulators or interest groups.  A few counterexamples do exist in the form of technologies to allow stakeholders to prevent certain kinds of unfairness even if the platform owner is accepting of unfair outcomes \citep{Nasr2020-td, Kulynych2020-nj}.

\subsection{Consumers}
\label{sec:space:consumers}
\index{consumer|(}
\intro[consumer fairness]{Fairness} towards "consumer" stakeholders may be grounded in different normative concerns: a goal of beneficence and the avoidance of various types of harms; a basic sense that the metric of system performance should include the broad distribution of its benefits; or, a more practical concern that unsatisfied users may go elsewhere.

We can characterize consumer fairness in various ways. The most straightforward is through quality of service, particularly in terms of error, user satisfaction proxies, or other performance metrics \citep[\cref{sec:consumer:utility};][]{Mehrotra2017-ns, Ekstrand2018-qm}. If particular classes of users less utility than others, whether measured in terms of prediction accuracy, ranking accuracy, or other measures, then we may say that the system is treating those users unfairly.

A system may also be unfair if its output is discriminatory in the content provided to different groups. For example, it is well-documented that real estate agents in the US have regularly steered minority home buyers to a limited set of neighborhoods; such incidents prompted the US Fair Housing Act, which explicitly disallows such activities~\citep{Rothstein2017-dc}. Not to run afoul of such laws, a real estate recommender would need to be sure that its lists of recommended properties contained a fair distribution of opportunities regardless of the buyer's minority status.
Thus, fairness may involve the specific content provided rather than differential performance on some error metric.
This issue can arise in many other domains as well, such as ensuring job seekers of different genders or ethnicities have access to comparable job listings; recommending lower-paying jobs to some groups of users than others would violate this principle, and in some jurisdictions may be illegal (case law is not yet settled on this point).
This phenomenon has been studied, for example, in Facebook's ad targeting platform \citep{Ali2019-nq}.

Finally, groups of consumers may be impacted if they have to incur disproportionate costs in order to use a system. Such costs might come in the form of information disclosure or effort. For example, a user with a disability may find they have to set up specific filtering rules to get the recommender system to provide acceptable hotel room recommendations and a able-bodied user does not, in spite of having provided the system with a similar amount of preferences or ratings.  Systems may perform better if users opt to share more data, or under-perform for users who are on low-quality connections and cannot provide as much data about their information needs (will note here that it is not necessarily possible to fix every potential fairness problem!).  It is also not clear that privacy-preserving recommender systems impose equal accuracy or quality costs on different groups of users \citep{Ekstrand2018-gw, Bagdasaryan2019-ib}.  Users who experience marginalization may also experience the tracking of activity and the generation of profiles that comes with personalized recommendation as considerably more threatening than others~\citep{Browne2015-sv, Burke2019-zh}.

These concepts of fair treatment do not necessarily correlate.
A system may look fair in that different groups of users receive comparable quality of service as measured by system effectiveness metrics, but perform poorly at presenting protected group users with the ``best'' inventory.
One specific way this can manifest is when using clicks to measure information satisfaction: users click on results presented to them, so the system may appear to be delivering satisfactory results, when in fact one user would prefer to receive the results another user is receiving.
Thus, a system designer looking to protect consumer fairness will need to think carefully about the types of harms user might experience and how to detect and measure them.

\index{consumer|)}
\subsection{Providers}
\label{sec:space:providers}

\index{provider|(}
There is a fundamental asymmetry between the various stakeholders participating in an information access system. Consumers come to the system to find information and are thus active in the information- or product-seeking process (`lean back' recommendation experiences aside).
They may be able to get multiple sets of results if they wish.
Content "providers", on the other hand, have a more passive role: their items are presented when and if appropriate users arrive, and they typically have little control over the recommendation or retrieval function. Despite this asymmetry, \intro[provider fairness]{fairness} concerns can arise in similar ways. Different groups may experience greater error when predictions of their items are made. For example, the system may systematically under-estimate user preference for books by minority authors \citep{Yao2017-vz}. 

More often, however, the concern will be about the exposure of items in results. 
This notion of provider fairness is concerned with how different providers, either individually or as members of protected groups, have their items appear (or not) in the rankings produced by a particular system. For example, in a search or recommendation tool to help recruiters find candidates for an opening, we want to ensure that the candidate lists it produces treat protected groups fairly \citep{Geyik2019-nh}. The personalized nature of recommendation (and many search systems), the fact that individual result lists are limited in size, and the rank-ordered nature of most information access systems means that we can only hope to achieve this kind of fairness over time and across multiple user visits \citep{Biega2018-zl, Diaz2020-oa}.
In some cases, a provider's items might be a poor fit to the users of a particular system. Consider a pianist on a job site whose primary user base is carpenters and electricians: there might not be many recruiters for whom such an applicant is relevant, and presenting their profile is likely unhelpful.
All of these considerations complicate the problem of measuring and ensuring provider fairness.

As we have used it so far, the term ``provider'' is a simplification of what can be quite complex systems of production and distribution that produce the documents or items in a system's inventory. The area of popular music is a good example, where the beneficiaries of the recommendation of a music track can be quite diverse, from the artist whose name is on the track, to the other musicians involved in the recording, the songwriter(s), the producer, the record label, etc. Fairness may not mean quite the same thing to each of these individuals; unfortunately, little data is currently available on how different provider groups perceive the fairness of recommender and information retrieval systems specifically. In one recent result, \citet{Ferraro2021-dm} report on interviews with musicians and their perceptions of fairness in music streaming platforms, with special attention on female artists. However, most literature focuses on the perceptions of recommendation consumers, and filling the gap to understand providers' experiences is an area in need of significant study.

There is significant conceptual overlap between provider-fairness and the diversity of search and recommendation results \citep{Ziegler2005-zo, Steck2018-st}. However, when diversity is invoked as a desirable property of an information access system, it is usually in the service of some user-oriented goal. For example, in information retrieval, query aspect diversity

Provider fairness in this sense also has a strong connection to ideas of fair allocation from welfare economics \citep{Moulin2004-zf, Thomson2016-jk}. The resource at issue is the opportunity for a provider's item to be made visible to a user, and the question is how to allocate that resource among various providers and what are appropriate desiderata. This is a cornerstone topic in social choice and has found practical application in a number of areas including allocating courses in schools \citep{Budish2012-us}, papers to reviewers \citep{Lian2018-gn}, and numerous other settings \citep{Roth2015-rl, Aziz2019-qr}.

\index{provider|)}

\subsection{Subjects}
\label{sec:space:subject}

Retrieved items can sometimes themselves be about individuals, which we call \intro{information subjects}. A well-known example arises in image search and recommendation: in many systems, image searches for terms like ``CEO'' turn up results that over-represent white men \citep{Metaxa2021-jj}. Female and non-white information subjects are not directly materially disadvantaged, but the results give a false impression that leads to the perpetuation of stereotypes \citep{Noble2018-vb} and possibly associated loss of opportunity. 
\cite{Karako2018-oz} examine this phenomenon and provide methods to address it, with a running example of seeking to provide a set of workout images that broadly represent the population.
Similarly, news recommendation may fail to give balanced coverage of issues affecting different groups, such as rural vs urban residents.

Another example of possible subject unfairness arises in medical information access.  If the studies returned in response to a query for current research on a medical condition a doctor is treating do not report on experiments whose subjects are not representative of the population --- or particularly do not include people sharing the patient's medically-relevant characteristics --- information may be missing for providing the best outcomes for the patient.
We are not aware of significant research on this particular potential problem, but proactive study of information access equity requires that we consider it.

Technically, subject fairness (being fair to information subjects) has a lot in common with both "provider fairness" and "diversity". In each case, it is the items over which representation is sought. However, in seeking subject fairness, it may be even more important that individual lists be diverse, as the goal of diverse representation is not necessarily satisfied by alternating between diverse and and non-diverse lists, something that might be acceptable in a provider fairness setting.

\subsection{Side Stakeholders}
\label{sec:space:side}

The concept of multistakeholder recommendation \citep{Abdollahpouri2020-survey} includes many stakeholders beyond those we have discussed, many of whom may have fairness concerns that a system should address. Indeed, in some settings, regulatory agencies may be the most important stakeholders for deciding the minimum legal standards with respect to fairness and / or non-discrimination that a system must meet. In other cases, the structure of a platform entails the participation of stakeholders who are neither consumers nor providers, but are still impacted by specific parameters of transactions on the platform. 

For example, consider a food delivery platform such as UberEats\footnote{\url{http://www.ubereats.com/}} discussed by \citet{Abdollahpouri2020-go}. This platform uses recommendations to match consumers with restaurants where they might order food to be delivered. The deliveries are made by Uber's drivers. There may be fairness concerns relative to the consumers or providers here, but there may also be fairness concerns over the set of drivers.
These individuals do not participate in the recommendation interaction but the recommendations may impact them.
For example, a goal might be to ensure that protected groups among the driver population do not receive fewer orders than others or do not receive a disproportionate number of difficult and/or low-tip jobs.

Another example arises in vehicle routing, which can be viewed as a kind of information access system where the items are routes or route segments.
The routing systems built in to mapping platforms such as Google Maps and Waze allow users to build routes for a variety of objectives, including ``beauty'' and avoiding heavily-traveled routes.  These systems can have the effect, however, of increasing traffic on side streets and through residential neighborhoods; residents have a significant stake in these kinds of changes to traffic patterns \citep{Johnson2017-qo, Fisher2022-mi}.

There has been comparatively little published work to date that considers such ``side stakeholders'' who are indirectly impacted by information access, but it is an important direction for future research.

\subsection{Joint Fairness}
\label{sec:space:joint}

So far, we have only considered each group of stakeholders in isolation.
This is the starting point for much machine learning fairness research, but it is a simplification.
In practical settings, multiple groups may experience harms and benefits through the actions of an information access system and therefore multiple simultaneous concerns may arise \citep{Mehrotra2018-fk}.

Information access systems are multisided platforms as noted above and therefore it is possible that the consumers of results and those provider side may each have fairness concerns. For example, consider a recommender system for rental apartment listings. Fairness concerns with respect to renters are well-established in housing anti-discrimination law; a system should not discriminate in the types of listings it provides on the basis of protected attributes like ethnicity or religion. But at the same time, there could be concerns relative to landlords. Hypothetically, if a system were found to be steering potential tenants with poor credit histories to properties owned by minority landlords and better prospects to other landlords, this would also be discriminatory.

Multiple fairness concerns can also arise on a single side of the information access interaction when there are multiple groups to consider. For example, a search engine may misrepresent both women and ethnic minorities in the results from image searches. The idea of fairness among multiple "protected groups" has seen some initial exploration under the label ``subgroup fairness'' \citep{Kearns2017-fo, Kearns2019-by, Foulds2020-cs}, but there is still much more to do. In particular, existing work as yet does not take into account the particular, compounded, challenges that may be encountered by individuals at the intersection of multiple protected categories~\citep{Cho2013-hn}.

\subsection{Cross-Group Harms}
\label{sec:space:xgroup}

In addition to unfairness for particular stakeholder groups directly harming that group's members, unfairness for one group may also harm other groups.

\index{job search}
If a system is provider- or subject-unfair, it may provide consumers with skewed perceptions of the space of content providers or subjects \citep{Noble2018-vb}.  A system that under-exposes job candidates from racial minorities may lead its users to believe such candidates are less common than they actually are.  It may also make it difficult for consumers to find content they particularly like, because such content is created by providers who are not well-represented by the system.

If a system is consumer-unfair, it may under-serve --- and thereby discourage --- a provider's primary audience, making it difficult for that provider's content to find an audience.

\highlight{
Measuring and countering unfairness in an information access system requires clearly identifying \textbf{who} is being considered in a particular evaluation or intervention.
Different groups have different concerns that will give rise to different metrics and techniques.
}

\index{stakeholders|)}

\section{Fair How?} 
\label{sec:space:how}

Another way of understanding fair information access is to look at \emph{how} different participants in the system may experience or be harmed by unfairness.
As with general algorithmic fairness, information access stakeholders may experience unfairness on \kl[individual fairness]{individual} or \kl[group fairness]{group} bases.

Classical welfare economics examines fairness in the form of distribution~\citep{Moulin2004-zf}: how to divide a resource fairly among individuals, all of whom have some claim to it. This type of fairness consideration can be considered \kl[distributional harm]{distributional}.
In the information access context, there are different multiple resources in question, depending on the stakeholder: providers and subjects receive exposure with its resulting material and reputational benefits, and consumers receive information that hopefully meets their information needs.

\citeauthor{Crawford2017-js}'s \kl{representational harms} also affect information access systems, when the system misrepresents a user, an item, or the information space.  We distinguish between a representational harm, where the harm itself is one of misrepresentation (e.g. misgendering a book author, or presenting results that discourage girls from seeing themselves as possible CEOs), and unfairness in the system's internal representations (e.g. user or product embedding spaces having a stereotyped component).  The latter may result in any of the kinds of harms we discuss in this section (either representational or distributional), or may be compensated for by other aspects of the system's behavior.

Sometimes harms, particularly with respect to legally-defined protected attributes, will be defined and proscribed by law; others will be a matter of policy for the designer of a particular system.  This breakdown is also not entirely crisp: some harms will fall under multiple categories simultaneously.  We submit that it nonetheless provides a useful way for understanding the ways in which discrimination and related problems manifest and harm the system's participants.

\subsection{Direct Misrepresentation}
\label{sec:space:misrepresentation}

An information access system can cause direct representational harms when it presents inaccurate information about items.

Direct misrepresentation of item or provider characteristics can harm both consumers and providers. Consumers are harmed because they obtain inaccurate information, and the misrepresentation may keep them from finding content through systems such as faceted browsing interfaces or detailed keyword searches.  Providers are harmed first because they are misrepresented, which can be harm in itself, but also may not have their product accurately discovered.  For example, if a children's book is not correctly labeled as such, then users may not find it when they are browsing or searching for children's books on a topic.  This specific harm is also an example of a multi-category harm, as it is also an unfair allocation of exposure to the book and its author and publisher.

Direct misrepresentation can itself be unfair (as in the example of misgendering) or it can happen in an unfair way (for example if some providers' content is more likely to be correctly represented than others in a systematic way).  It can also be either individual (if similar items do not have similar representation) or group (if socially-salient groups of items or producers are systematically misrepresented).  For example, as of 2020, a system using book author data from the Virtual Internet Authority File (VIAF) will exhibit group-based direct misrepresentation of transgender and non-binary authors, because multiple and non-binary gender identities are not accurately stored in the VIAF \citep{Ekstrand2021-iu}.

Misrepresentation can also harm broader sets of stakeholders.
For example, \citet{Nagel2021-bn} documents an instance of a search engine result page showing a carousel labeled ``Famous Cherokee Indians'' and displaying photos of several celebrities, most of whom have no documented Cherokee ancestry or affiliation.
This misrepresents the celebrities themselves, but likely does not cause them significant direct harm; its more substantial impact is misrepresenting what it means to be Cherokee --- and therefore the Cherokee nation --- to users of the search engine. 
This is also an example of a stakeholder that is impacted by an information access system but is not a producer, consumer, or subject of its items.
While the root cause of this problem is likely missing information in the underlying knowledge graph, it has an effect that compounds the difficulties already faced by indigenous communities; fairness-related problems can stem from any of the many issues in data or algorithms that affect other aspects of information access systems.

One additional potential harm that can arise from representation is that activating stereotyped perceptions a user may hold can affect their processing and assessment of information \citep{Bodenhausen1987-sj}; inaccurately representing producers or subjects, or even representing them accurately but unnecessarily, in a way that connects with users' negative stereotypes may impeded their ability to accurately and appropriately make use of the results the system provides, particularly in complex assessment situations.

\subsection{Unfair Result Set Composition}
\label{sec:space:result-sets}

An information access system can exhibit unfairness in the composition of its result sets and rankings.  This can have further downstream effects, as information access systems are often the first step in users' quests to gather information for other purposes.

One example is the previously-mentioned case of image search results for ``CEO'' \citep{Crawford2017-js}.  The set of results can affect users' perceptions of both the \emph{current state} and the \emph{possibilities} of the role of CEO (and it isn't immediately clear which one to emphasize if they differ --- should the gender representation in such a result set reflect the set of fortune 500 CEOs, all CEOs, or the general population?).  Further, if a student searching for ``CEO'' images for preparing a presentation, this not only affects who they see in the role of CEO, but affects the images available to them for communicating with their peers. This kind of downstream effect can arise in many settings, in education, business, and beyond. Any unfair representation that results in reinforcing stereotypes, either through the selection of items or explanations of results, may amplify --- or at least perpetuate --- societal biases.
\citet{Hoffmann2019-lg} argues that this kind of representation of what is ``normal'' has significant impact on how we understand and navigate the world and on the dignity of the people represented: in the situations we discuss in this monograph, the creators and subjects of information resources.
Such unfairness is also not limited to representation of the people involved; \citet{Raj2021-yg} notes that system results can also perpetuate gender stereotypes, which is of particular concern when used by children.

Unfair result sets can also arise through a personalized information access system's user profile, or its interactions with item representations.  A movie recommender may emphasize item relationships along gender-stereotyped lines, so that a user receives ``guy'' movie recommendations based on a few movies they've watched, instead of a set of recommendations more broadly reflective of their tastes.

Ways result sets can have unfair composition include \citep{Noble2018-vb}:

\begin{itemize}
    \item Reinforcing stereotypes (of users, content, providers, subjects, or any combination)
    \item Presenting an inaccurate picture of the information space
    \item Biasing users' sense of the possibilities of the information space
\end{itemize}

\subsection{Unfair Distribution of Benefits}
\label{sec:space:unfair-benefit}

Perhaps the most obvious way in which an information access system can be unfair is by being unfair or discriminatory in how it distributes benefits to its stakeholders.  Information access provides consumers with information that hopefully meets their needs, and providers with opportunities for their content to be discovered; this can have many repercussions, including financial (if access directly or indirectly results in revenue) and reputational (by their content being made broadly known), and can have material impact on providers' career prospects.

Distribution can be unfair at an individual level, if similar users do not receive similar quality of results, or if similar content does not have similar opportunity to be presented to users.
While similarity in the general case is often difficult to assess, in information access we often have some estimate (or even measurement) of content's relevance to an information need. Relevance assessments provide a useful basis for similarity-based evaluation of the distribution of opportunities for user attention to content providers is individually fair. If two content providers create documents that are comparably relevant to a particular query, and they do not receive comparable exposure in result lists or engagement from users, we may say that the system violates such an individual fairness objective \citep{Biega2018-zl}.

"Utility estimates" themselves may be unfair, if individual items with comparable characteristics with respect to their ability to meet the user's information need do not receive the same score. Such cases may require additional attention to ensure individual fairness.  Individually-fair distribution of result quality is, to some extent, already addressed by information access evaluations that consider the distribution or order statistics of quality and accuracy metrics, but differences in the rate at which attention and relevance drop off mean that static rankings generated by the Probability Ranking Principle are not necessarily fair, whereas stochastic ranking policies that attend to the distribution of exposure or attention can correct this discrepancy \citep{Diaz2020-oa}.

Group-based distributional discrimination may arise in different forms and for different stakeholders. On the consumer side, systematically underserving groups of users and failing to capture their perspectives or interests is a form of distributional group unfairness \citep{Mehrotra2017-ns, Ekstrand2018-qm}; we discuss this in \cref{sec:consumer:utility}.
It can also show up in less obvious ways, such as failing to properly interpret queries from children \citep{Dragovic2016-au}.

Provider-side group-based distributional unfairness has the clear manifestation of under-presenting results from particular, often disadvantaged, groups of content providers, such as authors who are members of ethnic or gender minorities.
\Cref{sec:provider} will cover this space in much more detail.

In addition to the kinds of protected and sensitive groups typically considered in fairness work, such as gender, religion, race, and ethnicity, information access applications bring additional sets of groups towards which we may want to ensure distributional fairness.  The work on \emph{cold start}\index{cold start} in recommender systems \citep{Schein2002-mk} can be framed as ensuring that new users and items are fairly treated. We may want to ensure new or independent authors, artists, or studios aren't crowded out by more established sources, reducing the ability of up and coming providers to thrive. Ensuring good quality across different languages or regions can also be beneficial in information access applications.

There are important distinctions between distributional fairness of different resources for different stakeholders that affect how we measure and provide fair distributions.
\index{subtractability}
One key distinction is the \textit{subtractibality} (or rivalrousness) of the resource in question \citep{Becker1995-zo}: does one person's use of the resource affect the ability of others to enjoy it?
For provider-side fairness, positions in result "rankings" are clearly subtractible: in any given ranking, only one item can be placed in the first position, and placing it there denies the position to other items.
The system may provide fairness overall by placing different items in the first position in different rankings, possibly even in response to the same information need, but any individual opportunity for user exposure (defined as a particular ranking position in an information access transaction) is a subtracible good.
\citet{Chakraborty2017-us} lean on this formulation by adapting existing algorithms for sharing limited resources --- specifically CPU time --- to help fairly allocate recommendation opportunities in sharing platforms.

On the consumer side, subtractibility is less clear.
System utility is typically not subtractible: one person obtaining high-quality, useful responses to their information need typically does not generally prevent others from receiving similarly relevant results.
Many classes of items are also non-subtractible, including digital content (web pages, streaming music, etc.) and plentiful physical goods where user demand is not likely to exhaust supply.
Other items, however, are subtractible: online auctions, for example, often have very limited stock, and only a small number of applicants will actually receive any particular job; while simply presenting the item to one user in a result list does not itself reduce the ability of others to benefit from it, successfully consuming the item does.
This is also a significant concern for reciprocal recommendation contexts such as those in matchmaking platforms for dating, mentorship, and other involved relational commitments \citep{Pizzato2010-ik}, as ``people have limited availability, so one person should not be recommended to too many others''.
\citet{Patro2020-ky} apply the concept to local business recommendations, which often have limited physical space that is further reduced by distancing requirements for public health; if too many people are recommended the same restaurant, they may not all be able to enjoy it safely, but spreading out the recommendations can help more people get to a restaurant that they can enjoy.
Whether or not it is necessary to reduce the resource distributed to some stakeholders in order to improve it for others is a key concern in describing the distributive harm that an information access system should avoid.

As much of the existing literature in fair information access focuses on distributional aspects of unfairness, we go into this point in much more detail in later chapters of this work (in particular, consumer distributional fairness in \cref{sec:consumer:utility} and provider fairness in \cref{sec:provider}).

\highlight{
Fairness in information access also requires careful attention to \textbf{how} the stakeholders may be harmed.
Again, different (potential) problems require different approaches.
}

\section{Fair from What Vantage Point?}
\label{sec:space:viewpoint}

Orthogonally to the stakeholder group, we can consider different vantage points for measuring fairness. As noted above, there is a range of different metrics and methodologies for measuring information access system performance and these give rise to different ways of thinking about fair outcomes. 

Since information systems often have an underlying predictive element, we can measure the accuracy of its predictions: does the system predict what items a user will like and how much? Accurate results deliver utility to users (helping them find items of interest) and also to item providers, because it means that their items are reaching appropriate targets. As an element of fairness, we might ask then if the system delivers different degrees of accuracy to different groups of users, serving some well and others poorly, and/or if the system differs in its prediction accuracy across protected groups.

There may be other aspects of information access performance of interest, depending on the recommendation application. In some settings, it may be possible to rank items on some objective scale of desirability. For example, credit card offers with lower interest rates and lower fees, and offer larger credit limits are better than those that charge higher interest and fees for less credit; all other things equal, a job with a higher salary is better than a lower-salary job.
The objective quality of the contents of a delivered results list is therefore an element of utility, especially for consumers. We may want to measure the comparative quality of such lists across protected and unprotected groups to identify possible discrimination. On the provider side as well, some products may be more profitable than others and unfairness may take the form of presenting low profit items for one group and high profit items for another.

In other cases, we may not distinguish between the qualities of items but rather their relative frequency of appearance on lists. A provider whose items appear very infrequently on result lists may feel that the system is being unfair in not promoting their products. This may also take the form of unfair representation as discussed above. From a consumer's point of view, differential distribution of item appearances may also have an element of unfairness: consider, for example, a recommender system that presents science toys on lists shown to boys but not on those shown to girls \citep{Raj2021-yg}.

\highlight{
Fairness concerns may go beyond the distribution of quantitative exposure or utility.
}

\section{Fair on What Time Scale?}
\label{sec:space:time}

\index{time}
Typical off-line evaluation methodologies for information access systems treat the test set over which outcomes are measured as if all the results are generated at the same time; this corresponds to the \textit{simultaneous} assumption articulated by \citet{Mitchell2020-mt}.
This tests how the system state induced by the training data produces results for all stakeholders at a point in time, but does not necessarily reflect how users, producers, and other stakeholders actually experience the system's effects.
\citet{Lathia2009-me} and others have looked at evaluation over time, and \citet{Sun2020-xn} advocate using time to split training and test data to provide a more accurate picture of system performance, but we have not yet seen significant applications of these concepts to studying fairness.

Since information access systems in practice deliver results lists to users who arrive at the system over time, it is important to examine fairness as a property of information access outcomes delivered over  time~\citep{Biega2018-zl}.
For example, we might ask whether a particular fairness metric has been achieved within a set of results delivered over some time window $\Delta t$. A system might look back over such an interval, determine whether a fairness metric has been met or not, and try to adapt its algorithm to deliver improvements over the next interval~\citep{Sonboli2020-kd}. 

\index{feedback loop}
Finally, the question of the dynamic nature of information access delivery leads to the question of the impact of results on user behavior, the impact of that behavior on subsequent system learning (the feedback loop). As a simple example, we can consider the impact of popularity bias, the fact that many recommendation algorithms reinforce the popularity distribution across items~\citep{Jannach2015-bi}. A popular item is recommended, then experienced and rated, appearing more popular in the data, and getting recommended more often, etc. While the unequal distribution of popularity is natural, this kind of positive feedback loop can exacerbate such distributions and associated unfairness \citep{Chaney2018-us}. 

\highlight{
The repeated nature of information access and its evolution over time, particularly as the system learns and updates its models in response to user interactions, means that point-in-time analysis is not sufficient to fully understand the fairness-related behavior of the system.
}

\section{Fairness and the System Pipeline}
\label{sec:space:pipeline}

In addition to the various dimensions along which we can define what it means for an information access system to be unfair, this unfairness can arise from a variety of sources.
\Cref{sec:access:architecture} and \cref{fig:info-access} described several components of a typical information access system; any of these components introduce unfairness to the system, and we can consider both evaluations and fairness interventions at many stages.
In most stages, unfairness can arise from the underlying data involved in that stage, the computational models used to make access-relevant inferences from that data, or both.

\textbf{"Item understanding"} can affect both producer fairness and the ability to correctly locate documents to meet an information need. Unfair representation of documents or categories may introduce representational harm or have downstream distributional effects. In a job search scenario, the document collection used to build the information system may have predominantly male candidates; if there are gendered aspects to how candidates present themselves in their documents, the system may learn gender correlations to job capabilities. Unfairness may arise from calculated metadata as well: for example, unfair inference of sentiment scores from review documents may introduce unfairness to the system at the classification stage.  Although it is impossible to have a holistic understanding of the items in a retrieval system, it is necessary to be intentional in mitigating potential unfairness and highlight possible limitations. 

Representation learning, and possible unfairness or discrimination in learned representations such as content embeddings, is one specific way in which item understanding may contribute to unfair outcomes from an information access system.  For example, the presence of racial bias in reviews \citep{Speer2017-ue} or gender bias in a job description can lead to an unfair outcome to different groups of stakeholders due to the bias in representation.
\citet{Bender2021-yp-parrot} provide an extensive discussion of the problems that can arise specifically when using language models for item understanding, a number of which touch on fairness.

\textbf{"User understanding"} most directly affects consumer fairness. The system needs to understand who its users are, their information needs, and their capabilities and preferences; the user modeling dimensions of information access particularly focus on this, as does the query understanding component of information retrieval.  The system does not necessarily learn these characteristics in an unbiased way.

\textbf{Retrieval} and \textbf{rendering}, often including ranking, are central to the observable output of the information access system.  In \cref{sec:provider} we will go into more detail on the problem of \emph{fair ranking}, which often connects to re-ranking ideas from information retrieval. Other aspects of rendering, such as result presentation, are less explored from a fairness perspective. Several biases may get introduced in the system purely from the user experience of the result page. Much more research is needed to understand how different design elements may introduce unfairness to otherwise fair retrieval results. 

\textbf{Behavior understanding} is how the system improves itself, either through automatic learning or feedback to system designers. However, as noted previously, that opens up the possibility of creating a feedback loop of that reinforces and amplifies any unfairness in the system. The varied nature of behavioral feedback between user groups may also affect the system's ability to accurately learn to provide relevant results towards different groups. Relevance feedback and click-through data are common ways of understanding user behavior and improving information access systems, but the data underlying them can easily be biased with respect to any of the stakeholders. Such feedback loops can easily amplify small differences, such as one item being slightly more relevant than another, into large differences in exposure or attention \citep{Ensign2018-bt}.

\textbf{"Evaluation"} of information access systems, as discussed in \cref{sec:access:eval}, is inherently different from the evaluation of classification systems. Where classification has a rather fixed notion of decisions and outcomes, evaluating retrieval systems relies on understanding the underlying user model \citep{Singh2018-zy, Biega2018-zl,Sapiezynski2019-qi}.  Evaluation based on biased relevance judgements or user response data can result in incorrect design decisions with fairness implications, or in selection of models or parameters that are unfair; evaluation is also a crucial place for assessing the fairness of an information access system.  One approach to this latter goal to evaluate fairness and relevance separately \citep{Yang2017-tu, Das2019-vo}. There is a scope for novel metrics that incorporate different notions of fairness as well as relevance.

\highlight{
Unfairness can arise from both data and models at any stage of the information access process.  Much research is needed to understand the role each plays in the overall fairness-related behavior and impacts of an information access system.
}

\section{Fairness and Other Concerns}
\label{sec:space:concept-relations}

The relationship between fairness and other concerns and concepts in information access is not straightforward.  As we have argued, some historical concerns can be framed as certain kinds of fairness problems: recommender system cold-start work, for example, seeks to ensure that new items and users are given a fair opportunity for exposure or quality from the recommender system, and long-tail recommendation looks to prevent the system from being unfairly biased towards popular items disproportionate to their utility to users.  Length normalization attempts to decrease systems' unfair preferential treatment of long documents \citep{Singhal2017-zm}.  While these do not deal with the kinds of socially-salient groups often considered in fairness research, they still represent the core logic of anti-discrimination: items should be retrieved based on their utility to the user's information need, not other incidental factors such as popularity, newness, or length (except when those factors are directly related to the need).

Beyond the desire to root out unwanted incidental biases, fairness research can be viewed as extending this logic from biases stemming from endogenous properties of items and their representation in the system to biases stemming from exogenous properties that relate to the broader social location of items, providers, users, and other entities.

Other concerns are related, and may have metrics and technical machinery that can be reused for fairness purposes, but flow from different normative concerns.  One frequently-mentioned example of this is "diversity": "provider fairness" and diversity look very similar, and systems providing more diverse search results or recommendations will probably often be more fair towards different providers.  However, they flow from different normative concerns and should therefore be assessed with metrics that reflect those concerns.  Diversifying techniques such as MMR \citep{Carbonell1998-qk} or xQuAD \citep{Santos2010-zy} can be used to improve fairness as done by \citet{Sonboli2020-nc}, but the results should be assessed on the basis of fairness.

Finally, some concerns may be in tension.  Work in both information access fairness and general ML fairness often discusses a tradeoff between fairness and accuracy.  \citet{Sonboli2022-tp} makes the terms of this potential tradeoff precise by framing it specifically as a tradeoff between fairness and accuracy in classical metrics used to measure recommendation accuracy in offline evaluations.  \citet{Wu2021-ex} treat this tradeoff itself as a fairness problem, arguing that decreasing consumer utility to improve fair provider exposure (\cref{sec:provider:exposure}) is unfair to consumers, but prioritizing consumer metrics with no consideration to equity of exposure is fair to providers (and they provide definitions and algorithms for navigating this multi-sided fairness).

However, the well-known disconnect between these offline metrics and online metrics of utility or user satisfaction \citep[see e.g.][]{Rossetti2016-wr, Kouki2020-nv} means that the relationship or tradeoff between fairness and utility may be very different than the relationship between fairness and offline accuracy.
Even in the offline setting, though, the tradeoff is not necessarily inherent, as \citet{Bigdeli2021-ru} show empirically through the existence of techniques improving both accuracy and fairness; \citet{Dutta2020-bo} argue that observed tradeoffs often arise due to bias in the data used to evaluate accuracy.
There is not currently enough research on the contours and limits of either fairness, accuracy, or utility to make definitive conclusions on their relationship in the general case.

\highlight{
Fairness may be in tension with accuracy or utility in some cases or experimental settings, but more research is needed to more fully understand and predict their relationship.  Fairness has significant overlap or complementarity to other concerns for information access, such as diversity and popularity bias.
}

\section{Contributing Back to ML Fairness}
\label{sec:space:facct}

In light of the significant differences between information access fairness and the fairness contexts typically studied to date, and the various challenges in the problem of ensuring fairness in information access systems, we also believe that information access has much to contribute back to the broader algorithmic fairness community.

One contribution is that the ways information access violates assumptions of classical fairness techniques (\cref{sec:space:breakdown}) can help make the limits of those techniques concrete.
We can point to specific applications where decisions can no longer be made and evaluated separately, or simultaneously, and study the implications of that violation for fairness metrics and methods.
Real life frequently violates these assumptions as well, but often on different time scales; while people apply for a relatively few jobs over their lifetime, information access systems make millions of decisions about how to rank items over short time periods.

Information retrieval and recommender systems also have well-understood data sets and strong community norms of demonstrating performance on benchmark data sets.
Many of these data sets are amenable to various forms of fairness analysis as well.
This data availability may enable the study of fairness concerns over different, and sometimes much more, data than are widely available for other applications.
The details will of course change when applying fairness research tested on these kinds of data to other applications, but information access may be a useful testing ground for mathematical or computational techniques with broader applicability.

Finally, information access represents a domain with substantial impact on lives, livelihoods, and perceptions of the world, but a very different impact than the often-studied domains of criminal justice or lending.
It may, therefore, present an opportunity to experiment with fairness measures, possibly even in real systems, where the human cost of getting it wrong or making the situation worse is quite     different.

While great care is required to avoid abstraction traps when attempting to translate from fair information access to other problem settings \citep{Selbst2019-hf}, we think there is much that information access has to say with implications for other domains.

\highlight{
Information access has the potential to improve fairness research more broadly, as lessons learned studying information access may be applicable in additional domains considered by fairness researchers as well.
}

\section{Navigating the Problem Space}
\label{sec:space:navigating}

Our goal with this section is to provide guidance to enable researchers, developers, students, and others building or affected by information access systems to consider potential harms, particularly harms related to fairness and discrimination, that can arise in information access.
We do not claim this treatment is complete, and problems do not necessarily fall cleanly into one problem or another; scholars may also disagree with some of our categorizations here.
Humanity is messy, and attempting to categorize the ways in which it can be harmed by technology is necessarily a messy and imprecise endeavor.
We find this framing useful, however, for organizing our own understanding of the subject, and will use it to organize our discussion in the remaining.
We submit that clearly describing and characterizing the harm(s) to be measured or mitigated in a particular work is more important than determining precisely which box it occupies in a rigid taxonomy, and we hope that our treatment provides a useful starting point for developing such clear descriptions.

In the rest of this monograph, we describe existing work and needed future research to address some of the harms cataloged in this section.
\chapter{Consumer Fairness}
\label{sec:consumer}

Having described the problem space of fairness in information access, we now turn to surveying the literature to date on various definitions, methods, and metrics for fairness, beginning with "consumer fairness".
As noted in \cref{sec:space}, information access systems are often in the position of mediating between providers of items or information and consumers of recommendations who are interested in those items. While fairness concerns may arise for any stakeholder in the system, these two groups have the most direct stake in the fairness properties of a recommender system and are the most widely-studied.
\Cref{tab:consumer-side} summarizes key work we cite in this \namecref{sec:consumer}.

\kl{Consumer fairness} is concerned with how an information access system impacts consumers and sub-groups of consumers, and whether those effects are fair or result in unjust harms.
For example, if a system is delivering recommended job postings to job seekers, it might be a fairness concern that different sub-groups of users, women, for example, could receive lower quality results than others.

\begin{table}[tb]
    \centering\footnotesize
    \begin{tabular}{p{0.32\textwidth}p{0.65\textwidth}}
    \toprule
    Measuring group fairness 
    & \citet{Yao2017-vz,Ekstrand2018-qm,Mehrotra2017-ns}.
    \\
    \midrule
    
    Enhancing group fairness
    &  \citet{Yao2017-vz,Abdollahpouri2020-go}.
    \\
    
    Recommended Items
    & \citet{Nasr2020-td, Kamishima2017-dx, Li2021-ci} \\
    
    Fair User Embeddings
    & \citet{Beutel2017-lx, Edwards2016-ex, Madras2018-at} \\
    
    \bottomrule
    \end{tabular}
    \caption{Summary of articles in consumer-side fairness.}
    \label{tab:consumer-side}
\end{table}

\section{Individual Fairness}
\label{sec:consumer:individual}

The distinction between group and \kl{individual fairness} is relevant for consumer fairness. As described in \cref{sec:fairness}, individual fairness considers how individuals are treated by the system and whether similar users have similar experiences or quality of service within the system.

One of the most basic outcome measures that can be applied is the accuracy of results produced. Typical evaluation measures such as recall or nDCG as described in \cref{sec:access:eval:simulated} can be used in offline experiments to determine the degree of accuracy that each user experiences in the system. While the central tendency of such measures form standard evaluation metrics for information access, the question of individual fairness calls for an examination of the distribution of utility across information requests, possibly marginalized to one dimension (such as user or query). In an extreme case, one might see a bimodal distribution of the evaluation metric, with some users getting accurate results and other quite inaccurate results. In such a case, the average performance is not capturing the user experience well; in particular, some users are being poorly served.

A form of evaluation that looks at the system's minimum performance would provide a form of corrective to this type of individual fairness problem. The ``fairness without demographics'' approach described by \citet{Hashimoto2018-jx} works to solve this problem by constraining performance for all users within a particular error region so that the discrepancy in accuracy across users can be controlled overall.  
We are not aware of work applying this form of individual fairness in information access systems.

Constraining the system's accuracy distribution is a rough form of individual fairness with a kind of Rawlsian logic. It amounts to an assertion that all users are similar to all others, and thus can be used to ensure that all users are getting some basic level of service from the system.  However, enforcing this type of fairness produces some challenges for information access systems, especially personalized ones. For example, User A with a small user profile (a \textit{cold start}\index{cold start} user) is generally expected to get less accurate recommendations than User B with a more extensive profile. It would not be a good solution to deliberately corrupt the recommendations for User B in order to equalize the accuracy of their results. 

It is possible to take a more textured view of individual fairness in keeping with ``similar users, similar results'' rubric. For example, we could control for profile size in comparing accuracy distributions, ensuring that we only compare the system's performance for User A against other cold-start users. Thus, profile size itself might become a source of unfairness, and this could well be be true of other features along which we might compare users to determine their similarity for the purposes of individual fairness. From this standpoint, it would be considered fair for fans of action movies to get similarly good results and devotees of documentaries to get similarly bad results from a recommender, an outcome that stretches what we might want a normative definition of fairness to provide.

\section{Group Fairness through Disaggregated Utility}
\label{sec:consumer:utility}

One major consumer-side "group fairness" problem is to determine whether the system provides comparable quality of service or utility to different groups of consumers, or whether there are groups --- especially protected groups --- whose information needs are systematically under-served by the system.
One way to assess this is by performing the same kind of utility-based "evaluation" that is usually used to evaluate the system's effectiveness, such as an offline accuracy evaluation or an online A/B test, and disaggregating utility by consumer groups.
That is, rather than computing an overall mean utility per user, computing average utility for each group.

In many cases, this utility can be operationalized through measures of the system's ability to meet information needs: click-throughs on search results or recommendations, ranking accuracy metrics such as nDCG or ERR, etc.
Consumer fairness studied in this way does not bring anything new to the problem of evaluating the system, except how the results are broken down and analyzed.  An overall metric $\bar\evaluationMetric = \frac{1}{n} \sum_\systemDecision \evaluationMetric(\systemDecision)$ is replaced by a per-group metric:

\begin{equation}
    \evaluationMetric_G = \frac{1}{n_G} \sum_{\systemDecision, \request : \implicitGlobal \in G} \evaluationMetric(\systemDecision)
\end{equation}

This metric can then be tested for significant between-group differences to assess whether some groups are experiencing better effectiveness (for whatever reason) than others.
It applies to both online and offline measures of effectiveness.

Some applications, as noted above, may have further dimensions of utility connected to objective qualities of an item.
For example, in a job opening recommender system, job listings have salaries.
If protected group users receive on average lower-salary listings, this could be considered unfair regardless of other personalization considerations or equal satisfaction metrics, depending on the goals and context of the application.

\citet{Mehrotra2017-ns} performed a  disaggregation of user satisfaction with a search system across multiple measures (graded utility, page clicks, query reformulations, and successful clicks), finding differences between user age groups and genders; the system was more effective for older users than younger users across all measures.
They further employed matching to control for query type and difficulty, to determine if differences in effectiveness were due to demographic differences in the queries issued; after context matching, both age and gender differences in satisfaction reduced almost to zero.

\citet{Yao2017-vz} focus on predicted rating fidelity, developing several unfairness metrics capturing different types of disparate prediction errors for protected and unprotected groups.
They measure overall disparate error as well as separately analyzing over- and under-predictions: does the system systematically under- (or over-) estimate some users' preference more than others?

\citet{Ekstrand2018-qm} disaggregated offline top-$N$ performance --- as measured by nDCG --- by age and (binary) gender for collaborative filtering algorithms trained on movie ratings (with the MovieLens 1M data set) and on music plays (with the Last.FM 1K and 360K data sets), finding statistically significant differences in utility between gender and (in some cases) age groups, although not always in the same direction.
They further showed that this difference was not explainable by differences in user profile size, and that resampling training data to have equal gender representation had the effect of substantially reducing cross-group utility differences.

\section{Disparate Effectiveness}

The studies by \citet{Mehrotra2017-ns} and \citet{Ekstrand2018-qm}  represent different approaches to --- and extents of --- understanding the reasons for observed differences.
Both begin with demonstrating the \textit{existence} of a disparity in group outcomes: some groups receive better quality of service, as measured by the result quality metric.
Such discrepancies bear similarity to disparate impact, in that there is a difference in outcomes for different groups, but differs in a crucial respect: unlike a typical classifier in the settings in which disparate impact is considered, an information system is not making decisions \textit{about} the consumers that differ by group.
We therefore refer to this kind of unfairness as \intro{disparate effectiveness}: the system is more or less effective (in this case capable of satisfying information needs) for different groups of users. Identifying this disparate effectiveness is relatively straightforward.
If there is a per-user or per-query measure of result utility, aggregating that over users' group membership and looking for (statistically significant) disparities detects the existence of disparate effectiveness.

That is only the beginning of studying fair utility, however.
Disparate effectiveness can be caused by biases in any portion of the information access pipeline (\cref{sec:access:pipeline,sec:space:pipeline}).
Narrowing down potential causes is crucial for identifying whether and how to address disparate effectiveness, particularly since --- as we will discuss shortly --- regularizing it away is often is not a desirable strategy.
There are several pathways that could give rise to disparate effectiveness:

\begin{itemize}
    \item The outcome measure may not satisfy \textbf{measurement invariance}\index{measurement invariance} with respect to consumer groups: that is, users in two different groups with the same subjective experience of satisfaction of their information need may still respond to the system in different ways, such that a behavior-based measure of satisfaction (such as clicks or session length) may measure satisfaction differently for them.
    \item The system's ability to model document relevance may depend on the \textbf{availability of training data}, and thus the system is not able to learn as effectively how to meet information needs distinct to minority groups of users.
    \item \textbf{Item relevance} to the needs of different groups may differ in a way that the model may hold constant across all users or groups of users, so minority users are forced to use a relevance model optimized for the majority group.
    \item There may be \textbf{mediating factors} between an information need and its satisfaction that systematically differ between groups that in turn affect the system's ability to deliver satisfactory responses.
\end{itemize}

\citet{Mehrotra2017-ns} address mediating factors through their context-matching design: by matching queries as closely as possible on multiple dimensions that affect their difficulty for the system, they are able to control for many of these factors.
The fact that this control eliminated most of the disparate effectiveness is evidence that groups' differing satisfaction is mostly a result of these mediating factors.
If a query has the same difficulty, groups tend to have the same satisfaction with the system's results for that query.
This does \textit{not} imply that observed differences are therefore not evidence of unfairness and do not need to be addressed; rather, it points to \textit{where} the differences are.
Younger users have lower satisfaction \textit{because} they issue queries that are more difficult for the system to satisfy.
Identifying the kinds of queries that present such difficulties, and are more frequently issued by under-served groups, provides a pointer to where engineering effort can be spent to improve quality of service for users currently receiving worse results.

\index{matching}
The matching design is very powerful for isolating these effects, as it allows for variables known to affect query difficulty to be held as constant as possible between groups.
It has the downside, however, of discarding a great deal of data (in this case, queries) that cannot be matched.
If the disparate effectiveness arises primarily from those queries that cannot be matched, a matching design may obscure a real inequity in quality of service.

\citet{Ekstrand2018-qm} targeted one specific mediator, profile size, with a linear model, and found that it did \textit{not} explain observed differences in recommendation quality.
They also investigated the impact of availability of training data, and found that it did have significant impact on the observed differences.
Downsampling is not necessarily an advisable approach in actual applications, because throwing away some of a group's data just because there is more of it is questionable machine learning engineering practice, but it is a useful strategy for narrowing down the causes of an observed discrepancy.

Neither of these approaches is strictly better than the other; they achieve different and complementary objectives through different means.
Studying and ensuring consumer group fairness requires a variety of tools, and the field is still so new that systematic understanding of the strengths and weaknesses of different methodologies has not yet been developed.

\section{Providing Fair Utility}
\label{sec:consumer:providing-utility}

Detecting and quantifying inequitable distributions of system utility is one thing; correcting them is another.
\citet{Yao2017-vz} introduce regularizers to remove discrepancies in rating prediction errors; regularizers can also be employed to target various other discrepancies.

Examples of post-processing approaches take the form of \kl{re-ranking} recommendations to improve their fairness properties. These approaches have typically focused on provider-side fairness, but \citet{Abdollahpouri2020-go} considers user groups based on their level of interest in popular items and shows that, across recommendation algorithms, users with an interest in less-popular niche items were not receiving recommendations in line with their interests.
\citeauthor{Abdollahpouri2020-go} also presented a re-ranking approach based on the idea of calibration \citep{Steck2018-st} to improve the fairness for these user groups.

Removing disparate effectiveness through an algorithmic intervention is not the obviously correct solution in many cases, however. Providing one user better results than another does not take anything away from the under-served user that they may otherwise have obtained and may not violate legal or ethical norms of fair treatment. This is different from classical fair decision making settings, where disparities such as a qualified borrower's chances of being approved for a loan differing on account of their race or religion are considered by many ethical and legal frameworks to be unacceptably discriminatory. It also differs from the provider fairness context, where giving one provider a recommendation slot \textit{ipso facto} prevents another provider from occupying that slot and obtaining the benefits thereof.
In particular, and in contrast to provider fairness, the well-served user's high-quality results are not usually the \textit{reason} the under-served user receives worse results, and decreasing their result quality in the name of fairness is itself arguably unfair.
Information access quality is not a rivalrous good\index{subtractable}, a fixed amount of which can be allocated across the users; we can improve experience for some users without hurting others at all.
We therefore recommend caution when using regularizers or other algorithmic techniques designed to simply reduce disparaties in system consumer-side effectiveness.

\index{process}
Another approach is to treat the inequity through engineering \textit{process}.
If an analysis of mediating factors identifies that an under-served group issues queries that are systematically more difficult to satisfy in an identifiable way, prioritizing efforts to improve the system's ability to handle those queries, instead of efforts that will primarily improve quality for users already receiving the system's best results, can address the inequity (and may improve service for the majority group as well).

\section{Fairness Beyond Accuracy}
\label{sec:consumer:beyond-acc}

While much of the work on consumer fairness focuses on the quality of recommendations, some work looks at other aspects of consumer experience that may be discriminatory, such as stereotyping or the specific items users receive.
\citet{Ali2019-nq} studied the distribution of ads on Facebook to understand potentially discriminatory impact in the visibility of different kinds of ads. They found that even when an advertiser wishes to have fair distribution of their ad, for example to ensure that an ad for a job opening is seen by people of all genders, the combination of relevance optimization and market dynamics results in disparate distribution of ads across racial and gender lines.
\citet{Nasr2020-td} describe bidding strategies to attempt mitigate such effects and ensure fair ad distribution even when the platform does not provide it.

More generally, \citet{Kamishima2017-dx} presented a probabilistic test for the independence of results from a user's (or item's) protected class.
Fairness, under their construct, is when the probability of a particular item being recommended is independent of the user's protected class.
They then incorporated this idea into a loss function for a matrix factorization collaborative filtering algorithm to optimize the system to produce independent results.
This can be useful in any context where users should not be recommended different types or sets of items on the basis of their group membership.
\citet{Li2021-ci} build on this independence objective in two ways: they allow fairness to be personalized, such that different users have different sensitive features they don't want affecting their recommendations; and they adopt a causal model to produce recommendations without causal pathways from the sensitive features to the recommendation lists.

Consumer fairness also extends beyond the items recommended, and can be applied to inner components of information. 
\citet{Beutel2017-lx} present an approach to learning fair representations in a way that can be applied to consumers, by learning embeddings (such as the user and item embeddings in a recommender system) in an adversarial setting set up to minimize the ability to predict a user's sensitive attribute, such as gender, from their embedding.
This has the potential to reduce stereotype effects in resulting recommendations, among other applications to both consumer- and provider-side fairness, although we have yet to see it deployed in this way.
Similar ideas are explored by \citet{Edwards2016-ex} and \citet{Madras2018-at}.

\section{More Complex Scenarios}
\label{sec:consumer:complex}

Most of the work to date on consumer fairness assumes rather limited group fairness settings.
In particular, it often assumes that only a single protected group, or a single dimension of sensitive attributes, needs to be considered for fairness; to the extent that they do consider multiple dimensions, these are considered separately (e.g. age and gender, but not combinations thereof).
But a job recommender, for example, may need to meet simultaneously meet constraints having to do with race, gender, religion, and other types of protected categories, as determined by applicable laws and organizational requirements. As noted above, the complexities of the interaction of multiple protected categories have been explicated by \citet{Crenshaw1989-km} and others under the framework of "intersectionality". In the fair machine learning literature, it has been studied under the topic of \emph{rich subgroup fairness} \citep{Kearns2019-by}. In recommender systems, there is some research involving subgroup fairness across providers~\citep{Sonboli2020-nc}. However, no existing work addresses the compound nature of the disadvantage that Crenshaw highlights as characteristic for individuals who find themselves at the intersection of multiple protected identities.

Another simplification in this model of consumer-side fairness is that it assumes categories are binary (protected vs unprotected) rather than constellations of attributes, including continuous qualities. It is possible that some existing models could be extended to handle continuous sensitive features (age or income, for example) but there is not any extant work in recommendation fairness along these lines as of this writing.
\chapter{Provider Fairness}
\label{sec:provider}

We now turn to the second of the two primary stakeholder typically considered in multistakeholder information access analyses: \kl{provider fairness} is concerned with fairness towards the "providers" (or \textit{producers}) of the content or items the system makes available to its users.
This primarily considers systems where content providers create and publish content that they want consumed, and for which they obtain some benefit from having users discover their content.
This may be a direct tangible benefit, such as subscription, advertising, or pay-per-play revenue; it may be indirect, such as the reputational benefits that accrue to journalists or academics for producing widely-read content; or it may be intangible benefits, such as the satisfaction of providing useful content to readers.
Under the definition of information access with which we opened \cref{sec:access}, this aspect of fairness considers the impact of the system on providers who gain utility from the system satisfying a user's information need with an item they provided.
\Cref{tab:provider-side} summarizes key papers we cite here.

\begin{table}[tb]
    \centering\footnotesize
    \begin{tabular}{p{0.32\textwidth}p{0.65\textwidth}}
    \toprule
    Measuring group representation 
    & 
    \citet{Das2019-vo,Zehlike2017-vj,Sapiezynski2019-qi,Deldjoo2019-ax,Yang2017-tu,Raj2020-om,Ekstrand2021-iu,Epps-Darling2020-eh} \\
    \midrule
    Enhancing group representation
    &  
    \citet{Celis2018-zh,Ekstrand2021-iu,Zehlike2017-vj,Garcia-Soriano2021-gx}  \\
    \midrule
    
    Measuring individual utility &
    \citet{Diaz2020-oa,Biega2018-zl} \\
    \midrule
    
    Measuring group utility &
    \citet{Biega2018-zl,Diaz2020-oa,Singh2018-zy} \\
    \midrule
    
    Enhancing group utility &
    \citet{Biega2018-zl,Diaz2020-oa,Singh2018-zy,Kamishima2018-nn,Burke2018-fm,Zhu2021-uu} \\
    \midrule
    
    Pairwise fairness &
    \citet{Beutel2019-mv,Narasimhan2020-hr} \\
    
    \bottomrule
    \end{tabular}
    \caption{Summary of articles in provider-side fairness.}
    \label{tab:provider-side}
\end{table}

These benefits are often abstracted under the notion of \textit{\kl{exposure}} (or \textit{attention}) \citep{Diaz2020-oa, Biega2018-zl}: an item, and therefore its provider, appears in result lists, and users have the opportunity to interact with these items.
We can treat result list opportunities as a resource, in which case the system is distributing these resources across the different providers (either individually or by groups), and we are concerned with whether or not that allocation is fair.
For example, in the job candidate search scenario, when an employer is looking for people to hire, different protected groups (e.g., gender and ethnic groups) should be treated fairly in terms of their members appearing in recommended candidate lists.
In music discovery, fairness would arguably require different artists whose work is equally relevant to a user's taste to have comparable exposure in their recommendations and streams.

In many scenarios, there are multiple parties who could be considered the provider of an item, at different levels and with different roles.
For example, in a news portal, both individual journalists and publication venues are providers of a news article.
In music recommendation, songs and albums have recording artists and record labels, as well as additional providers such as songwriters.
Movies and television shows typically have a long list of contributors who could be considered providers.
Studies of provider fairness typically focus on just one type of provider, but recognizing the diversity of provider relationships helps contextualize these concepts in the broader space of provider-side impacts of retrieval and recommendation.

Allocating utility is not the only way that an information access system can affect content providers, although it is the most commonly studied.
In this chapter, we will also discuss problems and research related to other ways information access systems may unfairly harm (or help) providers.

Finally, we note that "diversity" is often closely related to provider fairness --- indeed, one question we are often asked when presenting work on provider fairness is how this differs from diversity.
As noted in \cref{sec:space:concept-relations}, system modifications intended to enhance the diversity of results may be useful tools for improving the system's provider fairness, but diversity and fairness respond to different normative concerns and demand different metrics.
Diversity in recommendation and search results is mainly focused on consumer intent, intending to present results that meet a wide range of users' topical needs.
In contrast, provider fairness is motivated by justice concerns to ensure that different providers receive fair opportunity for their content or products to be discovered.

Our presentation here builds on the integration of provider fairness metrics for "rankings" provided by \citet{Raj2022-js}.
The literature to date differs in whether it establishes fairness \emph{metrics} or fairness \emph{constraints}; we here present the constructs in their original form, normalized for notation; constraints can often be converted to metrics, and \citeauthor{Raj2022-js} provide metric versions of some of these constraints.  \cite{Zehlike2022-wb} and \cite{Kuhlman2021-gk} provide additional comparisons and summaries of provider-fair ranking; in particular, \citeauthor{Zehlike2022-wb}'s treatment provides a particularly thorough discussion of the normative principles underlying different ranking metric decisions.

\begin{table}[tbp]
    \centering\scriptsize
    \begin{tabular}{lr}
Metric(s) & \\
\hspace{1.5em} Goal & Section(s) \\
\toprule
$\mathrm{PreF}_\Delta$ \citep[prefix fairness,][]{Yang2017-tu} \\
\hspace{1.5em} \textit{Each prefix representative of whole ranking} & \\
$\mathrm{AWRF}_\Delta$ \citep[attention-weighted rank fairness,][]{Sapiezynski2019-qi} \\
\hspace{1.5em} \textit{Weighted representation matches population} & 
\ref{sec:provider:rep-measure} \\
FAIR \citep{Zehlike2017-vj} \\
\hspace{1.5em} \textit{Each prefix matches target distribution} &
\ref{sec:provider:rep-measure} \\
\midrule
DP \citep[demographic parity,][]{Singh2018-zy} \\
\hspace{1.5em} \textit{Exposure equal across groups} & \ref{sec:provider:exp-group} \\
EUR \citep[exposed utility ratio,][orig. DTR]{Singh2018-zy} \\
\hspace{1.5em} \textit{Exposure proportional to relevance} &
\ref{sec:provider:exp-group}\\
RUR \citep[realized utility ratio,][orig. DIR]{Singh2018-zy} \\
\hspace{1.5em} \textit{Discounted gain proportional to relevance} &
\ref{sec:provider:exp-group}\\
IAA \citep[inequity of amortized attention,][]{Biega2018-zl} & \\
\hspace{1.5em} \textit{Exposure proportional to predicted relevance} &
\ref{sec:provider:exp-individual}, \ref{sec:provider:exp-group} \\
EEL, EER \citep[expected exposure \{loss, relevance\},][]{Diaz2020-oa} & \\
\hspace{1.5em} \textit{Exposure matches ideal (from relevance)} &
\ref{sec:provider:exp-individual}, \ref{sec:provider:exp-group}\\
EED \citep[expected exposure disparity,][]{Diaz2020-oa} & \\
\hspace{1.5em} \textit{Exposure well-distributed}  & 
\ref{sec:provider:exp-individual}, \ref{sec:provider:exp-group} \\
\midrule
Pair \citep{Beutel2017-lx, Narasimhan2020-hr} \\
\hspace{1.5em} \textit{Pairwise rank accuracy equal across groups} & \ref{sec:provider:pairwise} \\
\bottomrule
\end{tabular}
    \caption{Fair ranking constructs and their objectives. Adapted from \citet{Raj2022-js} by permission of the authors; metric names from that paper, except for Pair.}
    \label{tab:fair-rank-goals}
\end{table}

\section{Provider Representation}
\label{sec:provider:representation}

Many constructs for provider fairness are concerned in some way with \textit{representation}: are the providers of items returned representative of the broader population, or some other reference distribution of provider groups?
It is always a group fairness construct, as this kind of representativeness is meaningful to the extent that item providers are representative of their groups.
Unfortunately representation is an overloaded term; we are not concerned here with the internal or external representations of any individual provider, but rather with how the system represents the \textit{space} of providers to the user.

This kind of provider fairness can be concerned with a "representational harm", in that users who experience a skewed view of the population of item providers may develop (or have reinforced) an imbalanced view of who creates content; or it may be a proxy for a "distributional harm", as result lists in which particular provider groups are systematically under-represented are likely to result in unjust denial of exposure or utility to those groups.

These fairness constructs are usually independent of relevance: it is assumed that the lists in question are already optimized for utility, and their fairness is measured as a separate concern.

\subsection{Measuring Representation}
\label{sec:provider:rep-measure}

Provider group representation in result lists is typically operationalized through a distribution over provider groups.
There are therefore three components to a representation-based measurement of information access results:

\begin{itemize}
    \item A multinomial target distribution $\targetDist$ over provider groups $\groups$
    \item A distance function $\distance$ that computes the distance between two distributions over provider groups
    \item A means of computing group distributions $\distOfRank{\systemDecision}$ from the list and comparing them to the target distribution
\end{itemize}

When there are only two provider groups under consideration, such as a \kl[protected group]{protected} and unprotected group, the distributions reduce to binomials.

The simplest form of measuring provider representation is to compute a multinomial from a system decision $\systemDecision$ based on the number of times each item appears \citep{Das2019-vo}; for a single group $G \in \groups$, this is:

\begin{align*}
    \distOfRank\systemDecision(G) & \propto |\{\doc \in \systemDecision: \provider\doc \in G\}|
\end{align*}

Unfairness can then be defined using the distance $\distance(\distOfRank\systemDecision, \targetDist)$ (e.g., the Kullback-Leibler divergence, $\distance(\distOfRank\systemDecision, \targetDist) = \distance_\mathrm{KL}(\distOfRank\systemDecision \| \targetDist)$); a normalized (un)fairness metric can be computed by minimax-scaling the divergence \citep{Das2019-vo}.  Simple binomial fractions are also the family of metrics for which \citet{Kirnap2021-vf} developed sampling procedures for estimating with incomplete labels.

This approach is good for measuring overall list composition, but it does not take into account the relative visibility of different positions in the ranking.
Even when we are concerned with representation, not the distribution of utility, it is reasonable to expect the distribution among items at the top of the ranking to have a larger impact on the users' perception of the space than items further down the list.

There are two basic approaches in the literature to date for incorporating rank position into a representation-based fairness construct.
The first is to consider \textit{prefixes} of the ranking.
\citet{Zehlike2017-vj} presented a prefix-based approach for binomial group fairness that applies hypothesis tests to prefixes of the list of increasing length.
For each prefix $\prefix\systemDecision{k}$ (of length $k$) with $m$ items from the protected group, they compute the probability that a list of that length would have at most $m$ protected-group items under the null hypothesis that its item providers were independently drawn from the binomial target distribution ($\targetDist[\cdist](m|k)$, where $\targetDist[\cdist]$ is the cumulative distribution function of the target distribution).
If the null hypothesis is rejected (i.e. $\targetDist[\cdist](m|k) < \alpha$) for a prefix of the ranking, after correcting for multiple comparisons, the ranking is deemed to be unfair.
This ensures that a ranking cannot be considered fair unless provider groups are evenly represented throughout the ranking. 

Another approach is to employ a discount factor $\discountFactor(r)$ to down-weight representation further down the list.
\citet{Sapiezynski2019-qi} do this, so that:

\begin{align}
    \distOfRank\systemDecision(G) &
    \propto \sum_r \discountFactor(r) \indic(\provider{(\systemDecision_r)} \in G)
\end{align}

where $\indic(\cdot)$ is the ${0,1}$ indicator function.
This construct, called ``Attention-Weighted Rank Fairness'' (AWRF) by \citeauthor{Raj2022-js}, supports more than two groups, and can be extended to real-valued or mixed group membership weights $w(d,G)$, either to represent multiple group membership or uncertainty about the group alignment of an item.
If $\discountFactor$ forms a distribution such that $\discountFactor(r)$ is the probability of the user selecting the item at position $r$, then the distributions computed under this definition are the probability that the user will select an item provided by a member of a particular group.

As before, the distribution from the ranking can be compared to the target distribution using a suitable discount function; this can be Kullback-Leibler divergence \citep{Das2019-vo}, cross-entropy \citep{Deldjoo2019-ax}, or another suitable difference; in their study, \citet{Sapiezynski2019-qi} used the $Z$-approximation for the binomial test statistic.

So far, we have not discussed $\targetDist$: how do we determine the ideal target to which a ranking's group distribution should be compared?
Each of these approaches abstracts over this target; \citet{Sapiezynski2019-qi} call it the \textit{population estimator}, assuming that the goal is for the providers in a ranking to be representative of the broader population from which they are drawn; \citet{Deldjoo2019-ax} call it the \textit{fair distribution}, assuming we have some target we deem ``fair''.
The precise choice of target distribution will depend on the domain, application, and specific fairness goals.
Potential reasonable choices include:

\begin{itemize}
    \item Uniform \citep{Deldjoo2019-ax}
    \item The overall population of item providers
    \item The set of providers of items at least marginally relevant to the information need \citep{Yang2017-tu}
    \item An estimate of the distribution in society at large
\end{itemize}

\index{calibrated fairness}
\textit{Calibrated} fairness \citep{Steck2018-st} compares result lists to the user's past activity: under this definition, the group distribution in the user's reading, listening, or purchasing activity is used as $\targetDist$.
For example, this would consider a music recommender to be gender-fair if the mix of artist genders in each user's recommendations match the mix in their previous listening history.

There is not currently consensus, or even much study, of the relative strengths and weaknesses of these approaches.
\citet{Raj2022-js} and \citet{Kuhlman2021-gk} provide some direct comparisons, but papers typically measure a particular fairness construct without comparing with other metrics (aside from possibly variants on the same idea).
This isn't as bad as it seems, however, because construct validity --- measuring the intended fairness objective --- is a more important property for a fairness metric than consistency with prior results.

\subsection{Studies of Gender Representation}
\label{sec:provider:gender-studies}

Representation-oriented metrics have formed the backbone for studies of gender fairness in recommender systems that are focused primarily on understanding system behavior, not on developing fairness constructs.
Both of these studies operationalize gender fairness as the fraction of recommendations or interactions that are with items provided by women (``\

\citet{Ekstrand2021-iu}\footnote{An earlier version is provided by \citet{Ekstrand2018-um}.} studied this in the context of book recommendation, looking at gender representation in repositories, user reading or rating histories, and collaborative filtering recommendations across multiple book recommendation data sets.
Their work documents a large composite data set for studying fairness in book recommendation\footnote{\url{https://bookdata.piret.info}}, that is likely useful for many more search and recommendation studies, particularly ones looking at provider fairness.

\citeauthor{Ekstrand2021-iu} found that women were better-represented among the authors of books read or rated by users than they were in the Library of Congress catalog, and that users tended to read more men than women but were highly diffuse in their gender tendencies.
They employed a hierarchical Bayesian model to account for different user activity levels and produce smoothed estimates of users' gender biases, which they then used in a regression to examine whether collaborative filtering recommendation lists had gender balances that correlated with users' reading histories (\textit{calibrated fairness}\index{calibrated fairness}).
They found that collaborative filters did reflect users' biases with respect to author gender in their recommendation lists, although to different degrees.

\citet{Epps-Darling2020-eh} conducted a similar analysis of music listening activity on Spotify with respect to artist gender.
They found that male artists dominated streaming activity in both recommender-generated (``programmed'') and user-generated (``non-programmed'') activity; they also found, though, that increased prevalence of female artists in programmed streams was correlated with increased non-programmed listening activity for female artists.
They also looked for a difference in listening to women artists along user gender and age, but did not find an user demographic differences in the share of streaming activity that went to female artists.

In contrast to this lack of an interaction effect between artist gender and user demographics in music listening, \citet{Thelwall2019-ax} found in his analysis of GoodReads reviews and ratings that users are more likely to give high ratings to authors of their own gender.

While much of the work on provider-side fairness is concerned with defining metrics and optimization strategies, these studies provide more extended examples of studying the sources of bias and the propagation of such biases through standard recommendation algorithms.

\subsection{Ensuring Fair Representation}
\label{sec:provider:rep-ensuring}

Methods for ensuring fair representation often follow from the metrics that implement a fairness construct.
One common way to provide representational group fairness is through \kl{re-ranking}.
In binary-group settings, a greedy approach that selects the best item from the original ranking that does not violate the fairness constraint \citep{Celis2018-zh} or make representation worse \citep{Ekstrand2021-iu} can be effective.
\citet{Zehlike2017-vj} greedily process the list, picking the best item available (by the ranking's underlying relevance scores) if it would not make the protected group under-represented, and selecting the best protected group item if it is necessary to prevent under-representation.

These approaches, properly implemented, maintain \textit{in-group monotonicity} \citep{Zehlike2017-vj}: the order between items within a particular group is preserved in the fairness-enhanced re-ranking, and items are only reordered with respect to items from other groups.
\citeauthor{Zehlike2017-vj} further prove that their greedy approach results in the ranking with maximal overall utility subject to the binomial fairness constraint and in-group monotonicity, assuming the accuracy of the system's underlying utility estimates.
\citet{Ekstrand2021-iu} show empirically that greedy approaches need not result in substantial loss on utility-based evaluation metrics, at least in their experimental setting; \citet{Gomez2021-av} provide a similar result for geographic representation in MOOC course recommendations.

One concern often raised about group representation fairness is that majority-group providers of relevant content may be moved aside in order make room for the items needed to achieve group fairness goals.  Under some theories of equity, such as "anti-subordination", this is expected and acceptable. \citet{Garcia-Soriano2021-gx} address this concern by using randomness to ensure that it is not always the same items that are bumped aside, but that an individual fairness bound is preserved while meeting representative group fairness objectives.  Exposure, discussed in the next section, provides another perspective on relating group and individual fairness.

\section{Provider Exposure and Utility}
\label{sec:provider:exposure}

As noted at the beginning of this chapter, many provider fairness constructs are designed to ensure that providers have fair opportunity to realize the utility that arises from providing content responsive to users' information needs.
Even representational measures of provider fairness are often intended as a proxy for access to utility (see e.g. \citet{Ekstrand2018-um} and \citet{Sapiezynski2019-qi}). 

A more recent line of fair ranking constructs shifts this discussion in four important ways:

\begin{itemize}
    \item Assuming that measures of relevance produced by an information access system are good proxies for the value of an item to a user, such that the inclusion of a high-scoring item is worth more to the provider, as well as to the user.
    \item \AP Directly measuring \intro{exposure} (or \textit{attention}) as a resource that the system should distribute fairly among providers.
    \item Relating provider-side utility, abstracted through exposure, to consumer-side utility.
    \item Measuring fairness over repeated or stochastic rankings, rather than a fixed ranking in response to a single information need.
\end{itemize}

The first of these changes involves an aspect of the "WYSIWYG" assumption, namely that users' preferences, as filtered through the information access system and output as predicted utility or preference, are unbiased indicators of the value of a item. As opposed to the prior construct of representation, which assumes all list appearances have utility, the exposure construct considers utility to be a function of the match between user and item, as the system predicts it. This avoids one of the drawbacks of a purely representational approach: that the fairness metric can be satisfied with the inclusion of irrelevant protected group items, which are unlikely to attract user interest.

The second and fourth of these changes connect with the idea of "browsing models" used to evaluate information access systems (\kcref{browsing model}), and with common patterns in the presentation of result lists.
Because rankings as actually presented to users are often short, a single list contains only a small number of opportunities for exposure.
Further, because users are more likely to engage with items at the top of the ranking than the bottom, these slots are not of equal value: the first-ranked position (under most browsing models) provides more exposure to its occupant than the third or seventh position.

\index{stochastic ranking}
Therefore, provider utility is typically measured cumulatively (or in some cases amortized) over a sequence of recommendation results delivered to users, or as the expectation over a distribution of rankings defined by a stochastic ranking policy.
In important ways, it is often difficult --- if not impossible --- to fairly allocate exposure in a single ranking.
Considering fairness over sequences or distributions allows for a rich family of fairness constructs that are still achievable, at least in approximation. 
Note however that this kind of evaluation is only appropriate if aggregate utility over time is an appropriate scoring mechanism. Some fairness contexts, however, might still require that each generated ranking be fair with respect to protected groups: lists of job candidates in recruitment context are an example.

\subsection{Individually-Fair Exposure}
\label{sec:provider:exp-individual}

As noted in \cref{sec:fairness:individual}, the key idea of "individual fairness" is that similar individuals --- in this case, providers of items --- should be treated similarly \citep{Dwork2012-ai}; in exposure-oriented analyses of provider impacts of information access, that looks like receiving similar (opportunity for) exposure.
Information access's focus on utility or relevance to an information need provides a relatively natural basis for assessing similarity with respect to the task: two items are similar if they have similar relevance to the information need (either assessed by ground-truth relevance judgments or estimated by the system's relevance model).
Individual provider-side fairness is often computed at the item level, ensuring that "items" are treated fairly without aggregating to the provider level; in practice there is little difference between these concepts.

\citet{Diaz2020-oa} operationalize this by taking the \textit{expected exposure} over a stochastic ranking policy $\policy$.
Defining exposure based on a "discount model" $\discountFactor$, so that exposure $\eta(\doc|\systemDecision) = \discountFactor(\systemDecision^{-1}_\doc)$, the expected exposure for a item is:

\begin{align}
    \EE(\doc|\policy) & = \mdexpectover{\policy}{\eta(\doc|\systemDecision)}
    = \sum_{\systemDecision} \eta(\doc|\systemDecision) \fdprobfn_{\policy} (\systemDecision)
\end{align}

This exposure can then be compared to the exposure under a \textit{target policy} $\targetPolicy$; \citeauthor{Diaz2020-oa} used a policy that is uniform over all rankings that respect the relative relevance of documents to the information need as the target, and provide closed-form solutions for target exposure under two browsing models.
With a target policy, we can compute the \textit{expected exposure loss} as squared difference between actual and target exposure, over all documents:

\begin{align}
    \EEL(\policy) & =
    \sum_{\doc} \left(
    \EE(\doc|\policy) - \EE(\doc|\targetPolicy)
    \right)^2
\end{align}

If the policy $\policy$ distributes exposure comparably to the target policy, then the difference in exposure under the two different policies will be low, and thus the overall squared difference will be low.
This metric embodies the ``equal exposure'' principle: a fair ranking (policy) is one in which exposure is equally distributed among relevant documents.
Item exposure can be converted to provider exposure by aggregating over a provider's items.
Squaring just the expected exposure under the system policy yields \textit{expected exposure disparity} ($\EED$), a measure of how equally exposure is distributed among documents regardless of their relevance, a measure similar to Sapiezynski's discounted metric.

\citet{Biega2018-zl} similarly relate exposure to consumer-side utility by requiring a provider's exposure to be proportional to their utility, amortized over a sequence of rankings that may be in response to different queries:

\begin{align*}
    \CAttn(\doc) & = \sum_{\systemDecision} \discountFactor(\systemDecision^{-1}_d) \\
    \CRel(\doc) & = \sum_{\systemDecision,\request} \itemUtility(\doc|\request) \\
\end{align*}

Equitable attention is satisfied when $\CAttn(\doc) / \CRel(\doc) = c$ for all items $\doc$; \citeauthor{Biega2018-zl} quantified violations of this principle through the $L_1$ norm $\operatorname{IneqAttn} = \sum_{\doc} |\CAttn(\doc) - \CRel(\doc)|$.

One source of complexity in computing these metrics for a whole system, responding to multiple information requests, is determining how to aggregate over those requests.
\citet{Biega2018-zl} take the sum over all rankings, regardless of query, and do this sum \textit{before} comparing attention to relevance.
This results in a measure of the overall attention a document or provider receives from the system, taking into account the relative popularity of various information needs, which is useful for approximating provider utility when the goal is to ensure that providers obtain fair renumeration (e.g. ad clicks) for their production work.
Because attention and relevance are aggregated separately, however, a system can be fair by providing the correct exposure to items, but exposing them on the wrong queries.

\citet{Diaz2020-oa} go the other direction, and compute the metric over stochastic rankings in response to a single (likely repeated) information request.
This can be averaged over information requests, either with or without traffic-weighting (weighting a request by its relative frequency in the system logs).
Comparing actual and target exposure on a per-request basis binds an item's exposure to the information needs for which it is relevant, so the system cannot achieve fairness by exposing content in response to the wrong requests. That feature, however, makes this measure difficult to apply in a recommendation context where users' information needs are assumed to be personalization and more or less unique.

Neither of these metrics have a naturally-interpretable scale, and are not suitable for comparing across data sets or experimental settings; they are only effective for comparing the fairness of multiple systems on the same sequence or distribution of information requests.

\subsection{Group-Fair Exposure}
\label{sec:provider:exp-group}

These exposure concepts can be extended to "group fairness" by aggregating exposure over groups.
Both \citet{Biega2018-zl} and \citet{Diaz2020-oa} describe group-based aggregations of their amortized attention and expected exposure metrics; as presented, these consist of aggregating attention and relevance (for amortized attention) or exposure (for expected exposure) by provider group before computing the loss metric:

\begin{align*}
    \EE(G|\policy) & = \sum_{\doc: \provider\doc \in G} \EE(d|\policy) \\
    \CAttn_G & = \sum_{\doc: \provider\doc \in G} \sum_\systemDecision \discountFactor(\systemDecision^{-1}_d) \\
    \CRel_G & = \sum_{\doc: \provider\doc \in G}  \sum_{\systemDecision,\request} \itemUtility(\doc | \request)
\end{align*}

Unfairness can the be computed with the squared difference in groupwise exposure between system and target exposure, or absolute difference between group exposure and relevance.

\citet{Singh2018-zy} propose parity constraints and ratio-based metrics for fair exposure with respect to binary protected groups under stochastic rankings:

\begin{align}
    & \EE(G^+|\policy) = \EE(G^-|\policy)
    & \text{demographic parity} \\
    \operatorname{EUR}(\policy) & = \frac{\EE(G^+|\policy)/\CRel_{G^+}}{\EE(G^-|\policy)/\CRel_{G^-}}
    & \text{exposed utility ratio} \\
    \operatorname{RUR}(\policy) & = \frac{\mdexpectover{\policy}{\evaluationMetric(G^+|\systemDecision)}/\CRel_{G^+}}{\mdexpectover{\policy}{\evaluationMetric(G^-|\systemDecision)}/\CRel_{G^-}}
    & \text{realized utility ratio}
\end{align}

Demographic parity is a straight "statistical parity" constraint that ignores relevance and simply requires equal exposure; achieving this is equivalent to minimizing groupwise $\EED_\groups = \sum_{G\in\groups} \EE(G|\policy)^2$.
``Exposed utility ratio'' and ``realized utility ratio''\footnote{We use here the names provided by \citet{Raj2022-js}, as we believe they better reflect the general use of the terms "disparate treatment" and "disparate impact" than the original names of ``disparate treatment ratio'' and ``disparate impact ratio'' used by \citet{Singh2018-zy}.} realize a similar logic as equal exposure or equitable amortized attention: the system is fair if each group gets the exposure it merits by producing content relevant to users' information needs.
EUR is a direct ratio-based analogue to these metrics, while RUR incorporates the actual utility to the user in the numerator in addition to the overall utility in $\CRel_G$, through the use of the evaluation metric (a group-wise evaluation metric is the average of the metric value for documents provided by the group).
\citeauthor{Singh2018-zy} motivated this as an offline approximation of click-through rate, so that this metric is closer to the measuring the distribution of actual user \textit{engagement} instead of just exposure to users that \textit{may} lead to engagement.

These metrics all implement variants on the groupwise analog of the equal expected exposure principle: a system is provider group-fair if it distributes exposure to provider groups commensurate with their utility with respect to the users' information needs.
But as defined so far, they all have the drawback that they aggregate over groups \textit{before} computing whether the exposure is merited for a particular information request or not.
This is similar to the problem with aggregating an item's exposure and relevance separately across information requests and comparing total exposure to total relevance: not only may the system be able to achieve a good fairness score by exposing items to the wrong information requests, in group fairness it can be achieved by exposing the wrong items for a group.
So long as a group has some relevant items, and some items are exposed, there is nothing in the fairness metric (except for RUR) that requires that the relevant items are the ones exposed.
It can be achieved by randomly selecting items in a group-fair way with no attention to actual utility.
Combining it with a utility metric in a multi-objective analysis would help, but is a step backwards from the promise of exposure-based metrics to integrate fairness and utility directly.

One way to address this problem is to compute over- or under-exposure ($\EE(d|\policy) - \EE(d|\targetPolicy)$) on a item-by-item basis, and then take groupwise aggregates of this exposure difference.
The resulting metric will consider a system to be group-fair if no group's items are systematically under- or over-exposed more than another group's; this method has been adopted by the TREC 2022 Fair Ranking track.

\subsection{Ensuring Fair Exposure}
\label{sec:provider:exp-ensuring}

It is not sufficient to simply measure violations of fair exposure objectives; we often want to modify the system to provide results with greater fairness in their exposure.
We only outline the approaches here, referring the reader to the individual papers for details.

As with representational fairness constructs, \kl{re-ranking} can be a promising approach.
\citet{Biega2018-zl} describe a reranking stratgegy based on integer linear programming to ensure individual fairness of amortized attention.  \citet{Gomez2021-av} re-rank algorithms using minimal-cost swaps to reduce unfairness in both exposure and representation.
Given a stochastic policy $\policy$ represented as a doubly-stochastic matrix, \citet{Singh2018-zy} present a linear programming solution to produce stochastic rankings that satisfy their group fairness constraints.

\citet{Diaz2020-oa} directly use expected exposure loss as a learning-to-rank objective.
\citet{Singh2019-mc} similarly adopted a fair policy learning framework to learn stochastic ranking policies that fairly allocate exposure.
Their approach augments a standard utility maximization approach, that ensures the most relevant items receive the most exposure, with a lower-bounding inequality so that, for $\itemUtility(\doc_i) > \itemUtility(\doc_j)$, $\frac{\EE(\doc_i)}{\itemUtility(\doc_i)} \le \frac{\EE(\doc_j)}{\itemUtility(\doc_j)}$.
This ensures that while more relevant items get more exposure, the disparity in exposure doesn't outrun the difference in utility, thus addressing one of the drawbacks to "meritocratic fairness" \citep{Joseph2018-by} in which fairness can be achieved by giving the most relevant document all the exposure.

\citet{Kamishima2018-nn} present yet another approach to providing fair exposure, at least in a binary sense, to providers from different groups.
The provider-side element of their work formulates fair recommendation through statistical independence: $\fdprob{\doc \in \prefix\systemDecision{k} | \provider\doc \in G} = \fdprob{\doc \in \prefix\systemDecision{k}}$.
They then regularize the recommendation model to penalize violations of this independence objective.

Finally, \citet{Burke2018-fm} provide a more indirect approach that modifies neighborhood-based recommendation to ensure that neighborhoods are balanced between protected and unprotected groups, so that protected-group items have a good chance at being recommended.

Most techniques for provider fairness, including those described so far, focus on providing fair exposure in response to an established information need match: available query and document text for a search application or the normal steady-state case for recommendation.
\citet{Zhu2021-uu} examine provider fairnesss in \textit{cold-start} recommendation (recommending new items that do not yet have sufficient user interactions for typical collaborative filtering approaches to recommend them), and adjust the cold-start process to maximize the minimal exposure (expressed by discounted cumultative gain) of each new item to ensure that the system is fair to new items from different providers or provider groups.

\section{Fair Accuracy and Pairwise Fairness}
\label{sec:provider:pairwise}

Another way of conceptualizing provider fairness, that has very similar motivations to fair exposure but results in very different metrics, is to look at \textit{pairwise accuracy} as a basis for fairness.
As noted in Section~\ref{sec:access:ltr}, pairwise rank loss has long been used as a learning-to-rank objective for recommender systems \citep{Rendle2009-fo}.

\citet{Beutel2019-mv} and \citet{Narasimhan2020-hr} define fairness metrics based on pairwise accuracy. 
The key principle of these metrics is that a system is fair if it is not systematically more effective at correctly ordering relevant items from one group than it is from another --- that is, the probability that $\doc_+$ will be ranked above $\doc_-$ is conditionally independent of provider group given that $\doc_+$ has higher utility than $\doc_-$:

\begin{equation*}
    \fdprob{\doc_+ \succ_\systemDecision \doc_- | \itemUtility(\doc_+) > \itemUtility(\doc_-); \provider{\doc_+} \in G}
    = \fdprob{\doc_+ \succ_\systemDecision \doc_- | \itemUtility(\doc_+) > \itemUtility(\doc_-)}
\end{equation*}

In the case of a binary protected group, this can be further refined into \textit{intra-group} and \textit{inter-group} pairwise accuracy \citep{Beutel2019-mv}.  Intra-group requires that the protected and unprotected groups have the same pairwise accuracy for ordering items within the group (or, equivalently, each group has the same ROC AUC):

\begin{multline*}
    \fdprob{\doc_+ \succ_\systemDecision \doc_- | \itemUtility(\doc_+) > \itemUtility(\doc_-); \provider{\doc_+}, \provider{\doc_-} \in G^+} \\
    = \fdprob{\doc_+ \succ_\systemDecision \doc_- | \itemUtility(\doc_+) > \itemUtility(\doc_-); \provider{\doc_+}, \provider{\doc_-} \in G^+}
\end{multline*}

Satisfying this constraint ensures that the system is not more accurate at modeling relative preference for items created by one group than another; for age discrimination in job candidate search, for example, it would ensure that the system is not systematically more accurate at estimating the relative qualification of older candidates than younger ones.

Inter-group fairness requires that the groups have the same pairwise accuracy when compared with an item of the other group:

\begin{multline*}
    \fdprob{\doc_+ \succ_\systemDecision \doc_- | \itemUtility(\doc_+) > \itemUtility(\doc_-); \provider{\doc_+} \in G^+, \provider{\doc_-} \in G^-} \\
    = \fdprob{\doc_+ \succ_\systemDecision \doc_- | \itemUtility(\doc_+) > \itemUtility(\doc_-); \provider{\doc_+} \in G^+, \provider{\doc_-} \in G^-}
\end{multline*}

\citet{Beutel2019-mv} further extended these to leverage two-stage relevance feedback (e.g. clicking on an item, followed by a post-click signal of utility such as rating) to avoid simply optimizing to amplify click probabilities (a common signal for pairwise rank loss), and showed that group pairwise accuracy can be used as a regularization for a pairwise learning-to-rank algorithm like BPR \citep{Rendle2009-fo} to penalize group disparities in ranking accuracy.

Pairwise accuracy can be estimated by sampling and only requires group and utility data for the sampled items, as opposed to the exposure-based metrics which --- in their original form --- require data across the complete ranking.
This may make them more sample-efficient and/or easier to apply in partial data scenarios, but this potential benefit has not yet been well-explored.

\citet{Cui2021-qj} present a similar mechanism, re-ranking result lists to preserve within-group ordering and optimize AUC while balancing fairness and accuracy loss.

\section{Related Problem: Subject Fairness}
\label{sec:provider:subject}

As we noted in \cref{sec:space:subject}, "subject fairness" --- treating the subjects of items, such as the subjects of news articles or the population in a medical study --- has much in common with provider fairness, at least in terms of its structure.
Since subjects and providers are both entities associated with items, many of the same metrics and fairness mechanisms can be employed; the only change needed is the item attribute considered.
Subject fairness is also closely related to "diversity", even more so than provider fairness, as it aims to ensure that the results contain representation from a wide array of possible subjects.

However, depending on the information access context, subject fairness may require revisiting a key assumption behind many of the metrics discussed above, namely the assumption of cumulative utility. It may not be sufficient for subject fairness to allow fair results at time $t+1$ to compensate for unfair results at time $t$. For example, if we are concerned that image search results for ``CEO'' gives an unfair representation of the percentage of women in that position, we might not find it acceptable to mix 100\

Subject fairness introduces the challenge, also, of identifying the subjects.
While items are often annotated with their creators, they are not always annotated with the relevant aspects of their subjects, at least in a machine-readable manner.
More advanced content analysis techniques or extensive human annotation may be necessary to obtain the labels needed to pursue subject fairness, particularly for ensuring fairness to subject groups.

One sub-area of search that has seen significant work on subject fairness is in image search: \citet{Kay2015-gn} and \citet{Metaxa2021-jj} provide measurements and empirical techniques of gender and race biases by comparing representation in image search with estimated representation from the US Bureau of Labor Statistics; \citet{Singh2020-vc} provide another treatment. \citet{Otterbacher2017-dn} build on this with evaluation and design recommendations for improving such systems and their human effects. \citet{Karako2018-oz} provide a techniqe based on "maximum marginal relevance" (MMR), a diversification technique, for improving the diversity of a set of image results; \citet{Celis2020-tl} apply MMR and propose a new technique that does away with MMR's need to compare results with each other (improving the ranking effiency).  Further, \citeauthor{Celis2020-tl} use textual descriptions from off-the-shelf image summarization algorithms to improve diversity and subject fairness without needing explicit image labels.

Outside of image search, subject fairness is not extensively studied in the fair recsys and IR literature.
\citet{Rekabsaz2021-bj} provide one example, approaching subject fairness in retrieved text passages by measuring whether each retrieved document is neutral or unbalanced in its presentation of sensitive groups, and prioritizing the retrieval of neutral documents (ones that either do not include sensitive group information or are balanced in their representation of it, as determined by the relative frequency of group-related keywords).

Subject fairness is also implicated in many examples of search-related harms, such as the representational harms towards Black girls documented by \citet{Noble2018-vb}, but it has not yet received as much research attention --- that we are aware of --- in the research literature.

\chapter{Dynamic Fairness}
\label{sec:time}

In \cref{sec:consumer,sec:provider}, we have considered fairness for consumers and providers (and subjects) at a \textit{single point in time}: the current state of the system and its models should fare.  While stochastic policies act over time, the treatment in \cref{sec:provider:exposure} does not consider updates to the policy or changes to items or users.

Information access systems, however, operate in an iterated, changing environment.  They continuously make new decisions, gain fresh users, and lose established users. This dynamism is particularly salient in an application with high item churn such as news recommendation, where articles may be superseded by fresher stories in quick succession \citep{Karimi2018-ne}. But even when items have longer lifetimes, as in music, items and providers will come and go from the system over time. In addition, seasonal changes and longer-term trends means that historical profiles of users may lose utility over time. For example, a user who searches for an entry level job at one point in time may be looking for a different kind of position in the future. 

These dynamics are central to lines of research in recommender systems that consider the temporal aspects of markets and of user behavior \citep{Jambor2012-qz, Harman2014-pb, Campos2014-tg, Basilico2017-go, Zhang2020-qi}. Note that this area of research is distinct from work that treats the problem of recommendation as one that involves temporally-extended and dynamic learning behavior, as in multi-armed bandit or reinforcement learning formulations. In these settings, the recommender system changes its policy over time in a process of exploring user preferences and item qualities, but for the most part, the items and users are considered a static aspect of the environment over which learning takes place.

Understanding the dynamics of information access systems in general is a relatively recent project (although there are historical examples, such as that of \citet{Fleder2009-bi}), and ML fairness research is also only beginning to scratch the surface of dynamic fairness \citep{DAmour2020-zd}. The need to study information access fairness over time is clear, but there is so far relatively little work on it.
In this \namecref{sec:time} we provide pointers that researchers wishing to explore this vital topic.

\section{Feedback Loops}

\index{feedback loop|(}
The dynamics of recommendation contexts have also been considered in the context of recommender system fairness. One of the most troubling aspects of algorithmic bias generally is the potential for destructive positive feedback loops within the system~\citep{ONeil2017-bf}. Credit redlining provides an example. If a particular geographic area is determined to be too risky for lending, not only are current applicants impacted, but future ones as well. The system will not gather counterexamples that would help it identify the borrowers within the region that are actually good risks. 

\citet{Hashimoto2018-jx} study feedback loops in production systems from a fairness perspective.  The authors model a population of users iteratively engaging with a system that trains using behavioral data.  The model and supporting experiments in the context of predictive typing demonstrate that, over time, machine learning algorithms pay more attention to dominant subgroups of users as they lose under-represented subgroups of users.  The authors propose applying techniques from distributionally robust optimization to achieve more balanced performance, resulting in broad user retention.  \citet{Zhang2019-ue} extend this work by analyzing the dynamics of fairness in sequential decision-making.  Finally, in the context of predictive policing, \citet{Ensign2018-bt} theoretically demonstrate how to filter feedback to improve the fairness of decision-making systems learning from feedback loops.

A well-known effect in information access systems is that of various presentation-related biases \citep{Joachims2017-wz, Yue2010-hp}. User are more likely to experience and rate items that the system itself suggests, and their interaction may be affected by where and how it presents those results. This is by design: the system is presenting items that it regards as ones users will want to interact with. However, this bias can cause a form of positive feedback, in which presented items gain in popularity, leading to greater bias towards presenting them, at the expense of other items \citep{Chaney2018-us, Fleder2009-bi}. Positive feedback loops are inherently antithetical to fairness: they magnify small initial differences between item rating frequency into large ones as time goes on. It is also very difficult for new entrants to break into a market with positive feedback effects since they would have to gain traction against well-entrenched competition. Recommender systems therefore tend naturally towards unfairness, a tendency that a fairness-aware recommender system will need to continuously counter.

Feedback loops in recommender systems have been studied in a number of recent works. \citet{Chaney2018-us} examined the homogenization of recommendations in iterative environments. They find that recommendation systems, especially those based on machine learning, increase the consistency in recommendations across different users but also tend to increase the inequity of exposure across items. This phenomenon was termed ``bias amplification'' in work by  \citet{Mansoury2020-uh}. Similar effects were found an information retrieval context in \citep{Sun2018-nk}. Multi-agent simulation techniques were used to provide a theoretical basis for such findings in \citep{Jiang2019-mx}.  Some work in online learning contexts looks to rectify these biases and prevent bias amplification through the feedback loop; for example, \citet{Morik2020-tz} use separate fairness and utility estimators to improve group fairness in dynamic learning-to-rank settings.
\index{feedback loop|)}

\section{Dynamic Evaluation}
In its most basic form, accounting for recommender system dynamics means moving away from a recommendation experimentation model that has the form of batch training followed by batch testing. Instead we need to incorporate the cycle of user arrival, recommendation generation, user response, and periodic system re-training. Off-line evaluation takes on the character of simulation of a recommender's evolution over time. 

This type of evaluation has become standard in recommendation approaches that make use of reinforcement learning, in which the whole point is to develop algorithms that are compatible with a dynamic environment. \citet{Li2010-lt} and \citet{Zheng2018-ko} provide for more details about this algorithmic approach. By necessity, such training requires the application of \textit{off-policy evaluation} because ground-truth user responses will only be available for a small subset of the recommendations that could generated and yet we need to model these responses as training input.

In practical deployments where fairness is a concern, the appropriate form of evaluation might be to consider the system's fairness properties over some particular time interval and the evolution of its fairness through multiple evaluation cycles. However, we note that methodologies in this area are still emerging.

\section{Opportunities in Feedback Loops}

Feedback can also be harnessed to adjust system performance towards greater fairness. \citet{Sonboli2020-kd} present an adaptive recommendation approach to multidimensional fairness using probabilistic social choice to control subgroup fairness over time. In this model, deviations from fairness observed in a particular time window are addressed by adjusting the system's fairness objectives over the next batch of recommendations produced. 

\citet{Biega2018-zl} also account for time in their reranking strategy; while their algorithm does not directly use relevance feedback, it considers past rankings so that the ranking at time $t$ improves the aggregate fairness the system achieved up to time $t$.

\begin{oldcontent}
Analyzing the fairness of decisions at one point in time can overlook temporal patterns of equity.

A similar problem is well-known in recommender systems. Recommended items are more likely to be purchased, clicked on, or otherwise chosen by users. This is, after all, the intent of recommendation -- to provide shortcuts to items likely to be of interest to the user. Items that appear in such lists are therefore more likely to garner user feedback and are more likely to be recommended in the future. Thus, even small differences in the distribution of recommended items at an initial time can lead to large future distributional differences. 

\whatshard{Recommender systems are designed to incorporate user feedback and inherently prone to positive feedback effects that can lead to unfair outcomes.}
\end{oldcontent}

\chapter{Next Steps for Fair Information Access}
\label{sec:future}

Fair information access is a relatively new but rapidly growing corner of the research literature on information retrieval, recommender systems, and related topics.
The work in this space draws from concerns that have long been of interest to information access researchers, such as those motivating long-tail recommendation, the study of popularity bias, and examining system performance across a range of query types and difficulties, but connects it to the emerging field of algorithmic fairness and its roots in the broader literature on fairness and discrimination in general.

But while general literature on algorithmic fairness and fair machine learning is a crucial starting point, information access systems present particular problems and possibilities that make the straightforward application of existing concepts insufficient, as we have shown in \cref{sec:space:breakdown}.
In particular, the multisided nature and ranked outputs of many information access systems complicate the problem of assessing their fairness, as we must identify which stakeholders we are concerned with treating fairly and develop a definition of fairness that applies to repeated, ranked outputs, among other challenges.
The work of fair information access often requires data that is not commonly included with recommender systems or information retrieval data sets, particularly when seeking to ensure results are fair with respect to sensitive characteristics users or creators, such as their gender or ethnicity.
This work must also be done with great care and compassion to ensure that users and creators are treated with respect and dignity and to avoid various traps that result in overbroad or ungeneralizable claims.

We argue that there is nothing particularly new about these requirements, but that thinking about the fairness of information access brings to the surface issues that should be considered in all research and development.

\section{Directions for Future Research}

There are many open problems that need attention in fair information access.
Some of the ones we see include:

\begin{itemize}
    \item  \textit{Extending the concepts and methods of fair information access research to additional domains, applications, problem framings, and axes of fairness concerns.}
    Due to the specific and distinct ways in which social biases and discrimination manifest \citep{Selbst2019-hf}, we cannot assume that findings on one bias translate to another (e.g. findings on race may not apply to ethnicity or geographic location), or that findings on a particular bias in one application will translate (e.g. ethnic bias may manifest differently in recommendation vs. NLP classification tasks).
    Over time, generalizable principles may be discovered and give rise to theories that enable the prediction of particular biases and their manifestations, but at the present time we need to study a wide range of biases and applications to build the knowledge from which such principles may be derived.
    
    \item \textit{Deeper study of the development and evolution of biases over time.}
    Most work --- with the exception of fair policy learning and a handful of other studies --- focuses on one-shot batch evaluation of information access systems and their fairness.
    However, system behavior is dynamic over time as the system processes information requests, produces results, users respond to them, and the system learns from their feedback.
    This dynamicism means that an initially fair system may become unfair over time if users respond to it in a biased or discriminatory fashion, or that it may move towards a more fair state if users respond well to recommendations that increase overall fairness.  Tools such as T-RECS \citet{Lucherini2021-kz} may be valuable for such research.
    
    \item \textit{Define and study further fairness concerns beyond consumer and provider fairness.}
    We have identified subject fairness as one additional type of concern here, but we doubt it is the only additional stakeholder whose equity concerns should be considered.
    
    \item \textit{Study human desires for and response to fairness interventions in information access.}
    The first works are beginning to surface in this direction \citep{Smith2020-tf}, \citet{Harambam2019-ws} explored users' desired features and capabilities for recommendation with concerns that touch on fairness, and \citet{Ferraro2021-dm} studed provider perceptions, but at present little is known about what users or content providers expect from a system with respect to its fairness, or how users will respond to fairness-enhancing interventions in information access systems.
    
    \item \textit{Develop appropriate metrics for information access fairness, along with thorough understanding of the requirements and behavior of fairness metrics and best practices for applying them in practical situations.}
    For example, we believe expected exposure \citep{Diaz2020-oa} and pairwise fairness \citep{Beutel2019-mv} are useful frameworks for reasoning about many provider fairness concerns, but there is a much work left to do to understand how best to apply and interpret them in offline and online studies.
    
    \item \textit{Develop standards and best practices for information access data and model provenance.}
    \citet{Gebru2018-ml} presented the idea of \emph{datasheets} for data sets, arguing that data sets should be thoroughly and carefully documented so downstream users can properly assess their applicability, limitations, and the appropriateness of a proposed use.
    Information retrieval has a long history of careful attention to evaluation data through TREC, CLEF, and similar initiatives \citep{Voorhees2001-nm}, but until recently the evaluations in question have typically focused on overall effectiveness with some explorations of related issues such as diversity.
    Recommender systems has a substantial library of data sets, but has seen less attention to their careful documentation; \citet{movielens} provide a notable exception in their documentation of the MovieLens data set, addressing in advance several of the questions proposed by \citeauthor{Gebru2018-ml}.
    
    \citet{Mitchell2019-wz} built on this idea for reporting important properties of trained models, and \citet{Yang2018-qh} present a ``nutrition label'' for (non-personalized) rankings describing their data sources, ranking principles, and other information.
    These concepts need to be extended to information access, and to the complex integrated data sets that drive many search and recommendation applications.
    New research continues to discover that long-standing data management decisions, such as pruning \citep{Beel2019-li}, may have deep implications for experiment and recommendation outcomes, emphasizing the need for careful study of the properties of recommendation data, models, and outputs that should be documented.
    
    \item \textit{Engage more deeply with the multidimensional and complex nature of bias.}
    Most of the existing literature on fair information access --- and indeed all of algorithmic fairness --- focuses on single attributes in isolation, often restricting them to binary values.
    However, the intersection of group memberships often gives rise to particular forms of discrimination and social bias that cannot be explained by any one of the groups alone \citep{Crenshaw1989-km}.
    Some recent work begins to engage with multiple simultaneous axes of discrimination or fairness \citep{Yang2020-ki}, but as with many mathematical formulations of social concepts, multidimensionality does not fully capture the dynamics invoked by the concept of intersectionality \citep{Hoffmann2019-lg}.
    Further, many social categories are complex, unstable, and socially constructed, and algorithmic fairness is only just beginning to reckon with these complexities of human social experience.
    \citet{Hanna2020-ln} present a treatment of some of these issues in the context of algorithmic fairness, but much work remains to respond to that call and make fairness --- both generally for machine learning and specifically for information access --- responsive to these realities.
    
    \item \textit{Participatory design and research in information access.}
    Participatory design \citep{Schuler1993-lv} has a long history in human-computer interaction and user-centered design, but it is difficult to find examples of it applied to the design, evaluation, and study of modern, large-scale information access systems.
    \citet{Belkin1976-em} observe that ``it is necessary to establish and maintain an effective social relationship between [information] science and those whom it affects, so that the latter have a means of judging the implications of the former's activities''; this is true in general, but particularly for the concerns of this monograph.
    The field is accumulating many techniques for measuring and providing different kinds of fairness, but a serious understanding about what affected people actually want is currently wanting.
    One notable exception is the work of \citet{Harambam2019-ws}, who studied what Dutch news consumers want in terms of the control their news recommendation service provides.
    \citet{Smith2020-tf} and \citet{Sonboli2021-tu} studied users' opinions of fairness in recommendation, but but similar studies of producers, subjects, and other affected stakeholders are needed.
\end{itemize}

There is a lot of open space for research in fair information access, and this work has the potential for significant improvements to the human and societal impact of algorithmic systems for locating, retrieving, filtering, and ranking information.

\section{Recommendations for Studying Fairness}

Finally, we wish to leave our readers with some suggestions for how to approach research, study, and practice in fair information access, based on the work and concepts we have synthesized in this monograph.

\paragraph{Define the goal.}
Effective work on fair information access begins with a clear social goal: what specific fairness-related harms are to be avoided?
What is the legal, ethical, or other basis for understanding and defining those harms?
We hope our map of the space in \cref{sec:space} helps in that definitional work.
This is crucial for many reasons, but one is to avoid abstraction traps \citep{Selbst2019-hf} by keeping the work grounded in specific applications and risks; effective and appropriate generalization, in our opinion, flows from clear, contextualized findings.

\paragraph{Clearly operationalize the goal.}
With a specific harm in mind, select a metric that plausibly captures the kind of harm to be avoided.
Project writeups, whether as formal research papers or internal reports, need to clearly and specifically describe how (un)fairness and its resulting harms are being measured, and justify why it is an appropriate means of measuring the target concept.
The work we have cited in \cref{sec:consumer,sec:provider} provides examples of doing this for various fairness objectives.
\citet{Jacobs2021-ec} provde a more thorough treatment of the complexities of measuring subjective, contestible constructs like fairness.

\paragraph{Use appropriate data.}
Data is one of the major challenges for fairness research, in part because group fairness work often requires sensitive data that is often not collected with normal information retrieval or recommender systems data sets.
Some data sets provide group annotations, such as the data from the TREC Fair Ranking tracks \citep{Biega2020-gk} and certain older MovieLens data sets \citep{movielens}.
For some content creators, library data can be a source of author demographic information \citep{Ekstrand2021-iu}.
As noted in \cref{sec:intro:limits}, we advise against statistical inference techniques for annotating individual people; there has been work, however, on using background distributions to estimate metrics \citep{Kallus2020-mg}.

\paragraph{Carefully report limitations.}
Any research study has limitations, and fairness studies are no exception.
It is crucial to carefully and thoughtfully report the limitations of the data, metrics, and methods in order to help readers appropriately interpret and generalize the results.
Bracing honesty in discussing what any particular work can and cannot do is key to making true progress in this space.

\section{Concluding Remarks}

As we said at the outset (\cref{sec:intro:reading}), it is our hope that this monograph provides readers with an information access background who wish to learn about algorithmic fairness, and people grounded in algorithmic fairness and curious about what is happening on fairness in information retrieval, with a good starting point to understand the complexities, pitfalls, and possibilities in the rich and high-impact problem space of fair information access.
This field is still young; far too young to provide a comprehensive, retrospective treatment of its key ideas and findings.

What we have sought to do instead is to collect the work so far and integrate it into a prospective map of the space.
Much of this map is still incomplete, and the next years of research will fill in many details and likely unlock entirely new dimensions to consider.
We look forward to seeing the field grow and reading the many papers to come, and remember, please cite who we cite, not just us.

\printbibliography

\printindex

\appendix
\chapter{Resources for Fair Information Access}
\label{sec:resources}

In this appendix, we collect pointers to several resources for studying and working on fair information access.  We have made every effort to ensure these links are current as of the time of publication, but they may degrade more quickly than the references in the rest of the publication.

\section{Data Sets}
\label{sec:resources:data}

\begin{itemize}
    \item The TREC Fair Ranking track (launched in 2019) provides data sets for provider fairness in search rankings, both in academic search (2019--2020) and Wikipedia article search (2021).  The data is available in TREC (\url{https://trec.nist.gov/results.html}), with the track web site at \url{https://fair-trec.github.io}.
    
    \item The PIReT Book Data Tools at \url{https://bookdata.piret.info} provide tools to integrate book recommendation data sets (including from BookCrossing, Amazon, and GoodReads) with publicly-available book and author metadata to study provider fairness in book recommendation, as used by \citet{Ekstrand2021-iu}.
    
    \item \citet{Ghosh2021-ww} develop a number of data sets for fair ranking, using various methods and studying the errors of demographic inference for data augmentation.
\end{itemize}

\section{Software}
\label{sec:resources:software}

There are not yet widely-distributed open-source software for fair recommendation and retrieval; the available code is mostly embedded in published experiment scripts, or general-purpose systems repurposed for fair information access.

\begin{itemize}
    \item Terrier (\url{http://terrierteam.dcs.gla.ac.uk/research.html}) provides xQuAD, a diversification technique that has been successfully applied for fair search ranking \citep{Mcdonald2020-ek}.
    
    \item Experimental scripts are available for the fair recommendation studies of \citet{Ekstrand2021-iu} (\url{https://md.ekstrandom.net/pubs/bag-extended}) and \citet{Ekstrand2018-qm} (\url{https://md.ekstrandom.net/pubs/cool-kids}).
    
    \item librec-auto (\url{https://librec-auto.readthedocs.io/en/latest/}) provides automated support for running recommender systems experiments, including fairness metrics.
\end{itemize}

\end{document}